\RequirePackage{fix-cm}
\documentclass[smallcondensed,natbib,final]{svjour3}     
\smartqed  
\usepackage{graphicx}
\usepackage{color}
\usepackage{fancybox}
\usepackage[pagebackref=true,colorlinks=true,linkcolor=blue,citecolor=blue,urlcolor=blue]{hyperref}

\newcommand{\vv}{\vec{v}}

\newcommand{\beq}{\begin{equation}}
\newcommand{\eeqn}[1]{\label{#1}\end{equation}}

\newcommand{\greq}{\begin{equation}\left\{ \begin{array}{l}}
\newcommand{\egreqn}[1]{\end{array}\right. \label{#1}\end{equation}}
\newcommand{\disp}[1]{\displaystyle #1}
\newcommand{\dtau}[1]{\frac{\partial #1}{\partial \tau}}

\newcommand{\pr}{Pr}
\newcommand{\prm}{Pm}
\newcommand{\Ro}{Ro}
\newcommand{\vg}{{\vec g}}
\newcommand{\vu}{{\vec u}}

\newcommand{\na}{\vec{\nabla}}
\newcommand{\ez}{\vec{e}_z}

\newcommand{\re}{Re}

\newcommand{\RA}{{Ra}}

\journalname{Living Rev. Sol. Phys.}
\begin{document}

\title{The Sun's supergranulation}

\author{Fran\c{c}ois Rincon \and Michel Rieutord}
\institute{F. Rincon \and M. Rieutord\\
IRAP, Universit\'e de Toulouse,
14 Avenue Edouard Belin, F-31400 Toulouse,
France\\
IRAP, CNRS,
14 Avenue Edouard Belin, F-31400 Toulouse\\
\email{\{frincon,mrieutord\}@irap.omp.eu}}

\date{Received: date / Accepted: date}

\maketitle

\begin{abstract}
Supergranulation is a fluid-dynamical phenomenon 
taking place in the solar photosphere, primarily detected in the
form of a vigorous cellular flow pattern with a typical horizontal
scale of approximately 30--35~megameters, a dynamical evolution time of
24--48~h, a strong 300--400~m/s (rms) horizontal flow component and a much
weaker 20--30~m/s vertical component. Supergranulation was discovered
more than sixty years ago, however, explaining its physical origin and most
important observational characteristics has proven extremely challenging ever
since, as a result of the intrinsic multiscale, nonlinear dynamical
complexity of the problem concurring with strong observational and
computational limitations.  Key progress on this problem is now taking
place with the advent of 21st-century supercomputing
resources and the availability of global observations
of the dynamics of the solar surface with high spatial and temporal
resolutions. This article provides an exhaustive review of
observational, numerical and theoretical research on supergranulation,
and discusses the current status of our understanding of its origin
and dynamics, most importantly in terms of large-scale nonlinear
thermal convection, in the light of a selection of recent findings.
\end{abstract}

\setcounter{tocdepth}{3}
\tableofcontents

\section{Introduction}
The story of solar supergranulation started in Oxford in 1953 when Avril
B. Hart reported the existence of a ``noisy'' fluctuating velocity
field on top of the mean solar equatorial rotation speed that she was
attempting to measure \citep{hart54}. It is in fact very probable that
this ``noise'' was detected as early as 1915 by \cite{plaskett16}.
In a second article, \cite{hart56} confirmed her initial discovery and
gave an accurate estimate of 26~Megameters (Mm, or 1000~km) for
the typical horizontal length scale of these ``velocity fluctuations''
(sic). Supergranulation was subsequently recognised as a
characteristic, and essentially statistically steady dynamical feature
of the surface of the quiet Sun (the majority of the solar surface
characterized by relatively weak and statistically homogeneous
magnetic fields) after the seminal publication by \cite{leighton62} of
the first Doppler images of the solar surface (also resulting in the
first detection of the five-minute solar acoustic oscillations). This
work was soon supplemented by another paper by \cite{SL64}
establishing a strong spatial correlation between supergranulation and
the magnetic network of the quiet Sun. 

More than sixty years after its initial discovery, there remains a
significant theoretical uncertainty about how
supergranulation originates, what makes it particularly stand out
among other solar surface motions, what its exact interactions with
solar surface magnetic fields and solar rotation are, whether it
is a universal feature of solar type stars or of stellar surface
convection, and whether it plays a role in the local or global solar
dynamo process. As will become clear in this review, these
difficulties stem both from a combination of observational and numerical
limitations and from the intrinsic dynamical complexity of the problem. 
In recent years, however, the solar physics community has gained
unprecedented access to a large amount of high-resolution data
collected by space observatories, and to large supercomputers that
allow for increasingly realistic numerical simulations of the problem,
so that a clear resolution of these questions now look increasingly
possible. The main purpose of this document is to support this ongoing
effort with a large review of observational, theoretical, and
phenomenological research on this problem, with a
particular emphasis on some of the latest findings.

The review is divided into six parts including this introduction.
The next section provides some introductory material on the physics of
convection in the Sun and on the essence of the supergranulation problem
(Sect.~\ref{supergranulationpuzzle}). Section~\ref{obs} is
dedicated to observational results. Section~\ref{theory} 
reviews classical convection models and phenomenological
turbulence arguments historically put forward to explain the origin 
of supergranulation. Numerical simulations of the problem are reviewed
in Sect.~\ref{numerics}. Section~\ref{discussion}, finally, offers a
recap of the main observational results, a discussion of the physics
and dynamical phenomenology of supergranulation in the light of our
current theoretical, numerical and observational knowledge, and a brief
outlook on desirable and expected future research on the problem.

\section{The supergranulation puzzle}\label{supergranulationpuzzle}
\subsection{The dynamical landscape of the quiet Sun}

\subsubsection{The solar convection zone}
The outer 30\% in radius of the Sun are commonly referred to as the
solar convection zone
(SCZ, see, e.g., dedicated reviews by \cite{miesch05} and \cite{nordlund09}).
In this region, the solar luminosity (heat flux) originating in the
nuclear fusion reactions taking place in the core cannot be
evacuated by microscopic heat diffusion alone, but is instead
essentially transported by fluid motions driven by thermal buoyancy.
Physically, the strong non-adiabatic radiative cooling of the
surface layers, like the top cold plate of a convection experiment,
imposes a strong negative entropy gradient in the first
hundred kilometers below the surface. This gradient is likely
very small at larger depth due to the efficiency of the
convective mixing of entropy. Internal structure calculations
taking into account the ionisation properties of the gas 
and using local convective transport prescriptions suggest
that this gradient remains weakly superadiabatic down to the
depth of $\sim0.3\,R_\odot$, marking the transition with an internal
radiative zone. There is still some uncertainty about the
precise entropy stratification of the SCZ though, and, as we shall
discuss in detail in Sect.~\ref{simlargelocal2}, this question may be
an important piece of the supergranulation puzzle. Finally, the SCZ,
unlike the standard fluid Rayleigh--B\'enard convection system, is strongly
stratified in density: the density ratio between the bottom and top of
the SCZ is of the order of $10^6$ \citep[see][]{stixbook}.

With the notable exception of the solar differential
rotation, which is accurately determined by helioseismic inversions,
the internal dynamics of the SCZ is not very well constrained
observationally \citep{hanasoge12,gizon12,toomre15,greer_etal15}. While
there are many promising ongoing efforts to better characterize this
dynamics (some of them will be reviewed in Sect.~\ref{obs}), much
of the information we have still comes from observations of the photosphere
and chromosphere. Historically, it is important to remember that it
is this surface view which has defined the terms of the
supergranulation problem.
 
\subsubsection{Granulation}\label{granule}
The most prominent, and best understood dynamical feature visible at
the surface of the quiet Sun is a photospheric intensity and cellular
flow pattern called granulation, which paves the entire surface
of the quiet Sun and is characterized by an intensity contrast of
around 15\%, typical horizontal length scales ranging from
$\sim$0.5~Mm to 2~Mm \citep{R_etal10}, typical velocities ranging from
0.5 to 1.5 km/s \citep{title89}, and a typical lifetime/renovation
time of 5 to 10 minutes. Granulation is driven by buoyancy and
radiative effects in the thermal boundary layer formed in the strongly
superadiabatic surface region of the SCZ where the solar plasma
becomes optically thin (see Fig.~\ref{figure:plume_conv}). The
dynamical features of granulation are well reproduced by numerical
simulations of radiative hydrodynamics \citep{stein98,nordlund09}. 

\subsubsection{Supergranulation}
\begin{figure}[htb]
\centerline{\includegraphics[width=0.9\linewidth]{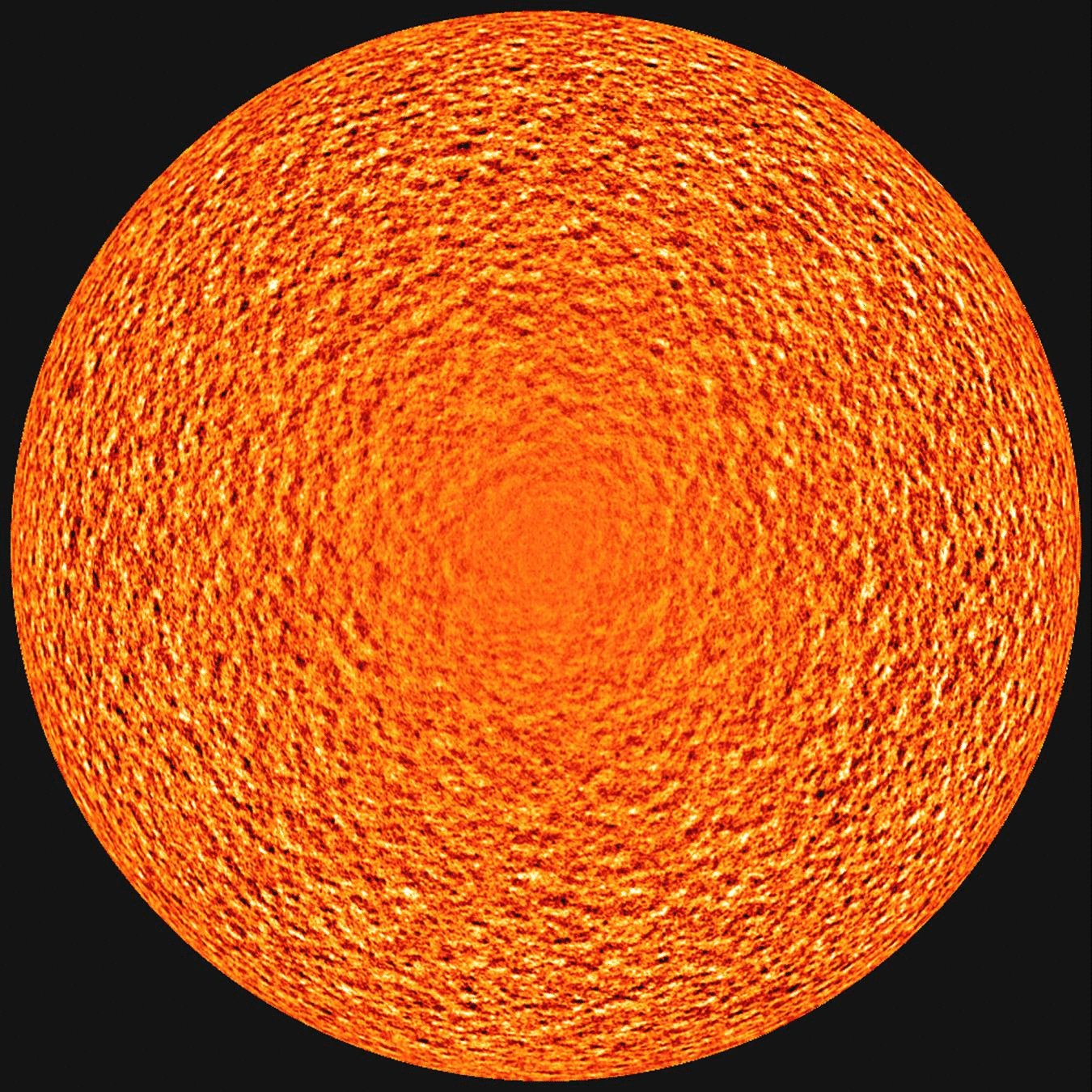}}
\caption{
  The supergranulation pattern as revealed by Doppler imaging of the
  full solar disc obtained with the MDI instrument onboard the SOHO
  satellite (image credits: SOHO/MDI/ESA).}
\label{figure:SG}
\end{figure}

As mentioned in the introduction, the existence of
supergranulation, a dynamical cellular flow pattern paving
the surface of the Sun with a typical horizontal scale of 30--35~Mm, was
first  established through Doppler measurements of solar surface
flows. This phenomenon is clearly illustrated by the Doppler image of
the solar disc shown in Fig.~\ref{figure:SG}. The much weaker
signal at the disc centre indicates that supergranulation flows
are essentially horizontal. The quest to understand what drives
these flows (and why they occur primarily at this particular
scale) is the central theme of this review. From a theoretical point
of view, this question may look
relatively easy at first glance: in most fluid systems, there are
usually just a few important physical processes at work, and as a
result only a small number of special scales can be formed from
dimensional and phenomenological
analysis. We will, however, shortly find out that, all ``simple''
known facts considered, supergranulation essentially resists a
straightfoward analysis of this kind. In particular, we will see that
while a thermal convective origin of supergranulation has long been,
and remains the most obvious and credible explanation, understanding
the exact nature and dynamics of the process in detail is extremely
challenging in practice.

\subsubsection{Mesogranulation}\label{obsmeso}
Mesogranulation has for a long time been thought to be a distinct
dynamical feature of solar surface convection taking place at 
intermediate horizontal scales between the granulation (1~Mm) and 
the supergranulation (30~Mm) scales. It was first reported  as such
by \cite{november81}, who identified a prominent pattern of vertical
motions with a typical horizontal scale of the order of 8~Mm in
time-averaged Doppler images, see also
\cite{november88,title89,chou92,ginet92}. However, several other
observational studies \citep{wang89meso,chou91,straus92}, including one
by \cite{hathaway2000} using high-resolution Doppler data from the SOHO
space observatory, found no evidence for a local maximum in the power
spectrum of solar convection (the scale-by-scale distribution of
energy) at such scales. 

While the topic remained controversial for a while
\citep{roudier99,rieutord2000,shine2000,lawrence2001}, a variety of
observational characterizations of the solar power spectrum derived from
high-resolutions, space-based observations with the SDO
observatory over periods of time of the order of 24h have now clearly
confirmed the lack of a distinctive spectral bump at mesogranulation
scales \citep{williams_pesnell11,hathaway_etal15,rincon_etal17}.
Observational analyses of magnetic field distributions at the
  solar surface  in the mesogranulation range also appear to support
  this conclusion \citep{yelles11,berrilli_etal14}.
Several authors had also previously made the case that
mesogranulation, understood as a singular dynamical feature, was
simply a ghost feature artificially generated by averaging procedures
\citep{matloch09,R_etal10}. As early as 1994, one of the authors of
the original mesogranulation study, \cite{november94}, wrote that the
term ``mesogranulation'' was misleading and instead suggested to
interpret this feature as ``the
vertical component of the supergranular convection'', while
\cite{straus97} argued that mesogranulation was a
mere powerful extension of granulation at large scales. 
Overall, it is clear that there is a lot of dynamics taking place 
at scales intermediate between granulation and supergranulation, but 
it does not seem that the dynamics at any particular scale in that
range particularly stands out. To avoid any possible misunderstanding,
we will, henceforth, refer to this range of scales as the mesoscale
range, without necessarily implying that there is anything physically
special about it.

\subsection{Physical scales in turbulent solar convection}\label{turbulentscales}
Having identified various dynamical phenomena taking place at the
top of the SCZ, is it now possible to come up with intuitive physical
explanations of their origin within the framework of thermal
convection theory? An important point here is that convection in the
SCZ is evidently strongly turbulent and nonlinear. Global
Reynolds numbers $\re=LV/\nu$, based on either the full vertical extent 
of the convective layer, or thermodynamic scales heights below the
surface, and on typical convective velocities estimated
from either surface measurements or mixing length theory, range from
$10^{10}$ to $10^{13}$ in the SCZ. To fix ideas, laboratory
experiments on turbulent convection are currently limited to $\re <
10^7$ \cite[e.g.,][]{Niemela_etal00}. We may then ask what kind of
scales can be baked in this kind of dynamical environment.

\subsubsection{Injection scale}
The injection range of a turbulent flow is the typical range of scales
at which kinetic energy is injected into the system by either a
natural or artificial forcing process. In turbulent convection, this
injection of kinetic energy is due to the work of the buoyancy
force. Within the framework of classical phenomenological turbulent
convection theories (the exact relevance of which for the problem at
hand will be discussed in Sect.~\ref{discussion}), the scale most
representative of the injection range is called the Bolgiano scale $L_B$
\citep{bolgiano59,oboukhov59,bolgiano62,lvov91,chilla93,rincon06}. It
can be shown, based on purely dimensional arguments and scaling
considerations for heat
transport in turbulent convection, to be almost always of the same
order (up to some order one prefactor) as the local typical scale height
\citep[][see also \cite{kumar14}]{rincon07}. In an incompressible
thermal convection
experiment, this corresponds to the distance between the hot and cold
plates, but in the strongly stratified SCZ, a more sensible estimate
is the local pressure or density scale height $H_p$ or H$_\rho$. Close
to the surface, the Bolgiano scale is, therefore, roughly of the order
of a Megameter, comparable to the typical scale of granulation
$L_\mathrm{G}$, but significantly smaller than the scale of
supergranulation $L_\mathrm{SG}$.

\begin{figure}[htb]
\centerline{\includegraphics[width=0.9\linewidth]{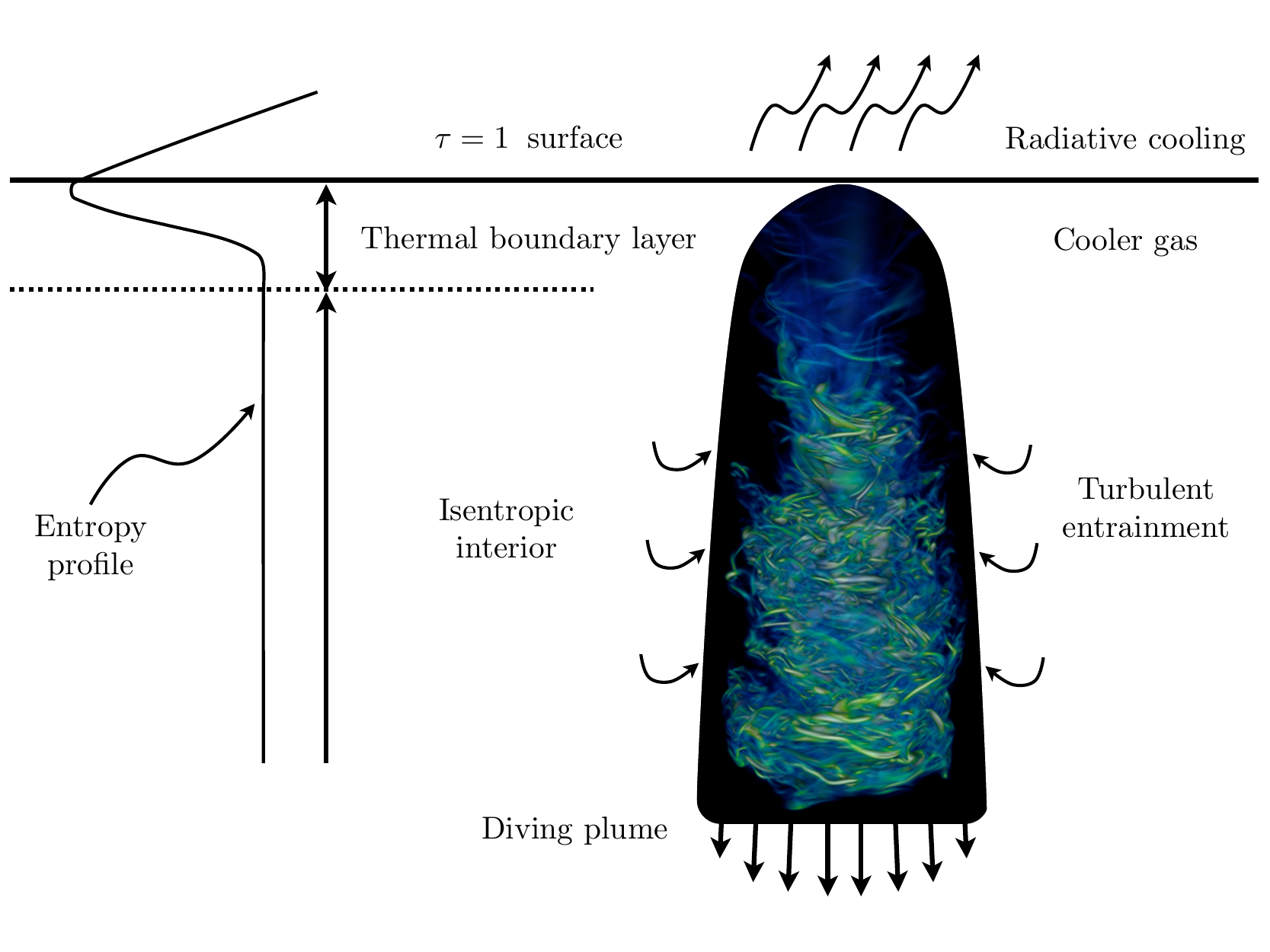}}
\caption{Left: the entropy profile as a
  function of depth, as estimated  by numerical simulations or
  one-dimensional structure models.
Right: numerical vizualisation a cool plume diving from the
surface. As it penetrates into the isentropic background, the plume
increases both its mass and momentum flux by turbulent entrainment
(represented by curly arrows) and expands horizontally. (image credits:
Mark Rast, see \citealt{clyne07}).
}
\label{figure:plume_conv}
\end{figure}

As we go deeper in the SCZ, the typical density and pressure scale
heights become larger and, therefore, so should the Bolgiano scale. 
As a result, the Bolgiano scale in the SCZ is expected to
increase with depth, ranging from 1~Mm close to the surface to 100~Mm
close at the bottom of the SCZ. From a physical point of view,
we can think of convection in the stratified SCZ as 
being driven by cold, low entropy plumes of fluid sinking from the
surface and expanding self-similarly throught the nearly isentropic
convection zone \citep{RZ95}. These large-scale plumes undergo
secondary instabilities along their descending trajectories, producing
a turbulent mixture of vorticity filaments \citep[see,
e.g.,][and Fig.~\ref{figure:plume_conv}]{rast98,clyne07,stein09}. 

\subsubsection{Dissipation scales}
As we are considering the turbulent motions of a non-ideal MHD fluid
with finite viscosity and finite magnetic and thermal diffusivities,
there are also three distinctive dissipation scales in the problem at
hand: the viscous dissipation scale $\ell_\nu$, the magnetic
dissipation scale $\ell_\eta$ and the thermal dissipation scale
$\ell_\kappa$. All of them are depth-dependent in the inhomogeneous SCZ. 

A rough estimate for the viscous dissipation
  scale $\ell_\nu$ can be obtained from the Kolmogorov phenomenology
  of turbulence \citep{frischbook} via the expression
  $\ell_\nu\sim\re^{-3/4}L$, where $L$ stands  for the injection scale
  and $\re=LV/\nu$, $V$ being the typical velocity at
  the injection scale.
In the SCZ, where $\nu\sim 10^{-3}$~m$^2$/s
\citep{R08}, we find $\ell_\nu\sim 10^{-3}$~m at the surface, 
taking $L\sim L_\mathrm{G}\sim1$~Mm and $V\sim 1$~km/s for the the
typical granulation scale and velocity. At the bottom of the SCZ,
where the injection
scale is much larger, one can estimate similarly that $\ell_\nu\sim
0.1$~m. Hence, $\ell_\nu$ is everywhere extremely small 
 compared  to the resolution of current observations and to the
 dynamical scales of interest.

 In MHD, the relative value of the magnetic dissipation scale
 $\ell_\eta$ with respect to the viscous scale  $\ell_\nu$ depends on
 the ordering of the dissipation coefficients of the fluid \citep[see
 e.g.][]{schekochihin07}. When the magnetic
  diffusivity $\eta$ is much larger than the kinematic
  viscosity $\nu$, as is the case in the Sun, one may use
  $\ell_\eta/\ell_\nu\sim \prm^{-3/4}$ \citep{moffatt61},
  where $\prm=\nu/\eta$ is the magnetic Prandtl
  number, and we have again assumed a Kolmogorov scaling for the
  velocity field.
The magnetic diffusivity in the subsurface layers of the Sun is
$\eta\sim 10^2$~m$^2$/s \citep{spruit74,R08}, hence $\prm\sim
10^{-5}$ and $\ell_\eta \sim 100$~m  close to the surface
  \citep[see also][]{graham09}.  This is also small compared 
  to the resolution of current observations and to the dynamical scales of
  interest.  Close to the bottom of the SCZ, $\prm\sim
  10^{-1}-10^{-2}$, so $\ell_\eta\sim 1$~m.

  Finally, we have  the thermal dissipation scale $\ell_\kappa$, which
  is the largest of the three dissipation scales in the solar
  context. In the SCZ, the thermal diffusivity $\kappa$ is everywhere
  much larger than the kinematic viscosity $\nu$, so the thermal
  Prandtl number $\pr=\nu/\kappa$ is very small. Under these
  conditions, we may estimate $\ell_\kappa$ from the expression
  $\ell_\kappa/\ell_\nu\sim \pr^{-3/4}$, once again assuming a
  Kolmogorov scaling for the velocity field.
  Thermal diffusion in the Sun is directly supported by photons and,
  thus, depends strongly on the opacity of the fluid and, therefore, on
  depth. In the deep SCZ, $\pr\sim 10^{-4}$--$10^{-6}$, so
  $\ell_\kappa\sim500$~m. In the optically thin surface layers, on the
  other hand, $\ell_\kappa$ is much larger, and comparable to the scale
  of granulation $L_\mathrm{G}\sim1$~Mm. As explained earlier, thermal
  cooling is a key ingredient of granulation. The typical granulation
  scale and flow turnover time can in fact be explained as a the
  result of a  balance between the advection time and cooling times of
  buoyant, high-entropy fluid particles. To summarise, the ordering of
  the characteristic spatial scales close to the solar surface is (a
  similar ordering holds for timescales):
$$
\ell_\nu \ll \ell_\eta \ll \ell_\kappa \sim H_p \sim H_\rho \sim L_{B}
\sim L_\mathrm{G} \ll L_{\mathrm{SG}}\ll R_\odot~.
$$

\subsection{What about supergranulation?}
While there is a clear coincidence between some of these scales and
the scale of granulation, we see that we cannot as easily construct 
a scale comparable to that of supergranulation at the solar surface
from the standard phenomenology of turbulent MHD convection.
Of course, we have argued that the injection scale of turbulent convection
should increase self-similarly with depth in the stratified SCZ. We
are, therefore, in principle allowed to speculate that supergranulation
could be associated with convective motions originating deeper into
the SCZ (say at $\sim 30$~Mm). But in that case we are still
left with the question of finding a process that would single out this
particular depth physically, among the available continuum of
injection scales.  At the minimum, we have to conclude that
understanding the origin, and determining what sets the scale of
supergranulation from phenomenological considerations, requires a
somewhat more sophisticated line of thinking than that outlined above.

Absent a straightforward answer (and given the historical difficulties
to simulate numerically the dynamics in the corresponding range of scale,
as will be explained in Sect.~\ref{numerics}), many different potential
clues and explanations, including the effects of changes in chemical
composition or ionisation, the effects of shear, rotation or/and
magnetic fields on convection, or nonlinear dynamical effects such as
inverse cascades, have long been sought through either observational
detection programmes or more or less rigorous theoretical proposals and
models. These different lines of research will be reviewed and
discussed at length in Sects.~\ref{obs} and \ref{theory},
but it is important at this stage to acknowledge that none of them
has led a comprehensive, predictive and falsifiable theoretical
explanation of supergranulation so far. On the other hand, we will see
that drastic improvements of both observational capacities and
computing power over the last ten years are now leading
to the emergence of new observational and numerical evidence strongly
supporting the idea that the supergranulation scale at the solar
surface is a special scale at which the dynamics is indeed first and
foremost driven by buoyancy forces. The possible physics
phenomenologies seemingly underlying these various results will be
discussed in detail in Sect.~\ref{discussion}.

\section{Observational characterization}\label{obs}
This section offers a wide review of the
observational characterization of supergranulation. After an
introduction of the principal methods of detection/inference of solar
surface flows (Sect.~\ref{obsmethods}), we review the numerous
observational characterization of the scale of supergranulation
(Sect.~\ref{obsvel}), measurements of supergranulation-scale
intensity variations (Sect.~\ref{obsintensity}), the
inferred depth of the pattern (\ref{obsdepth}) and its 
interactions with rotation (Sect.~\ref{obsrot}) and magnetic fields
(Sect.~\ref{obsmag}).

\subsection{Flow measurement methods}\label{obsmethods}
Supergranulation is first and foremost detected in the form of a flow
at the surface of the quiet Sun. Three methods are
currently used to measure the corresponding velocity field: Doppler
imaging, granule tracking and local helioseismology.

\subsubsection{Doppler imaging}
Doppler imaging is the oldest technique used to monitor
supergranulation (the first detection by \cite{hart54} was on a
Doppler signal). A SOHO/MDI Doppler view of supergranulation has
already been shown in the previous section in Fig.~\ref{figure:SG}. 
Dopplergrams only provide the line-of-sight component of the velocity
field. Therefore, except at the disc centre or at the solar limb, this
signal consists in a mixture of the horizontal and vertical velocity
field components. As already mentioned in the previous section, one clearly
notices that the supergranulation velocity field is mainly horizontal,
as the signal almost disappears near the disc centre. 
Doppler imaging of solar surface motions has been tremendously
developed since \cite{hart54}, and is a key component of many modern
space solar observatories, such as SOHO, SDO and Hinode.

\subsubsection{Correlation and structure tracking}\label{tracking}
Another way to measure the velocity field at the photospheric level 
is by tracking structures visible at the surface.
The idea is that small-scale structures such as granules (see
Sect.~\ref{granule} below) are simply advected by large-scale
flows. Three variations of this technique are used:
the local correlation tracking (LCT), the coherent structure
tracking (CST) and the ball-tracking (BT). LCT determines
the motion of features on an image by maximising the correlation
between small sub-images \citep{november88}.  CST identifies coherent
intensity structures in the image by a segmentation process and then measures
their displacement
\cite[e.g.,][]{roudier99,rieutord07,tkaczuk07}. BT follows 
the displacement of floating balls over the intensity surface of
images. The motion of the floating balls follows the mean motion
of granules; this is presumably more effective computationally
speaking than LCT and CST \citep{potts_etal04}.

\begin{figure}[htb]
\centerline{\includegraphics[width=0.86\linewidth]{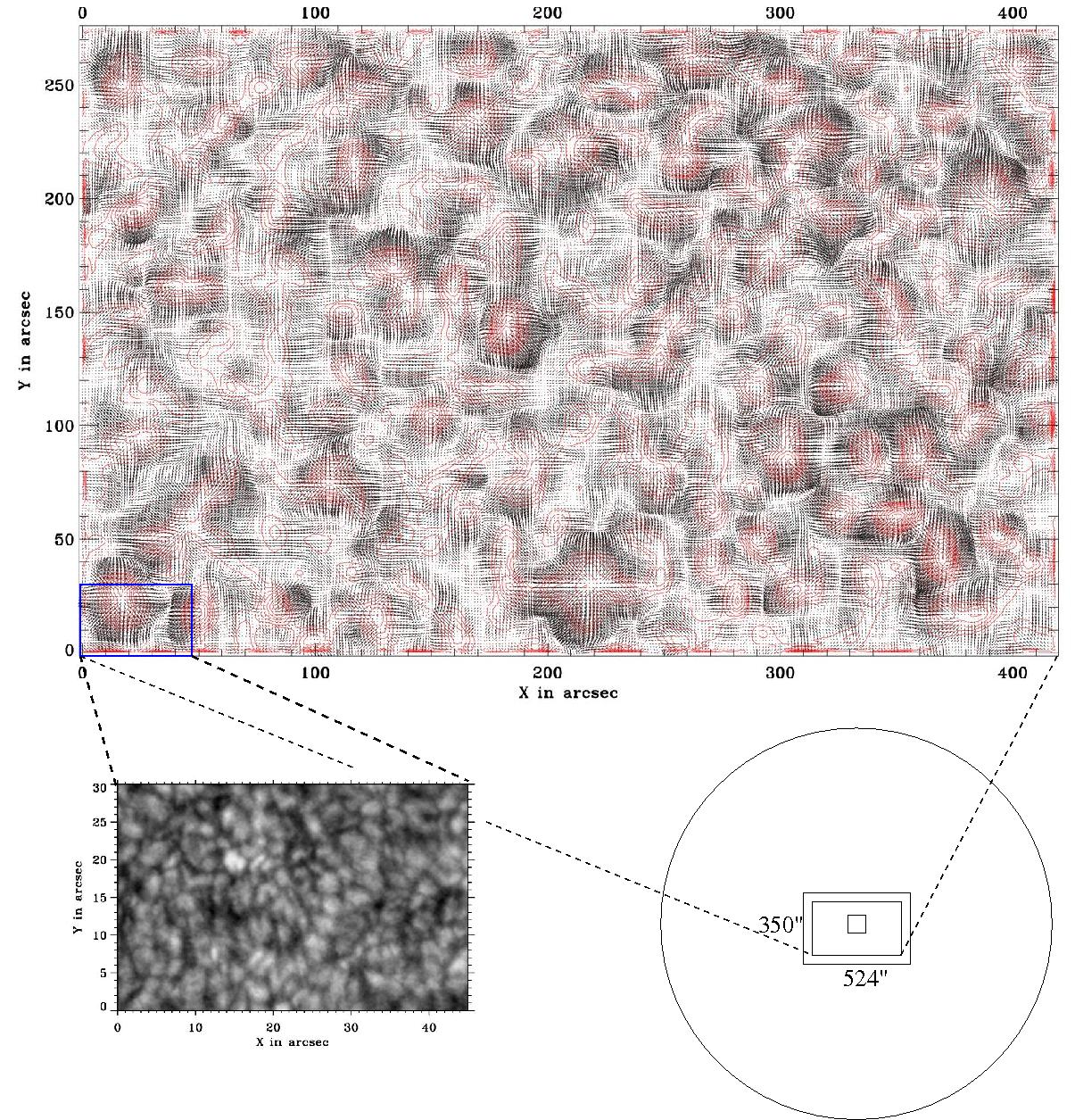}}
\caption{The supergranulation horizontal velocity field as obtained by granule
tracking \cite[from][]{rieutord08}.}
\label{figure:calas}
\end{figure}

The principles and accuracy of granule tracking with LCT or CST 
have been tested by \cite{rieutord01} using synthetic data extracted
from numerical simulations. Flows at scales larger than
2.5~Mm are well reproduced by the displacements of granules. At
shorter scales, the random motion of granules (which are dynamical
structures) generates a noise that blurs the signal. This 2.5~Mm lower
limit has been confirmed by \cite{R_etal10} using Hinode/SOT
observations. The 2.5~Mm resolution is significantly smaller than the
supergranulation scale, therefore, these methods are well adapted to
derive the horizontal components of the supergranulation flow. Unlike
Doppler imaging, they do not suffer from a projection effect, but only
measure the horizontal component of the flow. We will see a bit later
that a combination of Doppler and tracking techniques now appears to
make it possible to separate the horizontal and vertical components of
the flow (albeit in a limited range of scales). An example of the
horizontal velocity fields using this technique applied to
ground-based sequences of wide-field images obtained at Pic du
Midi is shown in Fig.~\ref{figure:calas}.

\subsubsection{Local helioseismology}
Helioseismology uses the propagation of acoustic or surface gravity
waves ($f$-modes) to determine the velocity of the medium over which they
propagate. If the wave velocity is $c$ and that of the fluid
is $V$, a plane wave travelling downstream shows a velocity $V+c$
whereas the one travelling upstream moves with a velocity $V-c$. The
sum of the two measured velocities gives that of the fluid. However, the
phase velocity of the waves is not directly measurable: the observable
quantity is the local oscillation of the fluid which results from
the superposition of many travelling waves. A proper filtering is thus
needed to select the desired wave; this operation requires a true
machinery. The reader is referred to the reviews of \cite{gizon05},
\cite{hanasoge14} and \cite{hanasoge15} for more detailed
presentation of these techniques in the context of subsurface
solar convective dynamics. Let us here simply recall some basic
information about their output: the spatial resolution at which
velocity fields can be measured  is around 5~Mm, and the time
resolution for time-distance helioseismology is around 8~h. While
this is lower than what can be
achieved with other methods, this technique provides the only means to
probe the vertical dependence of the velocities and the
subphotospheric dynamics. Typically, vertical variations can
be evaluated down to 10--15~Mm below the surface, but the accuracy 
of measurements deeper than 10~Mm is still debated. A comparison
between the tracking and helioseismic reconstructions of large-scale
solar surface flows, showing good agreement between the two, can be
found for instance in \cite{svanda07} and \cite{greer_etal16}.

\subsubsection{Power spectra of solar surface flows}
\begin{figure}[htbp]
\raggedright(a)\\
\centering \includegraphics[width=0.7\linewidth]{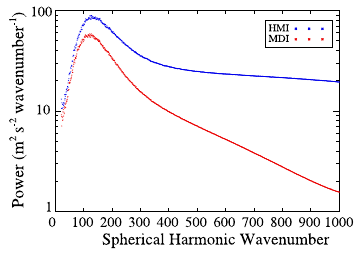}
\\
\raggedright (b)
\\
\centering\includegraphics[width=0.73\linewidth]{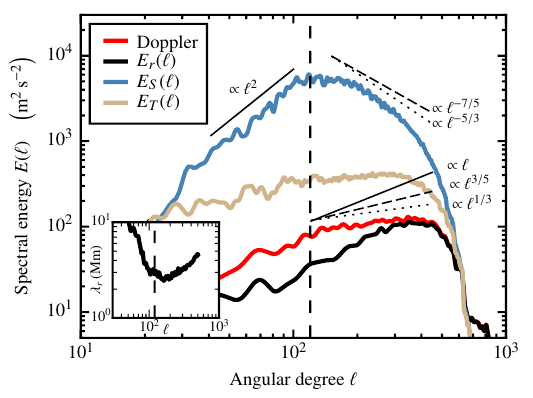}
\vspace{0.3cm}
\\
\raggedright (c)\\
\centering\includegraphics[width=0.7\linewidth]{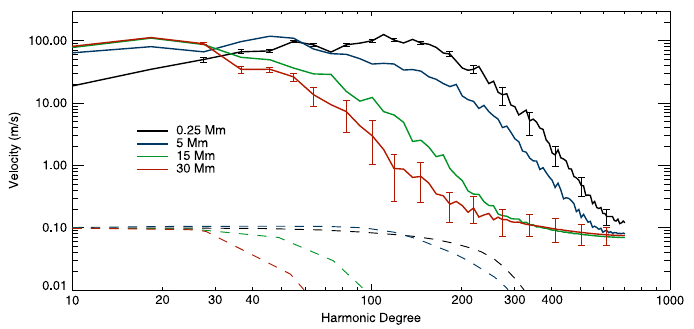}
\vspace{0.5cm}
\caption{
Kinetic energy and velocity spectra of photospheric and subphotospheric
flows derived using three different methods. (a) Global spherical harmonics
spectra of the line-of-sight Doppler velocity field obtained
with SOHO/MDI and SDO/HMI \citep{williams_etal14}. The
supergranulation peak around $\ell=120$ (SOHO) and $\ell=130$ (SDO) is
clearly visible. The granulation peak expected around $\ell=3000$ is
eroded and effectively shifted to larger scales  ($\ell\sim1500$) due
to time-averaging. (b) Global spherical harmonics power spectra of the
three components of the surface velocity field (radial $E_r$, spheroidal
$E_S$, toroidal $E_T$) inferred from a combined CST tracking/Doppler
analysis of 24 hours of SDO/HMI data. The inset shows the vertical
scale height of the flow inferred from the ratio between the
amplitudes of the horizontal and vertical velocity, as a function of
horizontal scale \citep{rincon_etal17}. (c)  Global spherical
harmonics spectrum of the
subphotospheric horizontal velocity field at different depths (full
lines), derived from a local helioseismic ring-diagram analysis
\citep{greer_etal16}. The peak scale of the flows shifts towards
larger scales with increasing depth. Note also that the sign of the 
vertical velocity gradient changes around $\ell=30-40$ in this
inversion (the vertical amplitude of the flow seems to increase with
depth at large scales, while it decreases at smaller scales).}
\label{figure:spectra}
\end{figure}

Given a local or global velocity field map obtained by one of the
means described above, power spectra (scale-by-scale distribution of
kinetic energy in Fourier or spherical harmonics space) are one of 
the most important tools to characterize the properties of the surface
and subsurface dynamics of the solar photosphere. These include
spectra of full-disc Doppler velocity maps obtained by SOHO/MDI
\citep{hathaway2000,hathaway2002} and SDO/HMI data
\citep{williams_pesnell11,williams_etal14,hathaway_etal15} and spectra
of horizontal velocity fields derived from tracking applied to
photometric ground-based wide-field data \citep{rieutord08},
Hinode/SOT data \citep{R_etal10}, SDO/HMI data \citep{langfellner_etal15},
or from helioseismic inversions of flows based on SDO/HMI Doppler data
\citep{greer_etal16}. A selection of recently published 
spherical harmonics spectra derived from SDO/HMI data using different
kinds of flow measurements covering large areas of the solar disc is
shown in Fig.~\ref{figure:spectra}. Figure~\ref{figure:spectra}(a) shows a
spectrum of the Doppler (line of sight) velocity field as measured by
the SOHO/MDI and SDO/HMI Michelson imagers
\citep{williams_pesnell11}. Figure~\ref{figure:spectra}(b)
shows the spectra of the three components of the photospheric
velocity field, determined from a new technique combining CST
(tracking) and Doppler velocity fields reduced from SDO/HMI
data \citep{rincon_etal17}. Finally, Fig.~\ref{figure:spectra}(c) shows
the velocity spectra of subsurface flows determined from a
ring-diagram analysis of SDO/HMI data \citep{greer_etal16}.

\subsection{The scales and structure of supergranulation}\label{obsvel}
\subsubsection{Horizontal spatial scale and morphology}\label{length_sc}

Several techniques are used to measure the horizontal scale of
supergranulation flows. The most popular one by far
is to estimate the peak scale of power spectra such as those shown in
Fig.~\ref{figure:spectra}. All studies of this kind basically find a
peak scale value of 30--36~Mm, corresponding to  $\ell=120$--$140$
spherical harmonics, with a typical width of 20--75~Mm. The comparison
shown in Fig.~\ref{figure:spectra}(a) between SOHO/MDI spectra and
SDO/HMI spectra computed for the same observation period suggests that
the lower spatial resolution of SOHO/MDI leads to a 10\%-larger
estimate for the size of supergranules \citep{williams_etal14}. 

A detailed analysis of the plots of \cite{rincon_etal17} in
Fig.~\ref{figure:spectra}(b) reveals that the spectra of the horizontal and
vertical components of the flow are very different. The
supergranulation peak around $\ell=120$ is essentially associated with
horizontal spheroidal motions corresponding to diverging/converging
flows. This dominance of the spheroidal spectrum over the
toroidal/vortical spectrum of horizontal flows  attests of the
predominantly cellular structure of supergranulation motions.
The weaker radial velocity spectrum increases monotically with
decreasing scale down to the 2.5--5~Mm  resolution of the velocity map
used in the study\footnote{Consistent
  use of the component-separation method combining CST and Doppler
  velocities requires downgrading the Doppler data to the 2.5~Mm CST
  resolution \citep{rincon_etal17}.} and only shows a
tentative inflexion at the supergranulation scale (a similar
conclusion applies to the toroidal/vortical component). 
This trend of the radial velocity field to increase with scale down to
the spatial resolution of the data has also been observed using a different
flow-component determination method based on a differentiation of
Doppler maps \citep{hathaway_etal15}. The very different  radial and
horizontal velocity spectra reflect the strong anisotropy
of the flow. We will shortly argue that this result can be
used to infer some information about the vertical scale height of 
the velocity field. Finally, the helioseismic data analysis shown in 
Fig.~\ref{figure:spectra}(c) reveals the interesting, albeit perhaps
not entirely surprising result that the scale of the spectral peak
increases monotonically with increasing depth, as one would expect from 
the self-similarity argument described in Sect.~\ref{supergranulationpuzzle}.

Other non-spectral techniques have been used to measure the horizontal
scale of supergranulation. In their seminal work, \cite{leighton62} and
\cite{SL64} calculated the auto-correlation length of Dopplergrams, 
and found 32~Mm. Other authors, like \cite{derosa2000} and \cite{derosa04}, used
local correlation tracking to determine the horizontal flows from the
Doppler signal of SOHO/MDI and identified supergranules with horizontal
divergences. From these data, they derived a rather small ``diameter''
in the 12--20~Mm range. Using a similar technique, \cite{meunier07c} 
found a mean value for supergranule diameters around 30~Mm. As
underlined in these papers, the exact size of supergranules very much 
depends on the smoothing procedure used in the data processing.
Another technique consists in estimating the supergranulation scale
from horizontal flow divergence maps derived from local
helioseismic analysis (this quantity is directly associated
with the difference between wave travel times in seismology and is,
therefore, more readily accessible than the velocity field itself).
Using this technique on SOHO/MDI data, \cite{delmoro04} computed the
statistics of the sizes of supergranules and found a mean diameter at
27~Mm, with a peak in the distribution at 30~Mm. Similar results have
been obtained  by \cite{hirzberger08} using an even larger set of data
(collecting more than $10^5$ supergranules).

Finally, several authors have used tesselation algorithms or
threshold-based identification techniques 
to capture individual supergranulation cells and subsequently study
their geometrical properties and spatial arrangement. Such techniques
have mostly been applied to maps of the chromospheric network
\citep[e.g.,][]{hagenaar97,schrijver97,berrilli98,chatterjee_etal17},
whose relationship to supergranulation is further described in
Sect.~\ref{obsnet}, and tend to give slightly smaller size
estimates.

\subsubsection{Time scale and lifetime}\label{time_sc}
As with horizontal spatial scales, there are several ways in which
the time scale of supergranulation can be measured. Historically, 
many such measurements rely on the statitics of coherent cellular structures.
\cite{worden76} suggested a supergranulation lifetime of 36
hours. Later, \cite{wang89} showed that supergranulation
lifetime estimates depended strongly on the choice of tracer or
proxy. They obtained 20 hours using Dopplergrams, two days using direct
counting techniques of supergranulation cells and  10 hours using the
tracking of magnetic structures (see also Sect.~\ref{obsmag}).  Here
again, SOHO/MDI data have dramatically increased the statistics and
thus the quality of the determinations. Using a wavelet analysis of MDI data,
\cite{parfinenko_etal14} found a time scale of 1.3 days, while 
\cite{hirzberger08} report a lifetime around 1.6--1.8 days using
helioseismic techniques. These latter estimates are somewhat
longer than the others, but they rely on very long time series and
large statistics enabling a better representation of long-living
supergranules. A recent SDO/HMI helioseismic analysis by
\cite{greer_etal16} finds a very similar coherence time of
supergranules in the first few Megameters below the surface, but also
concludes that the pattern visible at the surface at any given time
may subsequently propagate down to the base of the near surface shear
layer over a scale of a month. We will come back to this result when
we discuss the influence of rotation on supergranulation.

\subsubsection{Velocity scales}\label{velo_sc}
A typical horizontal velocity associated with supergranules can be
quickly derived from the ratio between the horizontal spatial scales
and time scales given above. Taking 30~Mm for the former and 1.7~days
for the latter leads to a velocity of 205~m/s, in reasonable agreement with
more direct inferences of the supergranulation velocity field from
observations: the original work of \cite{hart54} inferred 170~m/s,
\cite{SL64} mentioned 300~m/s and more \cite{hathaway2002}
evaluated this amplitude at $\sim360$~m/s from SOHO/MDI Dopplergrams.

A quantitative way to estimate the magnitude of the supergranulation
velocity field is through the spectral density of kinetic energy
$E(k)$ describing the relation between the scale and amplitude of the
flow. The dimensional value of the spectral density of horizontal
kinetic energy $E_h(k)$ was for instance derived  by \cite{R_etal10}
from granule tracking velocity measurements using Hinode/SOT data.
The spectral power density at supergranulation scales is
500~km$^3$/s$^2$, which is larger than
that at granulation scales.\footnote{At granulation scales, the
spectral power density is less than 300~km$^3$/s$^2$. We recall that
granules have a much larger typical velocity than supergranules though,
of the order 1--2~km/s (Sect.~\ref{granule}). The difference comes
from the definition of the spectral power density at wavenumber $k$,
$E(k)\sim k^{-1}V_k^2$, which introduces an extra $k$ factor.} This
energy density is related (dimensionally) to the velocity at scale
$\lambda=2\pi/k$ by the relation $V_\lambda=\sqrt{kE_h(k)}$.  This
leads to $V_{\rm \lambda=36~Mm}\simeq300$~m/s, consistent with the
direct Doppler measurements of the velocity field at supergranulation
scales.

It has been known from Dopplergrams for very long that the
supergranulation flow field is largely dominated by its horizontal
component. The 300--400~m/s estimates above refer to this
component. The vertical component of supergranulation flows is much
smaller and has proven much harder to measure, notably because the
corresponding signal is mixed with the much stronger 5~min signal of
acoustic solar oscillations, and polluted by the presence of
magnetic field concentrations at supergranule boundaries, where up and
downflows tend to be localized (see
Sect.~\ref{obsmag} below). \cite{november89,november94} advocated
that this vertical component was in fact the mesogranulation that he
detected some years before on radial velocities at disc centre
\citep{november81}. The rms value of this quantity was then estimated
to be 60~m/s. This quantity was more precisely evaluated using the
SOHO/MDI data. \cite{hathaway2002} derived an estimate of 30~m/s while
\cite{duvall_birch10} found a very low rms value around 4~m/s with
upwelling velocities of 10~m/s. Similarly, power spectra of
line-of-sight velocities from Hinode/SOT data derived by
\cite{R_etal10} point to a rms vertical velocity of 10~m/s at the
scale of supergranulation, however, with low statistics. 
The recent global analysis of SDO/HMI surface data by
\cite{rincon_etal17} aiming at disentangling the three components 
of the velocity and the corresponding spectra (Fig.~\ref{figure:spectra}b), points to a value of no more than 20--30~m/s
for the radial/vertical velocity field at supergranulation
scales. Only an upper bound can be inferred from such surface measurements
as it is very difficult to isolate the weak signal from this slow
flow component in the data. Recent helioseismic measurements by
\cite{greer_etal16} point to a typical vertical flow field of the order
40~m/s in the first few Mm below the surface. Whatever the exact
value, the results show that the vertical amplitude of the flow is at
best ten times smaller than the horizontal one, emphasizing once again
the strong anisotropy of the supergranulation flow.

\subsubsection{Depth and vertical structure}\label{obsdepth}
A typical scale height of subsonic flows in the supergranulation to
mesogranulation range can be indirectly estimated from velocities
measured at the photospheric level using the anelastic equation of
mass conservation\[ \partial_zv_z = -v_z\partial_z\ln\rho -\vec{\nabla}_h\cdot\vv_h~,\]
where the index $h$ refers to the horizontal quantities and $z$.
This equation can also be expressed as  $v_h/\lambda_h\sim v_z/\lambda_z$,
where $\lambda_h$ is the horizontal scale and $\lambda_z$ stands for
the smallest of the density scale height and typical vertical scale of
variation of the flow. We see that a separate measurement of the
vertical and horizontal velocities at given horizontal scale allows
for an estimation of the corresponding vertical scale of the flow. Combining
Dopplergrams and correlation tracking inferences, \cite{november94} argued
that the supergranulation flow should disappear at depths larger than
2.4~Mm below the visible surface. \cite{R_etal10} performed the same kind
of calculation  using divergences and velocity fields derived from Hinode data,
and found a vertical velocity scale height of $\sim1$~Mm, indicating a
very shallow structure. Finally, in their recent global spectral
analysis of the different components of the flow using SDO/HMI data,
\cite{rincon_etal17} found that $\lambda_z$ is
approximately constant and of the order of 2.5~Mm in the horizontal
range extending from the supergranulation scale down to a few Mm
(see inset in Fig.~\ref{figure:spectra}b). The conclusion that the
flow has a relatively small vertical scale of variation below the
surface appears to be an inescapable consequence of the measurements
of a large horizontal to vertical surface velocity ratio. This result,
however, does not imply that there is no flow below that depth.

There is obviously only so much one can do with surface
measurements to study the vertical structure of supergranulation.
Fortunately, the advent of local helioseismology in the late 1990s has
made it possible to start probing supergranulation-scale flows at
optically-thick
levels. Using MDI data, \cite{duvall97}, detected
flows at supergranulation scales only in the first few Mm below
the surface. \cite{duvall98} further estimated that the depth of
supergranulation was 8~Mm. \cite{zhao03}, on the other hand, found
converging flows at 10~Mm and estimated the supergranulation depth to be 15~Mm.
\cite{woodard07} reported a detection of a flow pattern down to 5~Mm
corresponding to the deepest layers accessible with their data set. Using
Hinode data, \cite{sekii07} found that a supergranulation pattern,
monitored for 12~h in a small field of $80\times40$~Mm$^2$,
does not persist at depths larger than 5~Mm. The existence of a return
flow at depths larger than 5~Mm was also suggested in that period but
remained unclear \citep{duvall98,zhao03}. There is a large
scatter in these early results, and a lot of ambiguity in what
they really represent. 

The helioseismic determination of the vertical extent
and structure of supergranulation remains a work in progress and a
difficult task. Various discussions of the early shortcomings,
difficulties and artefacts associated with the development of
helioseismic imaging of subsurface flows can be found in
\cite{braun03,lindsey04,gizon05} and more recently in
\cite{svanda15,bhattacharya_hanasoge16}. But, as these difficulties 
are progressively eliminated, helioseismic techniques 
are now increasingly  playing a key role in the study and understanding 
of supergranulation, and new helioseismic analyses of subsurface flows have
now been conducted with SDO/HMI data. \cite{duvall_hanasoge13} and
\cite{duvall_etal14} found suprisingly high-speed supergranular flows
at a depth $\sim2$~Mm, namely vertical upflows of 240~m/s at a depth of 2.3~Mm
and horizontal flows of 700~m/s at a depth 1.6~Mm. Such high values
have, however, been challenged by subsequent studies by
\cite{degrave_jackiewicz15} and \cite{greer_etal15,greer_etal16}. 
This tension may be due to the fact that \cite{duvall_hanasoge13} and
\cite{duvall_etal14} used a large skip distances and imaged vertical
flows, and also averaged over a larger number of supergranules, while
the local methods used in other studies typically image horizontal
flows using moderate skip distances. \cite{greer_etal16} also find an instantaneous
correlation depth of $\sim$7~Mm for the supergranulation pattern, but
suggests that the pattern actually propagates down to the bottom of the
near surface shear layer over a month at a vertical speed of the
order of 40~m/s. This propagation may help to explain some of the
deeper earlier estimates for the depth of supergranulation.
Note however that these determinations of the correlation
  depth of flows may be affected by the details of helioseismic
  inversion procedures, as suggested by the analysis of \cite{degrave14}.
Finally, \cite{greer_etal16} suggest that supergranules ``form[s] at
the surface and rains downward, imprinting [their] pattern in deeper
layers''. They also find that the pattern does not take the form of a
cellular flow like in the laminar picture of Rayleigh--B\'enard
convection, but is rather dominated by downflows. As mentioned
earlier, the spectral analysis of helioseismically determined
convective flows by \cite{greer_etal15} (Fig.~\ref{figure:spectra}c)
also suggests a monotonic increase of the energetically dominant
horizontal scale of fluctuations with depth.

\subsection{Intensity variations in supergranules}\label{obsintensity}
The surface thermal signature of supergranulation appears to be quite
small. Early measurements suggested that the intensity contrast
between the border and the centre of supergranules probably 
does not exceed a few percents in the infrared \citep{worden75}.
The photometric intensity contrast at supergranulation scales in
white light images is also much smaller and ambiguous than
that of granulation, which has been shown to be up to 27\%
at a wavelength of 450~nm \citep{WR09}. Several early studies
\citep{beckers68,frazier70,foukal84,lin92} found an
increase of intensity at the edge of supergranulation cells,
corresponding to a negative correlation between the supergranulation
horizontal divergence maps and intensity maps. These early results may
naively tend to rule out a convective origin for supergranulation but
are subject to caution. First, supergranulation vertices are strongly
correlated with magnetic bright points (see Sect.~\ref{obsmag} below).
To circumvent this difficulty, \cite{rast03a} considered only areas
with low magnetic fields and found a small decrease of intensity at
the edge of supergranules. The problem was reconsidered in detail by
\cite{meunier07b,meunier08} using MDI intensity maps. There too, the
influence of the magnetic network was carefully eliminated. In
contrast to most previous studies, they report a very small but
significant intensity decrease from the centre to the edge of
supergranulation cells (in the range 0.8--2.8~K). 
This conclusion was also reached by \cite{goldbaum_etal09} using
a different methodology.  In addition, \cite{meunier08} noticed that
the radial temperature profile at the surface of a supergranule is
very similar to that of a simulated granule. Using SDO/HMI
continuum intensity maps centered around 617~nm,
\cite[][]{langfellner_etal16} recently derived a new estimate at
$\Delta T=1.1\pm0.1$~K of the temperature drop between centre and edge
of supergranules, in line with the preceding results. 

While such contrats are relatively weak, it has been argued that they
remain consistent with a convective origin of supergranulation
\citep{goldbaum_etal09}. Also, the
many complexities of radiative transfer at the surface, including
strongly temperature-dependent opacities \citep{nordlund09}, imply
that small surface intensity contrasts cannot alone easily be accepted as
smoking-gun evidence for weak buoyancy/entropy driving below the surface.
Helioseismic analysis \citep[e.g.,][]{duvall97} notably seem to
point to larger relative temperature contrasts of the order of a few
percents below the surface. We will also see in Sect.~\ref{numerics}
that a strong buoyancy driving is observed at scales larger than
granulation in large-scale simulations, but does not generally
translate into a strong surface intensity pattern in simulations
incorporating realistic radiative physics.

\subsection{Effects of rotation}\label{obsrot}
\begin{figure}[htb]
\vspace{5pt}
\centerline{\hspace{0.05\linewidth}
(a)\hspace{10pt}\parbox[t]{0.42\linewidth}{\vspace{-10pt}\includegraphics[width=1.\linewidth]{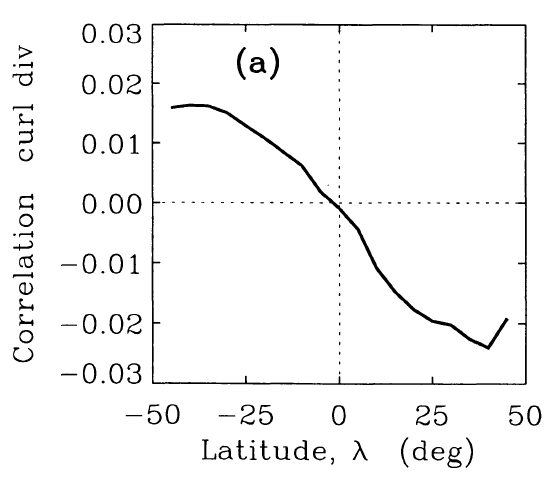}}\hspace{0.06\linewidth}
(b)\parbox[t]{0.45\linewidth}{\vspace{-15pt}\includegraphics[width=1.\linewidth]{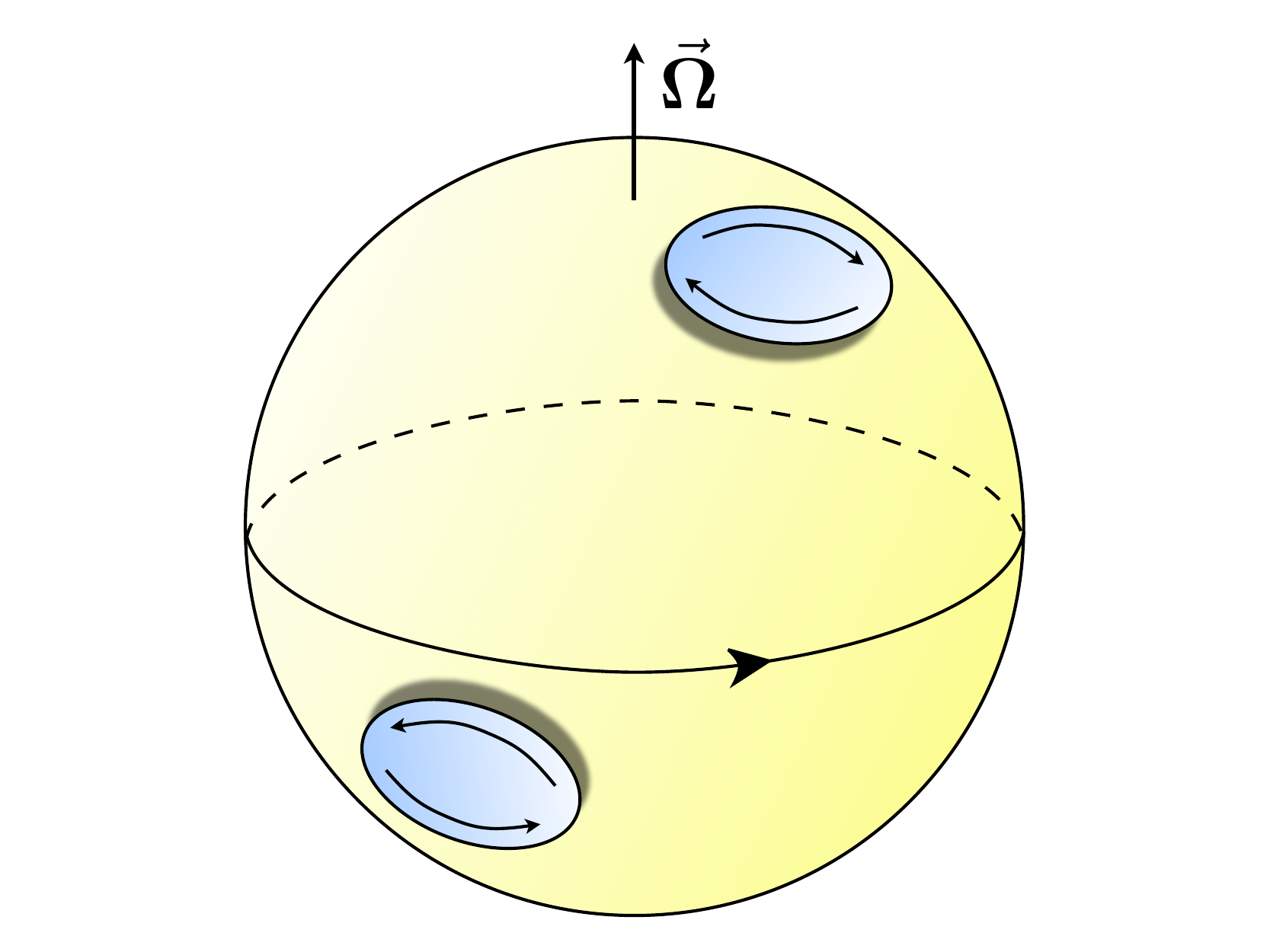}}
}
\caption{(a) Correlation between the horizontal divergence and
  vertical vorticity of the supergranulation flow as a function of latitude
  \citep[image credits:][]{gizon03a}. (b) Schematic view of
  anticyclones at the surface of the rotating Sun.}
\label{figure:rot}
\end{figure}

The influence of the global rotation $\Omega$ of the
Sun on the dynamics of a structure of size $L$ and typical velocity $V$
is measured by the Rossby number:

\[ \Ro = \frac{V}{2\Omega L} = (2\Omega \tau)^{-1}~.\]
The second expression uses the lifetime of the structure
$\tau=L/V$. In
numbers, taking $\tau_{\mathrm{SG}}=1.7$~days and a rotation period
of 25--30~days leads to $\Ro_{\mathrm{SG}}\sim2-3$. This is not a large
value, indicating that the Coriolis acceleration should affect
the dynamics of supergranules. Such an effect has been reported
by \cite{gizon03a}, who showed (Fig.~\ref{figure:rot}a) that the
correlation between vertical vorticity and horizontal divergence of
supergranules changes sign at the equator: it is negative in the northern
hemisphere and positive in the southern one. Hence, supergranules,
while essentially consisting of cells of diverging flows, behave like
weak anticyclones  (the vertical vorticity of anticyclones changes
sign at the equator, see Fig.~\ref{figure:rot}b). These anticyclones
are surrounded by cyclonic vorticity associated with downward flows;
because these downdrafts have a somewhat smaller scale, the cyclonic
vorticity is less conspicuous in measurements than the anticyclonic
contribution  of supergranules, but it has actually been singled out
in the work of \cite{komm07}. A recent study of
\cite{langfellner_etal15b} applying both time-distance helioseismology
and local correlation tracking to SDO/HMI data has recently confirmed
these conclusions with spatially-resolved measurements of the
supergranulation vorticity. They find a typical vortical flow component
of the order of 10~m/s in the diverging core of supergranules, much
weaker than the diverging horizontal flow component itself. This
result is consistent with the weakness of the toroidal flow component
with respect to the spheroidal flow component found in the spectral
analysis of \cite{rincon_etal17}.

Early studies of the rotational properties of supergranulation
focused on the rotation rate of the supergranulation pattern
\citep{duvall80,snodgrass90}.  Using Dopplergrams, they found,
surprisingly, that supergranulation is rotating 4\% faster than the
plasma. This phenomenon is usually referred to as the 
  superrotation of supergranules. Superrotation was seemingly
confirmed by \cite{duvall2000} using the time-distance technique
applied to $f$-modes detected with SOHO/MDI, and \cite{beck2000}
estimated that the supergranulation rotation rate is larger than the
solar rotation rate at any depth probed by helioseismology.

Analysing time series of divergence maps inferred from time-distance
helioseismology applied to MDI data, \cite{gizon03b} further found
that the supergranulation pattern had wave-like properties with a typical
period of 6--9 days, quite longer than the lifetime of individual
supergranules. They also found that the power spectrum of the
supergranulation signal close to the equator presented a power excess 
in the prograde direction (with a slight equatorwards deviation in
both hemispheres), thus explaining the anomalous superrotation
rate of the pattern. The dispersion relation for the supergranulation
``wave'' appears to be only weakly dependent on latitude
\citep{gizon04}. \cite{schou03} seemingly confirmed these findings with
direct Doppler shift measurements and found that wave motions were mostly 
aligned with the direction of propagation of the pattern.
These results brought interesting new light on
the supergranulation phenomenon and led to the conjecture that
supergranulation could be a manifestation of oscillatory
convection, a typical property of convection in the presence of
rotation and/or magnetic fields (see Sect.~\ref{theory}). 
This interpretation  of the observed power spectrum in terms of oscillations
was, however, questioned by \cite{rast04b} and \cite{lisle04}, who
instead suggested an explanation in terms of two
superimposed steady flow components identified as mesogranulation and
supergranulation advected by giant cell circulations. According to
\cite{gizon05}, this interpretation is not supported by
observations. They argued that the finding of \cite{lisle04} that
supergranules tend to align in the direction of the Sun's rotation
axis under the influence of giant cells can be explained naturally in
terms of wave dynamics. North-South alignments of supergranules have
also been found in the polar regions by \cite{nagashima_etal11} using
local helioseismology with Hinode/SOT data. A recent study by
\cite{langfellner18} using both time-distance helioseismology on
SDO/HMI data and correlation tracking appears to further confirm 
the results of \cite{gizon03b} on the wave-like oscillatory properties
of the dynamics in the range of scales $50<\ell<120$ comparable to or
slightly larger than the supergranulation scale.

Going back to surface measurements, \cite{hathaway06} argued that the
supergranulation pattern superrotation
inferred from Doppler shifts was due to projection effects on the
line-of-sight signal.  Using correlation tracking of divergence maps
derived from intensity maps \citep{meunier07c} and comparing it with
direct Doppler tracking, \cite{meunier07d} confirmed the existence
of projection effects with the latter method, but found that the
supergranulation pattern inferred from divergence maps was still
superrotating, albeit at smaller angular velocities than those inferred
by \cite{duvall80} and \cite{snodgrass90}.  
More recently, \cite{hathaway12} found that supergranules of
increasingly larger size seem to accurately track the solar rotation
rate at increasingly larger depths, down $\sim$50~Mm. This latter
estimate is quite large compared to the coherence length (7~Mm)
inferred by \cite{greer_etal16} but is consistent with the high speed
vertical flows suggested by the \cite{duvall_etal14} analysis. Also,
\cite{greer_etal16} note that the 7~Mm estimate is deceptive because
of the slow downwards propagation of the supergranulation pattern
diagnosed in their helioseismic analysis, and notably argue that the
slow propagation speed is consistent with a propagation of the pattern
down to the base of the near surface shear layer on a timescale
comparable to the solar rotation period. This suggests that
supergranulation is in dynamical interaction with the subsurface
differential rotation layer down to 30--50~Mm deep, and may actually
play an important role in the dynamical establishment of this layer.

For the sake of completeness, let us finally mention
the observations by \cite{kuhn2000} of small-scale 100 meters-high
``hills'' at the solar surface, which they interpreted as Rossby waves. 
It has been argued \citep{williams07} that these structures simply
result from the vertical convective motions associated with
supergranules.

\subsection{Effects of magnetic fields}\label{obsmag}
As explained in Sect.~\ref{turbulentscales}, the magnetic
dissipation scale at the solar surface is $\ell_\eta\sim 100$~m or
slightly less. Hence, convection at the solar surface is strongly
coupled to the Sun's magnetic dynamics at all observable scales,
including that of supergranulation. The aim of this section is 
to review the breadth of observational results on the interactions
between the two. We first look at the correlations between supergranulation and
the magnetic network and review the main properties of internetwork
fields, whose dynamics is directly related to the formation of
the magnetic network. We then review studies of the dependence of
supergranulation on the global solar magnetic cycle, and of its
interactions with active regions.

\begin{figure}[htb]
\centerline{\includegraphics[width=0.9\linewidth]{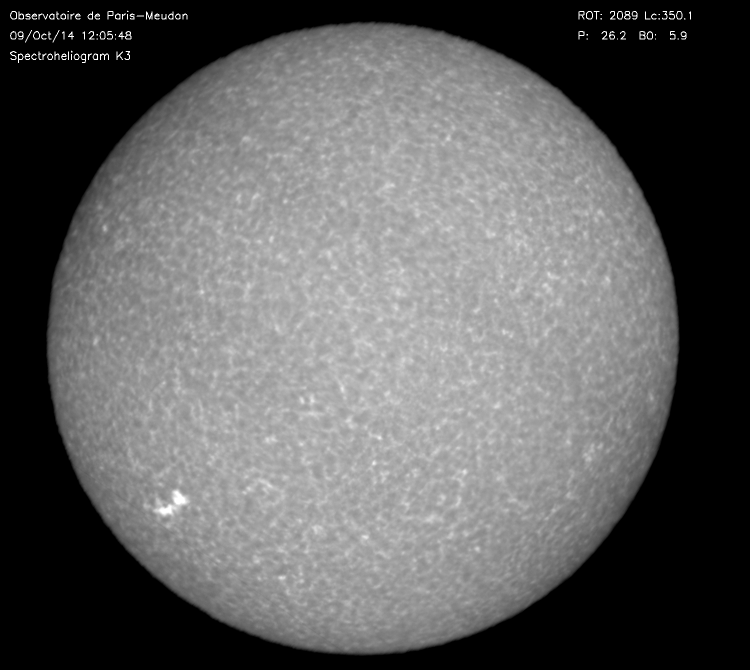}}
\caption{A view of the chromospheric network at the Ca$^+$~K3
  line at 393.37~nm (image credits: Meudon Observatory).}
\label{figure:spectrohel}
\end{figure}

\subsubsection{Supergranulation and the magnetic network}\label{obsnet}
The discovery of the chromospheric network in Ca$^+$~K
spectroheliograms (the K-line of Ca$^+$ at 393.4~nm) such as shown in
Fig.~\ref{figure:spectrohel} dates back to
\cite{deslandres1899}.  The first comparative analyses
between magnetograms, spectroheliograms and Dopplergrams 
by \cite{leighton62} and \cite{SL64} revealed a strong
correlation between the chromospheric network, the magnetic
field distribution of the quiet Sun and supergranulation. For
this reason, both magnetograms and spectroheliograms have been used as
a proxy to study supergranulation
\citep[e.g.,][]{lisle2000,delmoro07,tian_etal10}, but it should be
kept in mind that the dynamical interactions between magnetic
fields and supergranulation are actually still not well understood
(this problem will be discussed in Sect.~\ref{discussion}).

\begin{figure}[htb]
\centerline{\includegraphics[width=0.85\linewidth]{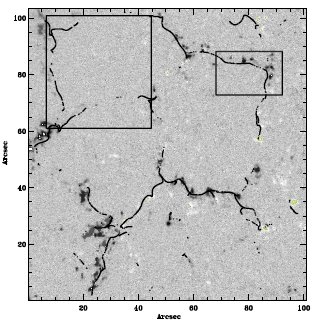}}
\caption{Distribution of magnetic field intensity (grey
  scale levels) on the supergranulation boundaries. The black dots show the final
positions of floating corks that have been advected by the velocity field
computed from the average motion of granules. The distribution of corks
closely matches that of the magnetic field. \cite[from][]{roudier09}.}
\label{figure:network_SG}
\end{figure}

The magnetic network refers to a distribution of magnetic field
concentrations (associated with bright points in spectroheliograms)
with typical field strengths of the order of 1~kG \citep[see reviews
by][]{solanki93, dewijn09}, primarily located on the
boundaries of supergranules \citep{simon88}, in downflow areas. Several
differences between supergranulation and the magnetic network have
been noticed, including a 2\% relative difference in
the rotation rate of the two patterns (see \cite{snodgrass90} and
Sect.~\ref{obsrot} above). The magnetic network is not regularly
distributed on the boundaries of supergranulation cells but rather
concentrates into localized structures (see
Fig.~\ref{figure:network_SG}). Estimates of the lifetime and
size of supergranules inferred from magnetograms or spectroheliograms
are significantly smaller than those based on direct velocimetric
measurements \citep{wang89,schrijver97,hagenaar97}.  For instance,
\cite{hagenaar97}, using correlations of maps of the chromospheric
network, obtained a typical size of 16~Mm. As far as the horizontal velocities
are concerned, the tracking of magnetic network elements gives values around
350~m/s, close to the estimates derived from granule tracking \citep{lisle2000}.
The spatial distribution of magnetic network fields can be
reconstructed quite well by letting passive magnetic elements
be advected by the surface flow field \citep{krijger03,roudier09}.
Moreover, tagging the granules and following their evolution and motion
leads to the so-called Trees of Fragmenting Granules in a space-time
diagram, a structure whose spatial boundaries also neatly match part of
the boundary of the embodying supergranule
\citep{roudier_etal16,malherbe_etal18}.

These results show that the formation of the magnetic network is
related to the large-scale dynamics of the surface
\cite[see][]{orozco_etal12,giannattasio_etal14,berrilli_etal14}.
The magnetic field-flow interaction occurs over a wide range of scales
extending up to the 35~Mm supergranulation scale, and several studies
with the Swedish Solar Telescope at La Palma observatory indicate that
strong correlations between flows at scales comparable to or smaller
than mesoscales (i.e., significantly smaller than supergranulation)
and intense magnetic elements exist \citep{dominguez03,dominguez03b}.
A study by \cite{roudier09}, combining spectropolarimetric
and photometric Hinode measurements, also established a very clear
correlation between the motions at mesoscales and those of the magnetic
network \citep[see also][]{dewijn_muller09}.  Note finally that the
network formation process may be influenced by an East-West
anisotropy now detected in supergranules
\cite[see][]{langfellner_etal15,roudier_etal16}.

\subsubsection{Internetwork fields}\label{obsINB}
One of the important advances on solar magnetism in the last two decades
has been the detection of quiet Sun magnetic fields at scales much
smaller than that of granulation
\cite[e.g.,][]{dominguez03b,berger04,trujillo04,rouppe05,lites08}. 
The ubiquity of these fields and their energetics suggest that the
dynamics of internetwork fields could also be a piece of the
supergranulation puzzle. The following summary is not meant to be
exhaustive. For a more detailed presentation, we refer the reader to the
reviews of \cite{dewijn09} and \cite{stenflo13}.
Internetwork fields refer to mixed-polarity fields that populate the
interior of supergranules. Their strength is on average thought to be
much weaker than that of network fields, but magnetic bright points
are also observed in the internetwork,
\citep[e.g.,][]{muller83,nisenson03,dewijn05,lites08}.
Besides, network and internetwork fields are known to be in permanent
interaction \citep[e.g.,][]{martin88}. In fact, in the light of nowadays
high-resolution observations and numerical MHD simulations (see Sect.~\ref{simmhd}),
the historical distinction between network and internetwork fields
cannot be easily justified on dynamical grounds, as the two appear to
be part of a complex multiscale dynamics involving some local
dynamical intensification process. Internetwork magnetism was originally
discovered by \cite{livingston71,livingston75} and subsequently
studied by many authors \citep[e.g.,][]{martin88,keller94,lin95} at resolutions 
not exceeding 1" (730~km). Observations with the solar telescope at
La Palma observatory revealed the existence of such fields at scales
comparable and even smaller than the granulation scale
\citep{dominguez03b,roudier04,rouppe05}. Studies based on 
Hinode observations \citep{orozco07,lites08}  reported magnetic field
variations at scales comparable to or smaller than 100~km.

The strength of internetwork fields, their distribution at granulation
and subgranulation scales and their preferred orientation are still a
matter of debate. Almost every possible field-strength value in the
5--500~G range can be found in literature
\citep{martin88,keller94,lin95,dominguez03b,trujillo04,lites08}. 
This wide dispersion is explained by several factors. Zeeman
spectropolarimetry, one of the most
frequently used tools to study solar magnetism, is affected by
cancellation effects when the magnetic field reverses sign at scales
smaller than the instrument resolution
\citep{trujillo04,dewijn09}. Hence, very small-scale fields
can partially escape detection via this method. Zeeman
spectropolarimetry estimates of the average field
strength based on Hinode observations \citep{lites08} 
are 11~G for longitudinal fields and 60~G for transverse fields
(horizontal fields at disc centre), but wide excursions from these
average values are detected and the observed signatures may also
be compatible with stronger, less space-filling magnetic fields.
Using Hanle spectropolarimetry, \cite{trujillo04} report
an average field strength of 130~G, with stronger fields in the
intergranular lanes and much weaker fields in the bright centres
of granules. The discrepancy between Hanle and Zeeman
  estimates can be reduced using techniques that circumvent
  cancellation effects, though, leading to estimates of 60-80~G
  \citep{danilovic10,danilovic16,danilovic16b}.

Some Zeeman estimates seem to indicate that small-scale internetwork
fields have a tendency to be horizontal \citep{orozco07,bommier07,lites08},
sometimes even bridging over granules, but other studies have 
come to the opposite conclusion that internetwork fields are mostly
isotropic \citep{martinez08,asensioramos09,bommier09}. 
Using Zeeman and Hanle diagnostics in a complementary way,
\cite{lopez10} came to the conclusion that internetwork
fields are mostly isotropic and highly disordered, with a 
typical magnetic energy containing scale of 10~km. We again refer the
reader to \cite{stenflo13} for a more detailed review of this problem.
The isotropy properties of the field also seem to be dependent
on the depth at which they are probed \citep{schussler08,rempel14},
and conditional on the amplitude of the signal polarization
\citep{lites17}.

Finally, at the scale of supergranulation, it has been pointed out
that regions of supergranulation up-welling (i.e., of positive
divergence) can be very quiet and have very little flux emergence
\citep{martinezgonzalez_etal12,stangalini14}. One possible explanation
is that supergranulation-scale diverging flows somewhat hinder small-scale
dynamo action, but this observation may also be a simple consequence
of the fast sweeping of the magnetic fields peppered by the
small-scale dynamo to the boundaries of supergranules.

\subsubsection{Magnetic power spectrum of the quiet photosphere}\label{obsmagspec}
The scale-by-scale distribution of magnetic energy and
the power spectrum of magnetic fields in the quiet photosphere
may also give us clues on MHD dynamics in the subgranulation to
supergranulation range. Several authors have notably argued that
solar magnetic fields, from the global solar scales to the smallest
scales available to observations, may have a fractal or multifractal
structure \citep{lawrence95,komm95,nesme96,meunier99,janssen03,stenflo02,
stenflo03a,stenflo03b,abramenko05}. 

Studies of the magnetic power spectrum of the quiet Sun are currently
mostly limited to the line-of-sight component of
the magnetic field.  Most spectra available in literature have been
obtained from either ground-based observations, SOHO / MDI and Hinode
magnetograms. The power spectrum of magnetic energy at scales below
1~Mm has recently been studied by \cite{stenflo12},
\cite{katsukawa_orozco12} and \cite{danilovic16} using Hinode/SOT
data. These different studies find a rather slow decay of the
magnetic energy ($\sim k^{-0.7}$--$k^{-1.4}$) at subgranulation scales
suggestive of vigorous  magnetic dynamics at such scales. At the
larger scales of most interest here, the magnetic power spectrum is
rather flat, but shows a maximum in the meso to supergranulation-scale
range. The spectrum appears to have a relatively shallow negative
slope below 10--20~Mm,
\cite[e.g.,][]{lee97,abramenko01,harvey07,mcateer09,longcope09}. There
have unfortunately been very few studies of supergranulation-scale
magnetic fields with SDO/HMI so far. A raw comparison of Hinode/SOT,
SOHO/MDI and SDO/HMI magnetic spectra obtained in this range of scale
has been made by \cite{abramenko11}.

\subsubsection{Supergranulation variations over the solar cycle}\label{cycle}
In view of the association between supergranulation and the magnetic
network, it is natural to wonder if and how the size of supergranules
varies with solar activity. Early studies did not directly focus on
supergranulation flows but rather on the cycle-dependence of the magnetic
network itself. \cite{singh81} studied spectroheliograms spanning a
period of seven solar maxima and found the typical size of the chromospheric
network to be smaller at the maxima than at the minima of the cycle. A similar
conclusion was reached by \cite{kariyappa94}, \cite{berrilli99}
and \cite{raju02}, but \cite{wang88}
and \cite{munzer89} both reported an increase of network cell sizes
in regions of stronger magnetic activity instead. Using MDI magnetograms
spanning the first half of Cycle 23, \cite{meunier03} found an increase
of the size of magnetic elements at supergranulation-like scales with
solar activity (\cite{berrilli99} also used data obtained at
the beginning of Cycle 23 close to the activity minimum). However, in
the upper solar atmosphere (quiet chromosphere and transition region)
there is a clear decrease of the supergranular scale when the magnetic
cycle decreases \cite[][]{mcintosh_etal11,chatterjee_etal17}. These somehow
contradicting results show that magnetic tracers must be used with care
for this kind of measurements. The results are sensitive to the
thresholds used to identify the various field components (e.g network
or internetwork), and disentangling these effects is difficult.

Other studies have attempted to use proxies independent of
magnetic tracers of supergranulation to measure its size, notably
velocity features like positive divergences.
\cite{derosa04}, using two data sets obtained at periods of
different levels of magnetic activity, found smaller supergranulation
cell sizes in the period of high activity. A similar conclusion was
reached by \cite{meunier08}.  \cite{meunier07a} found a decrease of the
typical cell sizes with increasing field strength within supergranules,
but noted that larger supergranulation cells were associated with
stronger network fields at their boundaries. Hence, it seems that a
negative or a positive correlation can be obtained, depending on whether
the level of magnetic activity is defined with respect to internetwork or
network fields. \cite{meunier07a} also reported the absence of large
supergranulation cells for supergranules with large internetwork
magnetic field strengths, suggesting that internetwork fields (whose
existence is most likely rooted in a small-scale dynamo mechanism
independent of the solar cycle) do have a dynamical influence on
supergranules. We refer the reader to \cite{meunier07a} for a more
exhaustive discussion of the previous results and of the possible
shortcomings and biases of the various methods.

The last marked and long minimum that occurred in
2008 between solar cycles 23 and 24, prompted \cite{williams_pesnell11} to
compare the characteristic time scale, length scale and velocities of
supergranulation with the ones of the preceding minimum in 1996. Only
slight differences have been found. Finally, on the helioseismic side,
the dispersion relation for the supergranulation oscillations found by
\cite{gizon03b} appears to be only weakly dependent on the phase of
the solar cycle \citep{gizon04}. However, the same authors reported a
decrease in the lifetime and power anisotropy of the pattern from
solar minimum to solar maximum.  Overall, it is therefore fair to say
that a possible dependence of supergranulation on the solar cycle, if
any, appears to be relatively weak, and does not appear to
drastically affect its dynamics.

\subsubsection{Supergranulation and flows in active regions}\label{obsactive}
Let us finally consider the properties of surface flows at scales
comparable to supergranulation
within active regions and in the vicinity of sunspots. The reason for
doing this is two-fold. First, we may wonder how supergranulation
evolves locally during the formation or decay of an active
region. Second, the properties of flows around sunspots may give
us some hints of the effect of strong magnetic flux concentrations 
on the flow dynamics in the quiet  Sun.

Information regarding the first point remains scarce. \cite{R_etal10}
reported the disappearance of the supergranulation spectral peak in
the kinetic energy power spectrum of solar convection during the
emergence of two magnetic pores. While the pores (of a size comparable
to that of a granule) are emerging, the supergranulation flow becomes
very weak just like if the surrounding magnetic flux associated with
the pores had a significant impact on the flow. A related observation
by \cite{hindman09} shows that the fairly regular tiling of the
surface of the quiet Sun associated with
supergranulation is somewhat disorganised and washed away within
magnetic active regions.

As far as the second point is concerned, many studies in the past have
focused on the detection and characterisation of intrinsic flows
associated with sunspot regions (see \citealp{solanki03} and \citealp{thomas08} for 
exhaustive descriptions of sunspot structure and dynamics) and significant
observational progress has been made on this problem in recent years
thanks to local helioseismology
\citep{lindsey96,gizon00,zhao01,haber01,braun03,haber04,zhao04,zhao10,hindman09,komm_etal11,komm_etal12,kosovichev12,jain_etal16,loptien_etal17}.
The general picture seems to be as follows \citep[][but see also
\cite{degrave14b} for a discussion of the possible limitations of
helioseismic techniques in sunspots]{hindman09}: an annular outflow
called the moat flow \citep{sheeley69} is observed at the surface,
close to the sunspot. There is a corresponding return flow at depths
smaller than 2~Mm, so the moat circulation is fairly shallow. In
contrast, further away from the sunspot umbra, larger-scale
circulations characterized by a surface inflow and a deep ($> 10$~Mm)
outflow are inferred from helioseismic inversions.

The structure of the moat flow has been probed using the 
Doppler signal \citep{Sheeley_Batnagar71,Sheeley72}, the tracking of surface
features, such as granules \citep{MullerMena87,shine87} or small-scale
magnetic elements \citep{Sheeley72,harvey73,hagenaar05}, and with
helioseismology \citep{gizon00}. The outflow appears to have
properties similar to those of supergranulation \cite[see
notably][]{brickhouse88}, albeit with a larger velocity $\sim
1$~km/s. It remains unclear whether this strong-field dynamical
behaviour has anything in common with supergranulation though. The
outflow in this case is centred on a strong field region  whereas it
is the supergranulation inflow vertices that coincide with magnetic
flux concentrations in the quiet Sun \citep[see,
e.g.,][]{requerey17}.

\section{Classical fluid theory and phenomenological models}\label{theory}
\subsection{Context}
The big problem with understanding supergranulation has always been
the important challenges that its numerical simulation and formal
theoretical description present. Until the mid-2000s and the advent
of the first large-scale 3D numerical simulations, our computing
arsenal was too limited to probe almost any aspect of the multiscale
dynamical nonlinear complexity of large-scale solar surface dynamics,
leaving theoretical astrophysicists almost naked with just linear or
weakly nonlinear theories from the 1960--1990 era at hand, plus a few
generic hand-waving turbulence concepts from the same period. 
All the classical phenomenological models of supergranulation 
from this ``pre-numerics era'' are extremely simplified, qualitative and
speculative. Most of them have consequently proven
unfalsifiable and, as we shall see with the review
of numerical results in the next section, are in fact most likely too
idealized. To be fair, similar limitations have plagued research on many,
if not all nonlinear astrophysical fluid dynamics problems.
It also has to be recognized that such models are not devoid of
meaningful physical insights either, and have played an
important role in shaping the generic astrophysical fluid dynamical
phenomenological landscape over the years. This section is, therefore,
perhaps best understood as a testimony of the historical development
of the many different possible phenomenological scenarios for
supergranulation dynamics. Modern numerical developments,  and ongoing
theoretical discussions inspired by them as well as by some of the
observational results reviewed in the previous section, will be
reviewed in Sect.~\ref{numerics} and Sect.~\ref{discussion}.

The classical phenomenological models of supergranulation are
essentially of two types: those which assume that supergranulation is
rooted in thermal convection (i.e., it is driven by thermal buoyancy)
and rely on classical fluid convection theory, and those which do not
and rely on more general phenomenological concepts of turbulent dynamics.
In the following, we therefore first briefly introduce the rotating MHD
Rayleigh--B\'enard problem (Sect.~\ref{RB}), which provides the
simplest mathematical description of rotating magnetoconvection in a
fluid. While this system is not entirely adequate to describe
convection in the strongly stratified SCZ and close to the optically
thin surface \citep{nordlund82}, it is sufficient for the purpose of
discussing  the generic phenomenology of linear and turbulent
convection, most classical convection models of supergranulation, 
as well as some aspects of the nonlinear dynamics observed in the
simulations that will be reviewed in the next section.
 We then review different simple ``linearized'' thermal
convection models of supergranulation (Sect.~\ref{theoryconv}), and
other possible fluid dynamical mechanisms involving nonlinear
turbulent interactions and collective dynamics of smaller-scale
turbulence (Sect.~\ref{theoryLSinst}).

\subsection{Rotating, MHD Rayleigh--B\'enard convection}\label{RB}
\subsubsection{Formulation}
\begin{figure}[htb]
\centerline{\includegraphics[width=.8\linewidth]{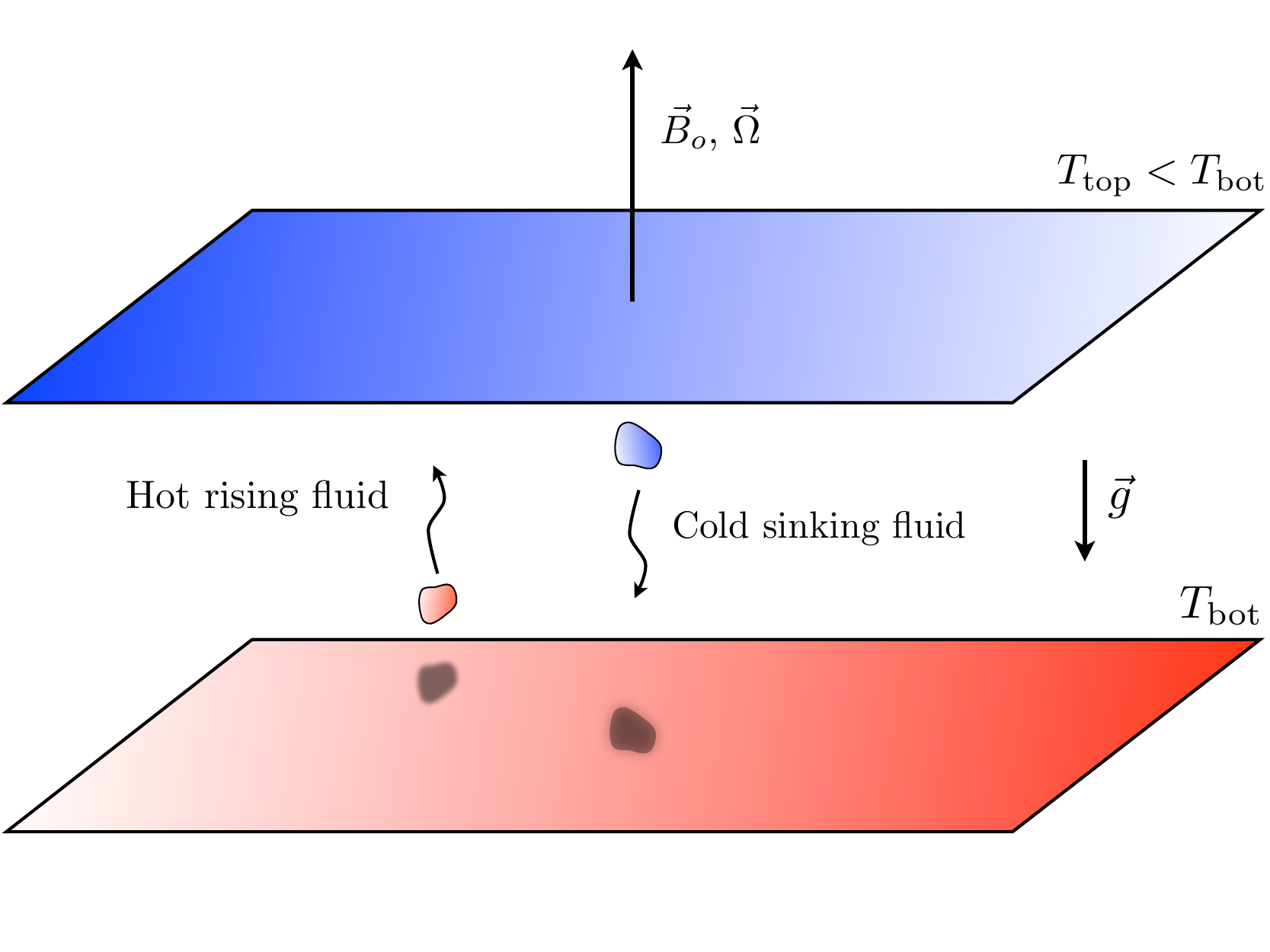}}
\caption{The rotating MHD Rayleigh--B\'enard convection problem.}
\label{figure:convec}
\end{figure}

The simplest mathematical formulation of the dynamical problem of
thermal fluid convection is the Rayleigh--B\'enard
problem describing convection between two differentially heated
horizontal plates, each held at a fixed temperature. This
model is derived under the Boussinesq approximation,
which amounts to assuming that the flow is highly subsonic and that
density perturbations $\delta \rho$ to a uniform and constant
background density $\rho_o$ are negligible everywhere except in the
buoyancy term $\delta\rho\,\vg$, where $\vg=-g\vec{e}_z$ stands for
the gravity \citep{chandra61}. The equilibrium background state is
a linear temperature profile with temperature decreasing from
the bottom to the top of the layer. 
Anticipating discussions of the effects of rotation and
magnetic fields on supergranulation, we consider the
case of an electrically conducting fluid threaded by a mean
vertical magnetic field denoted by $\vec{B}_o=B_o\vec{e}_z$ and
rotating around a vertical axis, with a rotation rate
$\vec{\Omega}=\Omega\,\vec{e}_z$. This set-up is pictured 
in Fig.~\ref{figure:convec}. 

In nondimensional form, the equations
for momentum and energy conservation, the induction equation, the
equations for mass conservation and magnetic field solenoi\-da\-lity
read

\begin{equation}
\label{bouss_nd}
\begin{array}{l}
\disp{\dtau{\vu}+\vu\cdot\na\vu +\sqrt{Ta}\pr\,\vec{e}_z\times\vu= -\na p + \RA\pr\,\theta\ez +
  Q\frac{\pr^2}{\prm}(\na\times\vec{B})\times\vec{B}+\pr\Delta\vu}~,\\
\\
\disp{\dtau{\theta} + \vu\cdot\na\theta- u_z = \Delta\theta}~,\\ \\
\disp{\dtau{\vec{B}} + \vu\cdot\na\vec{B} =
  \vec{B}\cdot\na\vu+\frac{\pr}{\prm}\Delta\vec{B}}~,\\ \\
\disp{\na\cdot\vu =0~,\quad \na\cdot\vec{B}=0}~,
\end{array}
\end{equation}
where the momentum equation has been written in the rotating frame,
lengths are measured in terms of the thickness of the convection
layer $d$, times are defined with respect to the  thermal diffusion
time $\tau_{\kappa}=d^2/\kappa$ ($\kappa$ is the thermal diffusivity),
the total magnetic field  $\vec{B}$ is expressed in terms of the Alfv\'en speed
$V_A=B_o/\sqrt{\rho_o\mu_o}$, temperature deviations  $\theta$ to the
initial linear temperature profile are measured in terms of the background
temperature difference $\Delta T=T_{\mathrm{top}}-T_{\mathrm{bot}}$
between the two horizontal plates enclosing the fluid in the vertical
direction.  Nondimensional velocity and pressure fluctuations are
denoted by $\vec{u}$ and $p$ respectively. This set of equations is to
be complemented by appropriate boundary conditions, most commonly fixed
temperature or fixed thermal flux conditions on the temperature, no-slip
or stress-free conditions on velocity perturbations, and perfectly
conducting or insulating boundaries for the magnetic field. Note
that the diffusive non-dimensionalization of the fluid equations used
above is a matter of convention. This formulation is faithful to the
historical development of the linear theory of rotating magnetoconvection
pioneered by \cite{chandra61}, and has been central to the formulation
of classical theories of many stellar convective phenomena ranging from
supergranulation to sunspot dynamics \citep{thomas08,weiss14}.

Several dimensionless numbers appear in the equations above,
starting with the Ray\-leigh number
\begin{equation}
 \RA = \frac{\alpha|\Delta T| gd^3}{\nu\kappa} = |N^2|\tau_{\nu}\tau_{\kappa}~,
\end{equation}
where  $\alpha$ is the thermal expansion coefficient of the fluid
defined according to $\delta \rho/\rho_o=-\alpha\, \theta$. Here,
$N^2=\alpha\Delta T g/d < 0$ is the square of the Brunt--V\"ais\"al\"a
frequency (negative for a convectively unstable layer) and
 $\tau_{\nu}=d^2/\nu$ is the viscous diffusion time, so the Rayleigh
 number measures the relative effects of the convection ``engine'',
 buoyancy, and of the ``brakes'', namely viscous friction and heat diffusion.
The second important parameter above is the Chandrasekhar number
\begin{equation}
Q=\frac{B_o^2d^2}{\rho_o\mu_o\nu\eta}=\frac{\tau_\nu\tau_\eta}{\tau_A^2}~,
\end{equation}
which is a measure of the relative importance of magnetic tension
($\tau_A=d/V_A$ is the Alfv\'en crossing time) on the flow in
comparison to magnetic diffusion ($\eta$ is the magnetic diffusivity,
$\tau_\eta=d^2/\eta$ is the typical magnetic diffusion time)
and viscous friction.
The relative importance of the Coriolis force in comparison to viscous
friction is measured by the Taylor number,
\begin{equation}
Ta=\frac{4\Omega^2d^4}{\nu^2}=(2\Omega)^2\tau_{\nu}^2~.
\end{equation}
Finally, $\pr=\nu/\kappa$ and $\prm=\nu/\eta$, where $\eta$ is the
magnetic diffusivity, stand for the thermal and magnetic Prandtl
numbers (see Sect.~\ref{turbulentscales}). 

\subsubsection{Linear theory}
In the simplest non-rotating hydrodynamic case ($Ta=Q=0$, no
induction), when the Rayleigh number is less than a critical value
$\RA_{\rm crit}$ that depends on the particular choice of boundary conditions,
diffusive processes dominate over buoyancy: the hydrostatic solution
is stable, i.e., any velocity or temperature perturbations decays. For
$\RA> Ra_{\rm crit} $, convection sets in as a linear instability
and perturbations grow exponentially in the form of convection rolls
or hexagons with a horizontal spatial periodicity $\lambda$ comparable
to the convective layer depth $d$ in most cases. For stress-free, fixed
temperature boundary conditions, $Ra_{\rm crit}=27\pi^4/4\simeq 657$ and
$\lambda_{\rm crit}/d=2\sqrt{2}$, while for rigid, fixed-temperature
boundary conditions,  $Ra_{\rm crit}\simeq 1707$ and $\lambda_{\rm crit}/d\simeq
2$ (the width of a individual convection roll is $\lambda/2$). Some effects
of magnetic fields and rotation on linear stability are discussed in
the next paragraphs.

\subsubsection{Turbulent renormalization of transport coefficients}
$\RA$ and $Ta$ computed from microscopic transport coefficients
(Sect.~\ref{turbulentscales}) are extremely large numbers in the
SCZ. Convective dynamics, therefore, takes place very far from the
actual convective linear instability threshold ($\RA/Ra_{\rm crit}\sim
10^{15}$--$10^{20}$). Therefore, using linear theory in this context to predict the
dominant scale of convection, for instance, does not 
\textit{a priori} seem appropriate. Is it possible to deal with this
problem simply theoretically?  A common argument is that the viscous,
thermal and magnetic diffusive transport properties of the plasma at
large scales are effectively set by the underlying vigorous turbulence
driven in the highly supercritical regime, so that transport
coefficients should be renormalized. In the context of large-scale
solar convection, turbulent transport coefficients may for
instance be estimated from the typical scale and velocity of
granulation, $\nu_T\sim L_GV_G$. The phenomenological rationale for 
doing this is that fluid systems driven strongly out-of-equilibrium react
dynamically in a way that effectively brings them back close
to their stability threshold by mixing the fluid in a way that
smoothes out unstable free-energy gradients. If we accept that the
renormalized turbulent $\RA$ (and perhaps also $Q$ and $Ta$) should be
close to its value at the stability threshold, then it makes sense
to use the simpler mathematical toolkits of linear and weakly
nonlinear analysis to describe the large-scale dynamics, rather than
solve the full, strongly nonlinear problem.

As we are about to see, linear and weakly non-linear models of
convection have proven very convenient and popular in the solar
context, mostly because they are solvable. Howevever, keep in mind
that they are at best only a quick, very approximate fix to make for
the lack of a better available dynamical theory of turbulence and
adequate numerics, and are ultimately fundamentally unsatisfying. 
In particular, their idealized nature implies that we have
almost no control over their accuracy in the dynamical regimes of
interest, apart from  an order of magnitude estimate of turbulent 
transport coefficients. While these models may be suggestive of
qualitative trends or effects, their actual predictive power 
is quite limited.

\subsection{Laminar convection theories of supergranulation}\label{theoryconv}
Following its discovery in the 1950s and further studies in the 1960s,
supergranulation was rapidly thought to have a convective origin,
very much like the solar granulation. Many theoretical models relying
on the basic ``convection cell'' phenomenology described in
Sect.~\ref{RB} have been devised to explain the apparently
discrete-scales regime of the dynamics of the solar surface
(namely the scales of granulation and supergranulation, but also that
of mesogranulation, see Sect.~\ref{obsmeso}). We will discuss a
few flavours of these laminar convection models, keeping in mind the
important caveats of the previous subsection.

\subsubsection{Multiple mode convection}
The simplest proposal for the emergence of a set of special scales is that
of multiple steady linear or weakly nonlinearly interacting modes of
thermal convection forced at different depths. The first theoretical
argument of this kind is due  to \cite{SL64}, who suggested that
supergranulation-scale motions corresponded to simple convection cells
driven at the depth of He$^{++}$ recombination and just advecting
granulation-scale convection. \cite{schwarzschild75}
invoked an opacity break, He$^{+}$ and H${^+}$ recombinations
as the drivers of supergranulation-scale convection. \cite{simon68} and
\cite{vickers71} suggested that deep convection
in the Sun had a multilayered structure composed of deep, giant cell
circulations extending from the bottom of the convection zone to
40~Mm deep, topped by a shallower circulation pattern corresponding to
supergranulation. \cite{bogart80} attempted to match a linear
combination of convective eigenmodes to the solar convective flux but
did not find that supergranulation came out as a preferred scale of
convection in this quasilinear framework.

\cite{antia81} argued that turbulent viscosity and diffusivity should
be taken into account in linear calculations, as they alter the
growth and scales of the most unstable modes of convection (but note
that  this assumption was already implicit in the laminar scenarios
described above). In their linear calculation with microscopic
viscosity and thermal diffusivity
coefficients replaced by their turbulent counterparts, granulation
and supergranulation show up as the two most unstable harmonics of
convection. Calibrating the amplitudes of a linear superposition of
convective modes to match mixing-length estimates of the solar convective
flux in the spirit of \cite{bogart80}, \cite{antia93} further argued that
they could reproduce the main characteristics of the power spectrum of
solar surface convection.

\cite{gierasch85} devised a one-dimensional energy model for the upper
solar convection zone from which he argued that turbulent dissipation
takes place and deposits thermal energy at preferred depths, thereby
intensifying convection at granulation and supergranulation scales.
On this subject, we also mention the work of \cite{wolff95}, who
calculated that the damping of $r$-modes in the Sun should
preferentially deposit heat 50~Mm below the surface
as a result of the ionisation profile in the upper solar convection
zone. He then argued that this process might lead to convective
intensification at similar horizontal scales.

\subsubsection{Effects of temperature boundary conditions}
A somewhat different phenomenological proposal was made by
\cite{vdb74}. He considered the case of steady finite-amplitude
thermal convection cells in the presence of fixed
heat flux boundary conditions imposed at the top and bottom of the
layer, and showed that the convection pattern in this framework has
much smaller temperature fluctuations than in the standard
Rayleigh--B\'enard model with fixed temperature boundary
conditions. This is interesting in the context of the
supergranulation problem, considering that intensity fluctuations at
supergranulation scales are somewhat elusive (see Sect.~\ref{obsintensity}).

Fixed heat flux boundary conditions naturally favour marginally stable
convection cells with infinite horizontal extent
compared to the layer depth, or  convection cells with a very large
but finite horizontal extent when a weak modulation of the heat flux is
allowed for \citep{sparrow64,hurle67,vdb74,busse78,chapman80,depassier81}.
This case is, therefore, very different from  the standard Rayleigh--B\'enard
case with fixed temperature boundary conditions, which gives rise to
cells with aspect ratio of order unity. The interesting qualitative feature
of this boundary-condition effect is that it naturally selects very flat,
anisotropic convection cells and, therefore, alleviates the need to invoke
convection at depths comparable to the typical horizontal scale of
supergranulation.

\subsubsection{Oscillatory convection and the relative role of dissipative processes}
The discovery by \cite{gizon03b} that supergranulation has wave-like
properties (Sect.~\ref{obsrot}) paved the way for new theoretical
speculations. In particular, it offered an opportunity to revive the
interest for theories of oscillatory convection dating back to the
work of \cite{chandra61}. Such a behaviour
requires the presence of an extra restoring force acting on the convective
motions driven by buoyancy. This force can be provided by Coriolis effects
(rotation) or magnetic field tension. The existence of oscillatory
solutions is known to depend very strongly on how various
dissipative processes (viscous friction, thermal diffusion and ohmic
diffusion) compete in the flow. This is usually measured or
parametrized in terms of the thermal Prandtl number $\pr=\nu/\kappa$,
where $\nu$ is the kinematic viscosity and $\kappa$ is the thermal
diffusivity, the magnetic
Prandtl number $\prm=\nu/\eta$, where $\eta$ is the magnetic
diffusivity, and the ``third'' Prandtl number
$\zeta=\eta/\kappa=\pr/\prm$. In the Sun, $\pr\sim 10^{-4}-10^{-10}$,
$\prm\sim 10^{-2}-10^{-5}$ (see Sect.~\ref{turbulentscales}) and
$\zeta \ll 1$ at the photosphere.

\subsubsection{Convection, rotation and shear}
As mentioned in Sect.~\ref{obsrot}, supergranulation-scale flows
are weakly influenced by the global solar rotation. In the presence of
a vertical rotation vector, overstable oscillatory
convection is preferred to steady convection provided that $\pr$ is
small \citep{chandra61}. Physically, an oscillation is only possible
if inertial motions are not significantly damped viscously on the
thermalization timescale of rising and sinking convective blobs. 
\cite{busse04,busse07} suggested on the basis of a local Cartesian analysis
that the drift of supergranulation could be a signature of weakly
nonlinear thermal convection rotating about an inclined axis and found
a phase velocity consistent with the data of \cite{gizon03b}, assuming
an eddy viscosity prescription consistent with solar estimates (based
on the typical sizes and velocity of granulation). Earlier work on the
linear stability of a rotating spherical Boussinesq fluid layer
heated by internal heat sources showed that the most rapidly growing
perturbations are oscillatory and form a prograde drifting pattern of
convection cells at low Prandtl number in high Taylor number regimes
corresponding physically to large rotation \citep{ZB87}. 

Related to this issue is the influence of differential rotation, or
shear, on supergranulation. \cite{green06} considered the possible
role of the solar subsurface shear layer \citep{schou98} by looking at the 
effect of a vertical shear flow on the onset of convection in a
strongly stratified Cartesian layer using linear theory, and 
found that steady convective modes become travelling when a weak shear
is added. Earlier studies had found that this behaviour is possible
either at low $\pr$ \citep{kropp91} or if some generic form of
top-down symmetry breaking is present in the system
\citep{matthews97}. As a linear shear flow alone does not break this
symmetry, it is likely that density stratification plays an
important role in obtaining the result. \cite{green06} also found
that the derived phase speeds of the travelling pattern were
significantly smaller than those inferred from the observations 
by \cite{gizon03b}.
 
Note that the relative orientations and amplitudes of rotation, shear
and gravity are fundamental parameters in the sheared rotating
convection problem. The results (e.g., the pattern phase
velocity and wavelength) of local Cartesian theoretical models of
supergranulation incorporating solar-like rotation effects are,
therefore, expected to depend significantly on latitude. This
is a  problem with local laminar models of sheared rotating
convection, because in practice there is no conclusive observational
evidence for a very strong latitudinal dependence of the scales or
propagation of supergranulation (see Sect.~\ref{obsrot}).  Laminar
spherical models do not necessarily suffer from this problem to the
same extent, as they predict global modes with a well-defined phase
velocity.

\subsubsection{Convection and magnetic fields}\label{magnetoconv}
The presence of magnetic fields in convection can have diverse
consequences, the most obvious of which is a coercive effect
on convective motions through magnetic tension. 
Such a coercive effect on the scale of supergranulation
was actually discussed very quickly after its discovery.
Much of the discussion at the time involved an equipartition argument
\citep[e.g][]{parker63,SL64,clark67,simon68,frazier70,parker74, frazier76}.
Many flux concentrations in the network are actually well above
equipartition with the supergranulation flow field. Using the typical
value for the velocity field at supergranulation scales given in
Sect.~\ref{velo_sc} and an order of magnitude estimate for the
plasma density in the first 1~Mm below $\tau=1$, we see that for the
kinetic and magnetic energy densities to be comparable in the
supergranulation peak range, an rms magnetic field
strength of 100~G is required:
\begin{equation}
  \label{kenergy}
E_{\mathrm{kin}}=45\,\left(\frac{\rho}{10^{-3}\,\mathrm{kg/m^3}}\right)\left(\frac{V}{300~\mathrm{m/s}}\right)^2~\mathrm{J
\ m}^{-3}~,
\end{equation}
\begin{equation}
  \label{magenergy}
E_{\mathrm{mag}}=40\,\left(\frac{B}{100~\mathrm{G}}\right)^2~\mathrm{J\
  m}^{-3}~.
\end{equation}
Hence, the magnetic energy density of strong network elements with kG fields appears
to be roughly 100 times larger than that of the supergranulation flow.  
This result first suggests that supergranulation-scale motions  cannot alone
generate these flux concentrations. Partial evacuation of  density and
vigorous localized motions such as granulation-scale motions
seem to be required to obtain superequipartition fields
\citep{webb78,spruit79,spruit79b,unno79,proctor83,hughes88,bushby08}.
Now, the question is whether such strong, but very localized and
inhomogeneous fields can have a strong dynamical impact on the
flow. Obviously, numerical simulations provide the best way to test
this, but some ``effective'' phenomenological theories have also been
considered in this context.  An interesting proposal of this kind is
due to \cite{longcope03}, whose (theoretical) calculations
suggest that the dynamical feedback of a distribution of magnetic
fibrils embedded into the solar plasma physically translates into a
large-scale viscoelasticity.

The forementioned idea of large-scale convection with fixed heat flux
boundaries was also further carried out with the addition of a uniform
vertical magnetic field threading the convective layer. Contrary to the
hydrodynamic case for
which zero-wavenumber neutral (zero growth-rate) solutions are
preferred linearly, convection cells with a long but finite horizontal
extent dominate in the magnetised case, provided that the magnetic
field exceeds some threshold amplitude.  The horizontal scale of the
convection pattern in this model is controlled directly
by the typical strength of the magnetic field. \cite{murphy77} was the first to
suggest that this model might be relevant to supergranulation, and the linear
problem in the Boussinesq approximation was solved by \cite{edwards90}.
\cite{rincon03,rincon04} further solved the fully compressible linear problem
numerically and revisited it in the context of supergranulation. Using
typical solar values as an input for their model parameters (density scale
height, turbulent viscosity etc.), it was shown that the average magnetic field
strength (measured in the nondimensional equations by the Chandrasekhar
number $Q$) required for compressible magnetoconvection with fixed heat
flux to preferentially ``select'' supergranulation-scale convection
was a rather reasonable 100~G (Sect.~\ref{obsnet}).

Magnetoconvection in a uniform vertical magnetic field is also known
to preferentially be oscillatory at onset provided
that $\zeta \ll 1$ \citep[e.g.,][]{chandra61,proctor82}, a phenomenon
known as magnetic overstability. This is also possible for non-vertical
magnetic fields \citep[e.g.,][and references
therein]{matthews92,hurlburt96a,thompson05}. Physically, field lines
can only be bent significantly by convective motions and act as a
spring if they do not slip too much through the moving fluid, which
requires, in this context, that the magnetic diffusivity of the fluid
be small enough in comparison to its thermal diffusivity.
Since $\zeta\ll 1$ in the quiet photosphere, oscillatory
magnetoconvection represents a possible option to explain the wavy
behaviour of supergranulation. On this topic, \cite{green07}
built on the work of \cite{green06} by considering the linear theory 
of sheared magnetoconvection in a uniform horizontal (toroidal) field
shaped by the subsurface shear layer. They found that the phase speed of
the travelling waves increases in comparison to the hydrodynamic 
case studied by \cite{green06} and argue that the actual phase speed
measured by \cite{gizon03b} can be obtained for a reasonable uniform
horizontal field strength of 300~G.

Yet another possible magnetic feedback mechanism is provided
by the interactions between magnetic fields and
radiation. Observations, theory and simulations
suggest that magnetic concentrations tend to depress the opacity
surfaces of the photosphere, which in turn is thought to channel
radiation outwards \citep{spruit76,vogler05}, and it has been argued
that strong magnetic concentrations at network scales may thereby 
alter the convection process at supergranulation scale and
consequently single this scale out in the energy spectrum.

\subsubsection{Dissipative effects}
Finally, it is known both theoretically and experimentally that even in the
absence of any effect such as magnetic couplings, rotation or shear,
the value of the thermal Prandtl number can significantly affect the
scales and time evolution of convection, both in the linear and nonlinear
regimes. Its value notably controls the threshold of secondary oscillatory
instabilities of convection rolls \citep{busse72}. At very low Prandtl
numbers, \cite{thual92} showed that a rich dynamical behaviour
associated with the interactions between the primary convection mode and
the secondary oscillatory instability takes place close to the convection
threshold. This includes travelling and standing wave convection.
Overall, most theoretical studies of supergranulation to date have
been either ideal (no dissipation) or for $\pr\sim\prm\sim 1$. For
this reason, some important thermal effects relevant to
supergranulation-scale convection may well have been overlooked until
now. Note that this limitation also applies to most numerical simulations.

\subsection{Large-scale instabilities, inverse cascades and collective interactions}\label{theoryLSinst}
A few other qualitative arguments and models have been put forward to
explain the origin of supergranulation besides thermal convection scenarios.
The generic proposal of these arguments is that supergranulation could
emerge from collective nonlinear interactions of small-scale
structures such as granules, for instance through a large-scale
instability reminiscent of inverse-cascade scenarios in flows with
specific spatial symmetries.

\subsubsection{Rip currents and large-scale instabilities}
The first work along this line of thought is due to
\cite{cloutman79}, who proposed to explain the origin of
supergranulation using the physical picture of rip currents on the
beaches of oceans: the repeated breaking of waves on beaches induces rip currents
flowing parallel to the coast line. On the Sun, he qualitatively
associated breakers with the rising flows of granules breaking into
the stably stratified upper photosphere.
This idea provides an illustration of a broader
phenomenological suggestion by \cite{rieutord2000} that the collective
interaction of solar granules may give rise to a large-scale
instability at supergranulation scales. The idea finds its root in
theoretical work on energy localization processes in nonlinear lattices
\citep{dauxois93} and large-scale instabilities (``negative eddy
viscosity instabilities'') of flows with particular spatial
symmetries, such as the Kolmogorov flow
\citep{meshalkin1961,sivashinsky85} or the decorated hexagonal flow
\citep{gama94}. Asymptotic theory on simple prescribed vortical flows can
be performed under the assumption of scale separation \citep{dubrulle91}
between the basic periodic flow and the large-scale instability mode. In
such theories, the sign and amplitudes of the turbulent viscosities
is found to be a function of the Reynolds number. For instance,
an asymptotic theory based on a large aspect ratio expansion 
was developed by \cite{newell90} for thermal convection. In this
problem, large-scale instabilities take on the form of a slow, 
long-wavelength modulation of convection roll patterns.
Their evolution is governed by a phase diffusion equation with
tensorial viscosity. In the case of negative effective parallel diffusion
(with respect to the rolls orientation), the Eckhaus instability sets
in, while the zigzag instability is preferred in the case
of negative effective perpendicular diffusion. In relation with our
earlier discussion of the renormalization of transport coefficients,
note that phase-instabilities like this one provide a very clear and
quantitative example as to why postulating  a generic positive
turbulent diffusion to parametrize the effects of small-scale motions
on large-scale dynamics is not always appropriate. 

\subsubsection{Plume and fountain interactions}
Yet another phenomenological dynamical argument put forward to explain the
origin of supergranulation is that the pattern results from the collective
interaction of ``plumes'' (buoyantly driven rising or sinking
flows). Plumes can be either laminar or turbulent, however, turbulent
structures have by far received most of the attention because of their
numerous applications \cite[see][]{turner86}. The first numerical
simulations of strongly stratified convection at high enough
Reynolds numbers \cite[e.g.,][]{nordlund89,RT93} quite clearly showed the
importance of vigorous sinking plumes in turbulent convection, and the
results prompted \cite{RZ95} to study the fate of these downdrafts in
some mathematical detail. Unlike the downflows computed in early
simulations, solar plumes are turbulent structures, which entrain the
surrounding fluid (see Fig.~\ref{figure:plume_conv}). As \cite{RZ95}
pointed it out, the mutual entrainment and merging of these plumes
naturally leads to an increase of the horizontal scale as one proceeds
deeper.

Toy models have been elaborated in this context to investigate
the properties of ``$n$-body'' dynamical advection-interactions between
plumes. For instance, \cite{rast03b} developed a model in which a
two-dimensional  flow described by a collection of individual divergent
horizontal flows (``fountains'') mimicking granules is evolved under a
simple set of rules governing the merging of individual elements into
larger fountains and their repulsion (this description was inspired by
an earlier kinematic model of flows in exploding granules by
\cite{simon91}). For some parameters
typical of the solar granulation ``fountains'', he argued that the
clustering scales of the flow after a long evolution of the system
resembled that of mesogranulation and
supergranulation. A similar model incorporating simplified magnetic field
dynamics was designed by \cite{crouch07}. They observed some magnetic
field organisation and polarity enhancement at scales similar to that
of supergranulation in the course of the evolution of the model. While
they may help to form an intuition of large-scale dynamical interactions 
in turbulent convection, the main caveat with models like this, of
course, is that they are not rooted in the actual fluid dynamical
equations.

\section{Numerical modelling}\label{numerics}
Our understanding of turbulent convection in general, and solar
convection in particular, has made huge progress over the last fourty
years thanks to numerical simulations. The numerical study of solar
convection was limited to granulation scales for a long time,
but there has finally been some important progress on
supergranulation-scale dynamics over the last fifteen years. The 
first part of this section provides an introduction to the potential
and limitations of numerical modelling, and describes a few
important results on the Rayleigh-Benard problem and granulation-scale
solar convection relevant to our discussion of supergranulation.
We then review the main achievements (and difficulties) of numerical
simulations of supergranulation-scale dynamics, and discuss their
implications for our current  understanding of the problem.

\subsection{Introduction to convection simulations}
\subsubsection{General potential and limitations}
Numerical simulations are the most important tool that we have to probe and
explore the nonlinear dynamics of hydrodynamic turbulent flows and
other complex fluid physics such a radiation hydrodynamics or MHD.
They are not a panacea though and face many limitations of their own,
including in the solar physics context. In particular, the finite
capacities of computers  make it completely impossible for any type of
simulation, even today, to approach dissipative flow regimes
characteristic of the solar surface and to span all the range
of time and spatial scales involved in the problem. It should,
therefore, be constantly kept in mind when discussing numerical
simulations of solar convection (and more generally of laboratory and
astrophysical turbulence) that we are not actually ``simulating
the Sun'', but rather a fairly quiet toy model of it. This being said,
large-enough simulations can give us important insights into
semi-quantitative trends about the nonlinear dynamics, statistics and
self-organization of turbulent flows, which is exactly the kind of
information needed to address the supergranulation problem.

\subsubsection{Turbulent Rayleigh--B\'enard convection vs.\ Navier--Stokes turbulence}
Despite its simplicity compared to the actual strongly
stratified solar convection problem (which also involves radiation),
the problem of incompressible, non-rotating turbulent convection
in the Rayleigh--B\'enard framework introduced in Sect.~\ref{RB} is
very important in the context of this review, because it contains both
buoyancy effects and all the nonlinearities underlying
the typical turbulent dynamics of a fluid. Numerical simulations of
this problem can, therefore, tell us a lot about the generic dynamics of
turbulent convection in the absence of any other diverting physical
effect.

The Boussinesq equations are very similar to the forced Navier--Stokes
equations: the main difference between the two is that the forcing
term in the Rayleigh--B\'enard problem is not an external body force, but is
determined self-consistently from the time-evolution of the temperature
fluctuations. Both experimental and numerical evidence at order one
thermal Prandtl number strongly
suggest that the basic phenomenology of Rayleigh--B\'enard turbulence
at vertical and horizontal scales smaller than the vertical
extent of the domain considered  should be similar to that of
Navier--Stokes turbulence in the inertial-range
\citep{rincon06,lohse10,kumar14}. In particular, 
the dynamics becomes increasingly isotropic at smaller scales.
To a very good first approximation, the numerical issues and
requirements to simulate the Rayleigh--B\'enard problem at very high
Rayleigh numbers are, therefore, the same as those pertaining to the
simulation of forced Navier--Stokes turbulence at high Reynolds
numbers. In particular, a strong scale separation between the
vertical size of the numerical domain and the grid size is required to
simulate the turbulent  cascade properly.  High-resolution simulations
using spectral methods \citep{canuto} remain the most adequate tool to
reach highly-supercritical regimes. The highest-supercritical
simulations to date ($\RA/\RA_{\rm crit}\sim 10^{11}$, see
\cite{verzicco03,amati_etal05,lohse10}) require numerical spectral
resolutions of the order of $\sim 512^3$ to $\sim1000^3$ grid points
to simulate turbulent convection in numerical domains of comparable
vertical and horizontal size (order one aspect ratios). This
is significant even by today's computing standards. An extra
difficulty of the turbulent Rayleigh--B\'enard problem, compared to
standard Navier--Stokes turbulence in a periodic box, is the need to
resolve very fine thermal boundary layers at the top and bottom
boundaries. These transition layers scale as a fractional power of
$\RA/\RA_{\rm crit}$ and, therefore, become increasingly difficult to
resolve at high $\RA$.

A critical implication of these strong numerical
constraints in the context of supergranulation modelling is that
simulations of turbulent convection at very high Rayleigh
numbers are still restricted to fairly low aspect ratios (the ratio
between the horizontal and vertical extents of the domain), typically
one-half or one. Simulations of turbulent convection dynamics at  horizontal
scales signicantly larger than the vertical size of the domain are 
possible but currently limited to mildly supercritical, soft
turbulence regimes, typically $\RA\sim 10^{5}$--$10^7$ ($\RA/\RA_{\rm crit}\sim
10^{3}$--$10^5$). Most simulations are also currently limited to
Prandtl numbers of order unity, which makes it difficult to
investigate the effect of scale separations between the various
dissipation scales of the problem.

\subsubsection{Solar convection models}
Numerical simulations of astrophysical convection are similar in many
ways to Rayleigh--B\'enard simulations and are, therefore, plagued by
the same numerical limitations. But they also include a variety
of extra physical effects relevant to the particular solar context,
which are not present in the Rayleigh--B\'enard,
adding a further layer of complexity. This includes
strong vertical density stratification, mild compressibility (up to
Mach number of order one), radiative transfer in the surface layers,
and the lack of a well-defined rigid bottom boundary. 
The first direct numerical simulations of strongly stratified convection
date back from the 1970s \citep{graham75}, and the first
numerical study of stratified convection incorporating radiative
transfer effects is due to \cite{nordlund82}. Since then,
numerical studies of astrophysical convection have split into two
``families'' of models that define the main trends in the field nowadays.

\textit{``Idealized simulations''} rely on simple models of 
stratified atmospheres such as polytropes, and implement the standard
incompressible, anelastic or compressible fluid equations, including
explicit viscosity, thermal and magnetic diffusivities in a 
domain usually bounded by walls in the vertical direction (just as in
the Rayleigh--B\'enard problem). These simulations do not explicitly
incorporate radiative transfer but instead rely on thermal diffusion
in the entropy equation to model its effects. This is of course not
totally appropriate in the optically thin surface layers, and makes it
difficult to directly compare the results with solar observations of light
intensity. Idealized simulations, on the other hand, are usually very
good at describing the nonlinear dynamics in highly-supercritical
regimes. They often rely on numerical spectral methods, which
remain the gold standard in simulations of incompressible homogeneous
turbulence \citep{vincent91,ishihara09}, but also face a few problems
when it comes to the simulation of compressible flows: for
instance, they cannot capture shocks easily. This is not a major 
issue at mild $\RA$ where the dynamics is quite subsonic,
but becomes a problems as $\RA$ increases \cite{cattaneo91}. 
A popular approach used in global spherical convection simulations
\citep{clune99} is to solve the anelastic equations instead
of the fully compressible problem. This approach filters out sound
waves, thereby removing a bunch of numerical problems and constraints.

\textit{``Realistic'' simulations} of solar convection
\citep{nordlund09}, on the other hand, aim at maximum astrophysical
realism by taking into account not only the stratified flow dynamics but also
other important physical processes in the solar context such as
radiative transfer, huge solar-like density stratifications and
realistic equations of state including Helium and Hydrogen
ionisations. Unlike idealized simulations, they usually ignore the
explicit physical dissipative processes like viscosity and instead
rely on grid dissipation or hyper-dissipation to avoid numerical
blow-up. These features, coupled to the use of ``handmade'' open
boundary conditions, makes this kind of simulations more realistic
to solar physics studies, in the sense that some of the physics
simulated is closer to that in the SCZ, and their results can be
directly compared with solar observations. However, grid-based
methods also generically offer less control over dissipative processes and
are usually more dissipative for a given numerical resolution than
spectral methods in simulations of homogeneous turbulence
characterized by space-filling fine-scale gradients. This limitation
does not appear to be an important issue for simulations of solar
granulation for which thermal radiation, which is well-accounted for
in such simulations, plays an overwhelming dynamical role. But, as we
shall see in Sect.~\ref{simRB}, understanding the large-scale dynamics
requires to understand not only the thermal physics, but also how
dynamical nonlinearities play out. Idealized simulations offer an
exact control over the supercriticality of the system and can reach
slightly more asymptotic regimes at comparable resolution. A second
potential problem of realistic simulations is that  they could be in
the wrong dissipative regimes, simply because they do not take into
account rigorously the disparity of time and length scales of 
different dissipative processes. This may notably be an issue
when addressing MHD effects in solar convection.

\subsection{Small-scale simulations}
We will now review a few  turbulent convection simulations in
``small'' order-one aspect ratio domains but whose results are also
directly relevant to the question of supergranulation-scale
dynamics.

\subsubsection{Turbulent Rayleigh--B\'enard convection}\label{simRB}
\begin{figure}[ht]
\vspace{0.1cm}
\centerline{(a)
\parbox[t]{0.28\linewidth}{\vspace{-10pt}\includegraphics[width=\linewidth]{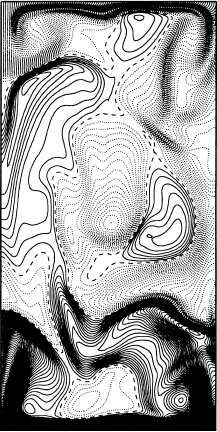}}
(b)
\parbox[t]{0.28\linewidth}{\vspace{-10pt}\includegraphics[width=\linewidth]{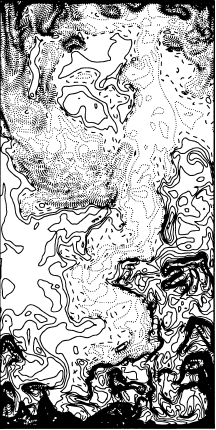}}
(c)
\parbox[t]{0.28\linewidth}{\vspace{-10pt}\includegraphics[width=\linewidth]{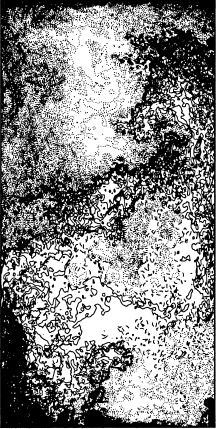}}
}
\caption{
Snapshots of temperature fluctuations in a vertical plane, from numerical
simulations of Rayleigh--B\'enard convection in a slender cylindrical
cell at $\pr=0.7$ and Rayleigh-numbers (a) $2\times 10^7$ (b) $2\times
10^9$ (c) $2\times 10^{11}$ \citep[image credits:][]{verzicco03}.}
\label{figure:verzicco}
\end{figure}

Turbulent convection dynamics in increasingly supercritical
Rayleigh--B\'enard simulations in a slender cylindrical cell of aspect
ratio one-half (for $\pr=0.7$ and up to $\RA=10^{11}$) is illustrated in
Fig.~\ref{figure:verzicco}. The figure shows temperature snapshots
in vertical planes and underlines two very important points. First,
there is a very marked evolution of the dynamical pattern from
moderate to very large $\RA$. An asymptotic large $\RA$ regime 
is not attained even at $\RA=10^{11}$. Such a value is way smaller 
than the $\RA$ in the SCZ, but also way larger than $\RA$ in any
current simulation of solar convection! This illustrates the problem
of accessing asymptotic dynamical regimes in numerical simulations.
Second, two large-scale circulations emerge from the
small-scale turbulent fluctuations as $\RA$ increase.  These
so-called thermal winds are also observed in laboratory experiments
on convection  \citep[e.g.,][and references
therein]{krishna81,sano89,niemela01,xi04}. As we will see in
Sect.~\ref{simlargelocal}, these structures are essentially
buoyant but they do not necessarily correspond to the most unstable
linear convection mode in the problem. They are instead a consistent
outcome of the nonlinear self-organization of the buoyant dynamics. It is
clear that these large-scale thermal winds are strongly constrained
laterally in low aspect ratio geometries, and are likely to
take more horizontal space in larger aspect ratio domains. This is 
a very important point in the context of supergranulation, and will
also be discussed in Sect.~\ref{simlargelocal}.

\subsubsection{Stratified convection simulations at granulation scales}\label{simgra}
The first realistic three-dimensional simulation of small-scale solar
surface convection of \cite{nordlund82} mentioned earlier was followed by
improved version at higher resolutions \citep{nordlund89, stein98}. 
These studies primarily focused on understanding the thermal structure and
observational properties of granulation. \cite{nordlund89} notably found
that convective granulation-scale plumes in a stratified atmosphere
merged into larger plumes at larger depth, producing increasingly
large convective patterns deeper and deeper. The results also
demonstrated the influence of stratification on the horizontal extent
of granules and on the typical plasma velocity within granules. 

Early idealized two-dimensional simulations of stratified convection
by \cite{graham75,chan82,hurlburt84} had already revealed an asymmetry
between up and downflows in two-dimensional simulations of
compressible convection in a stratified medium. The physical origin of
this phenomenon, known as buoyancy braking, was explicited by
\cite{massaguer80}. This kind of simulations were later expanded to 
Rayleigh numbers up to 1000 times supercritical in two dimensions
\cite{chan89,chan96}, and to three dimensions by \cite{cattaneo91}.
The results of these early idealized small-scale simulations
are qualitatively in line with those of \cite{nordlund89} as far as
the deep, large-scale dynamics is concerned. This suggests that an
explicit modelling of physical processes such as radiative transfer 
is not essential to study the turbulent dynamics and
scale-interactions. These simulations also showed that the presence or
absence of a solid bottom wall may be important, as the former tends to
generate more shear and large-scale recirculations in deep layers.

Readers interested in the particular problem of granulation-scale
convection modelling will find much more detailed information in the
reviews by \cite{nordlund09,Asp09}.  As explained earlier, the surface
features of granulation-scale convection are much better understood
with realistic simulations which are specifically tailored for this
purpose. However, these simulations are also very dissipative, and it
still remains a bit unclear whether this kind of modelling can
faithfully describe turbulent convection in strongly nonlinear,
supercritical regimes typical of the large-scale dynamics of in the SCZ.

\subsection{Large-scale simulations}\label{simlarge}
\subsubsection{Global vs. local simulations}
In the 1980s and 1990s, global (spherical) and local numerical
simulations of solar-like convection were respectively limited to the
study of global-scale convection dynamics (giant cells and larger)
and  granulation-scale dynamics. The numerical study of the
dynamics at scales larger than that of individual granules, but
smaller than global solar scales, only became possible in the
early 2000s. It is interesting in retrospect to recall as a short
anecdotal digression the following optimistic citation extracted from
an article of \cite{nordlund85}:
\emph{``There is a need for numerical simulations at the scale of
  supergranulation [\dots] This is probably feasible with present day
  computers and numerical methods.''}

There are two different possible approaches to the simulation of
supergranula\-tion-scale dynamics: local Cartesian simulations (taking a
small patch of the solar surface and ignoring curvature effects) and
global simulations in a spherical shell. 
In the local approach, the horizontal box size of the largest
simulations to date (i.e., the largest horizontal scale of the
simulations) is roughly a few times the
horizontal scale of supergranulation
\cite[][]{lord_etal14,cossette_rast16}. While numerical resolutions
of 512 to 1024 grid points in each spatial direction are now routinely
achievable, simulating the dynamics at 100~Mm horizontal scales still
requires to somewhat sacrifice the resolution of dynamical processes
at scales of the order $\sim 100~$km  only mildly smaller than that of
granulation, and way above the actual viscous dissipative cut-off at
the solar surface in any case. Note also that the dynamics at
supergranulation scales  in this kind of set-up remains slightly
constrained by (usually periodic) lateral boundary conditions. 

Global spherical simulations can model the dynamics at scales
significantly larger than supergranulation, but their problem is that
the smallest scales included in even the latest-generation simulations
of this kind are of the order of a few Mm, only mildly smaller than
scale of supergranulation. Hence, the supergranulation-scale
dynamics in these simulations is much more dissipative than in local
simulations. This is a significant issue because, as shown in
Sect.~\ref{obsvel}, the turbulent spectrum of solar surface
convection shows that supergranulation is located at the large-scale
edge of the injection range of turbulence, not in the dissipation
range. The dominant physical processes in these different regimes 
are obviously very different. Another potentially very important
weakness of global simulations in this context is the usually very
crude way in which they deal with the thermodynamic surface boundary
layer at the photospheric transition. As we shall see, the
thermodynamic structure in the first few Megameters below the 
surface now increasingly appears to play a significant role in setting
the horizontal scale of supergranulation-scale dynamics.

\subsubsection{Global spherical simulations}\label{simglobal}
Global spherical simulations of turbulent convection started to appear 
more than forty years ago. The first numerical model of 2D Boussinesq
convection in a spherical shell is due to \cite{gilman75}, who
used it to study the influence of rotation on convection, the problem
of large-scale circulations in the solar convection zone, and that of
the interactions between supergranulation and rotation
\citep[e.g.,][]{gilman79}. Spherical convection modelling was soon
extended to the anelastic approximation by \cite{gilman81} and
\cite{glatzmaier84,glatzmaier85}, and to the fully compressible fluid
equations by \cite{valde91}.
Most simulations in spherical geometry use the expansions of the
fields on spherical harmonics up to a given resolution $L_{\rm max}$
(the $\ell$ order of the smallest scale spherical harmonic).
The early simulations were restricted to fairly laminar regimes and
very large solar scales, namely $L_{\rm max}=32$ or, in terms of 
smallest resolved horizontal scale, $\lambda = 2\pi R_\odot/L_{\rm
  max} \simeq 130~{\rm Mm}$~, which is much larger than the scale of
supergranulation (36~Mm, or $\ell=120$, see Sect.~\ref{obsvel}).

\begin{figure}[t]
\centerline{\includegraphics[width=0.9\linewidth]{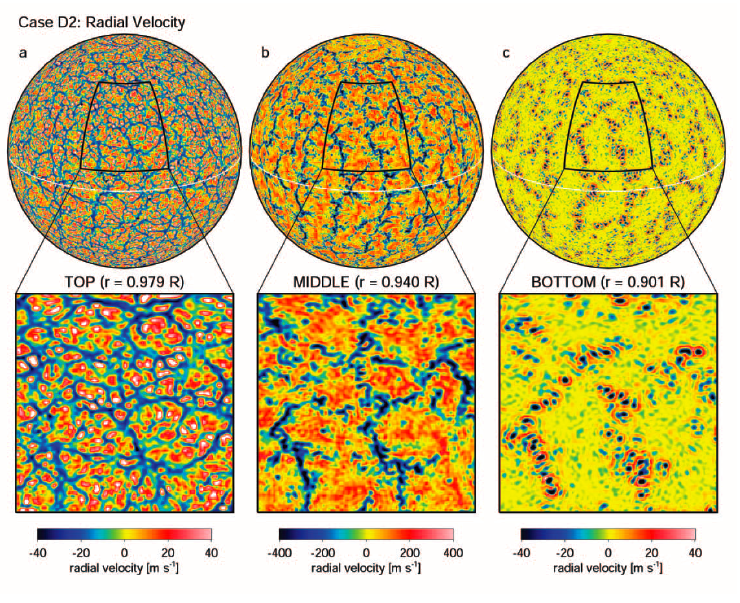}}
\caption{
Radial velocity snapshots at various depths in global simulations of
convection in shallow spherical shells, down to supergranulation
scales \citep[image credits:][]{derosa02}.}
\label{figure:derosa}
\end{figure}

The first dedicated attempt to study supergranulation-scale dynamics
through global spherical simulations
is due to \cite{derosaphd}, \cite{derosa01} and \cite{derosa02}, who
carried out idealized three-dimensional hydrodynamic simulations in thin
spherical shells with a horizontal resolution of $L_{\rm max}=340$,
corresponding to a smallest resolved horizontal scale of 13~Mm;
they used the popular high-resolution three-dimensional
spherical simulations ASH code \citep{clune99}. The simulations
exhibit structures at scales larger and comparable to that of
supergranulation (see Fig.~\ref{figure:derosa}). However, 
it was difficult to diagnose why supergranulation scales would play a
special role (except for being in the dissipative range) in these
simulations, and  the fact that the grid was so close to the
supergranulation scale  prevented them from drawing any robust
conclusion regarding the physical origin of
supergranulation. Simulations of solar-like convective shells by
\cite{miesch_etal08} at higher spherical harmonics resolution ($L_{\rm
  max}=682$) have further revealed the presence of intense cyclonic
downdrafts at scales comparable to those of giant cells, a very likely
signature of interactions between large-scale convection and rotation
such as described in Sect.~\ref{obsrot}. Looking at the spectrum of
these simulations though, it is clear that supergranulation scales are
still located close to the dissipative range.

\begin{figure}[t]
(a)\\
\centerline{\hspace{0.1\linewidth}\includegraphics[width=0.45\linewidth]{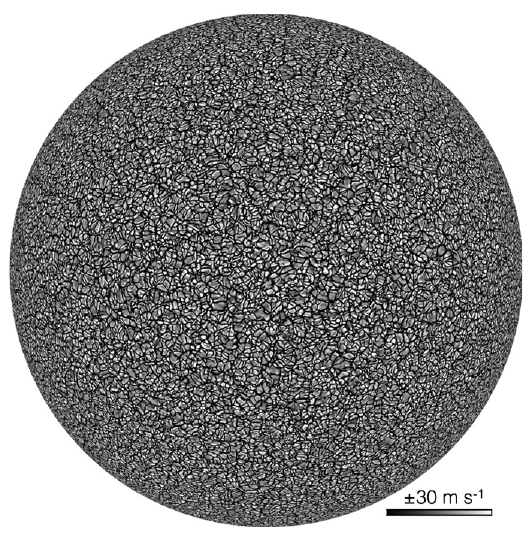}}\\
(b)\\
\centerline{\includegraphics[width=0.46\linewidth]{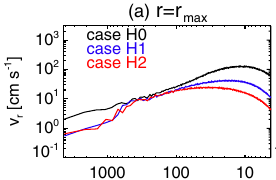}}\\
\centerline{\hspace{1cm}Horizontal scale (Mm)}\\
(c)\\
\centerline{\includegraphics[width=0.46\linewidth]{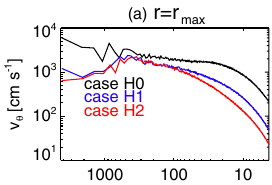}}\\
\centerline{\hspace{1cm}Horizontal scale (Mm)}
\caption{
(a) Radial velocity, (b) radial (vertical) velocity spectrum and
(c) latitudinal (horizontal) velocity spectrum at $r/R_\odot=0.99$
(case H0 in black) in global simulations \citep{hotta14}
(cases H1 and H2 only extend up to $r/R_\odot=0.96$ and have
  smaller density contrasts). The qualitative behaviour of both
spectra is very similar to the surface
radial and spheroidal spectra derived from observations by
\cite{rincon_etal17} (black and blue in
  Fig.~\ref{figure:spectra}(b)).\label{figure:hotta}}
\end{figure}

Even at the highest resolutions that can be achieved today in global
simulations, not all the relevant dynamical range required to quantitatively
address the problem of the dynamics in the supergranulation to
subgranulation range can be simulated adequately. However, there has
been some qualitative progress in that direction recently. A new set
of numerical
simulations of non-rotating convection by \cite{hotta14} with a grid
resolution of just a few Mm appears to resolve vigorous convective
dynamics at scales smaller than supergranulation. Some results appear
to be in good qualitative agreement with observations
(Fig.~\ref{figure:hotta}). For instance, the kinetic energy
spectra of the different flow components close to the surface appear
to follow the exact same trends as the observational trends reported
by \cite{rincon_etal17}, with the horizontal velocity spectrum peaking at
large scales and the radial velocity spectrum increases monotonically
down to the grid scale, with just a kink at the peak scale of
horizontal motions. Convective motions in these simulations extend
down to depths of at least $0.2\,R_\odot$, with the radial velocity progressively
increasing and peaking at a similar scale as horizontal motions at
increasing depths. The results also strongly suggest that the very
strong anisotropy of supergranulation peak-scale flows observed at the
surface is a surface effect shaped by the strong subsurface pressure
and  density gradients, and is somehow only the tip of the subsurface
convection iceberg.

An important difference with observations in the simulations of
\cite{hotta14}, though, is that the peak scale of horizontal motions
is a few hundred Mm, significantly larger than that of
supergranulation. We will soon encounter this scale-mismatch issue
again in local simulations, and will subsequently discuss different
physical factors possibly affecting the peak scale of the spectrum.

\subsubsection{Large-scale turbulent convection in local cartesian simulations\label{simlargelocal}}
Large-scale local cartesian simulations are also limited in terms of
dynamical regimes, but not quite as much as global ones when it comes
to the study of the dynamics at scales comparable to or smaller than
supergranulation. On current supercomputers, this family of models can
typically describe the nonlinear dynamics from scales slightly larger
than supergranulation scales (100~Mm) to subgranulation scales
(typically 10 to 100~km).

\begin{figure}[htbp]
\centerline{\hbox to -.7cm{}\includegraphics[width=0.43\linewidth]{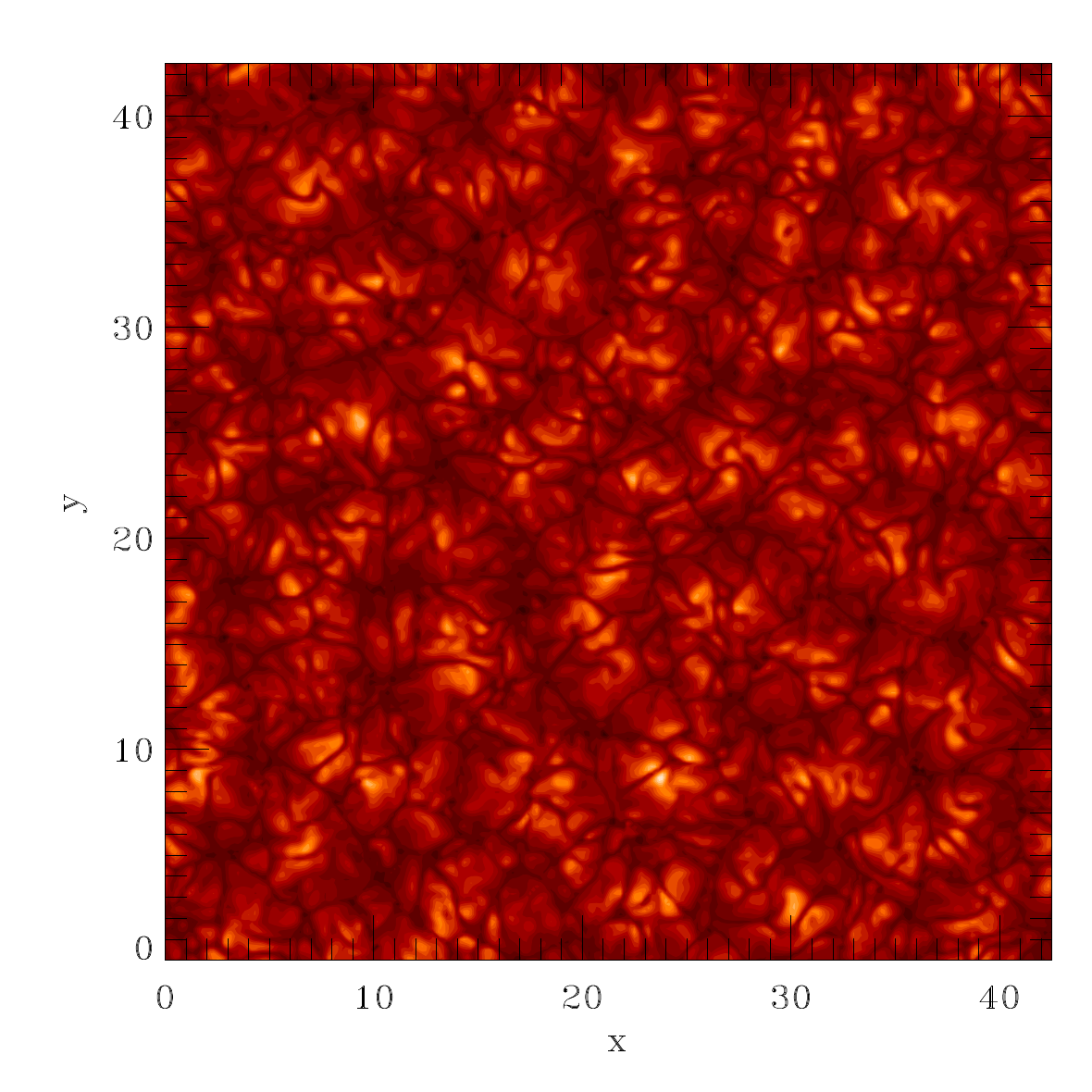}\hbox 
to -0.2cm{}
\includegraphics[width=0.43\linewidth]{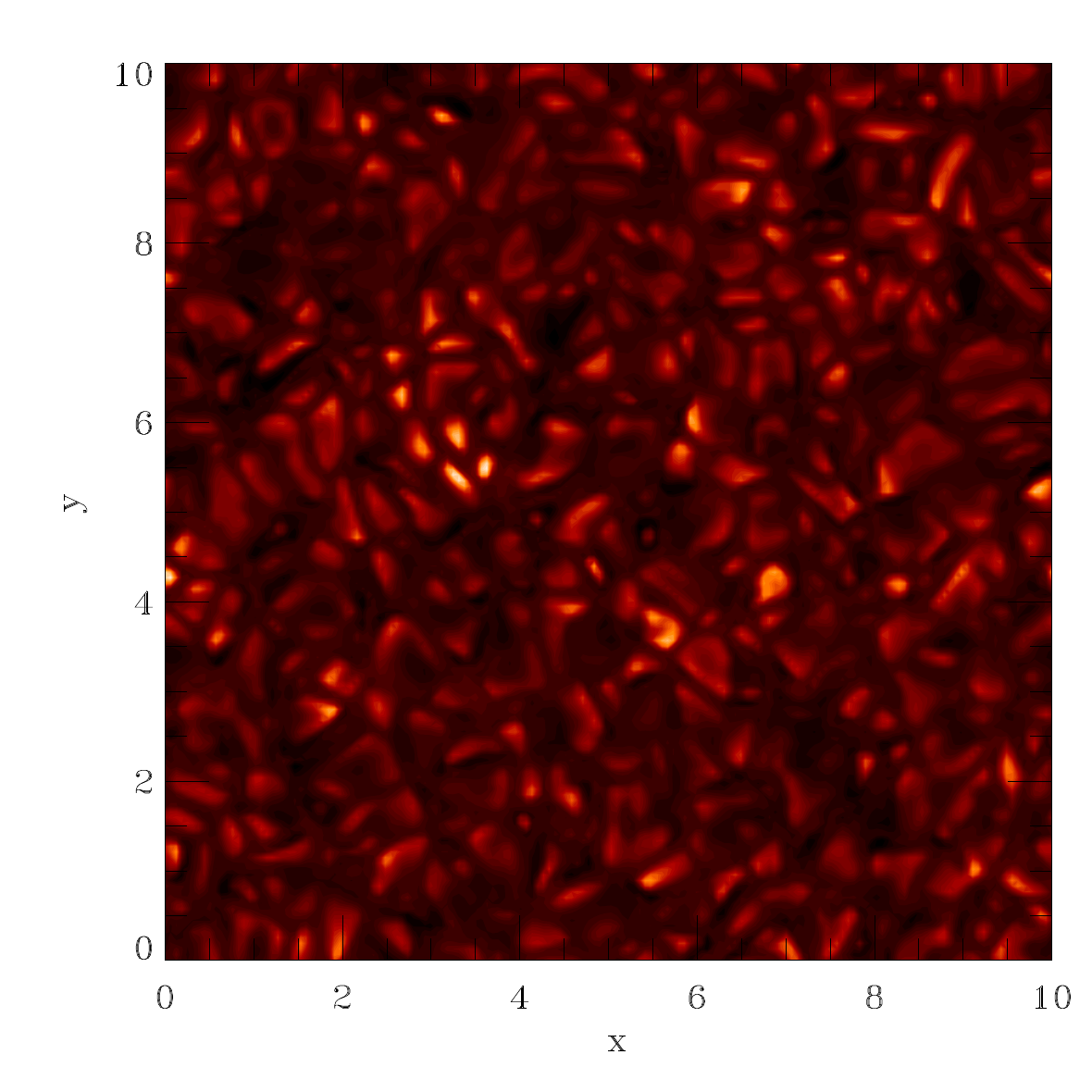}}
\vspace{-0.5cm}
\centerline{\hbox to -.7cm{}\includegraphics[width=0.43\linewidth]{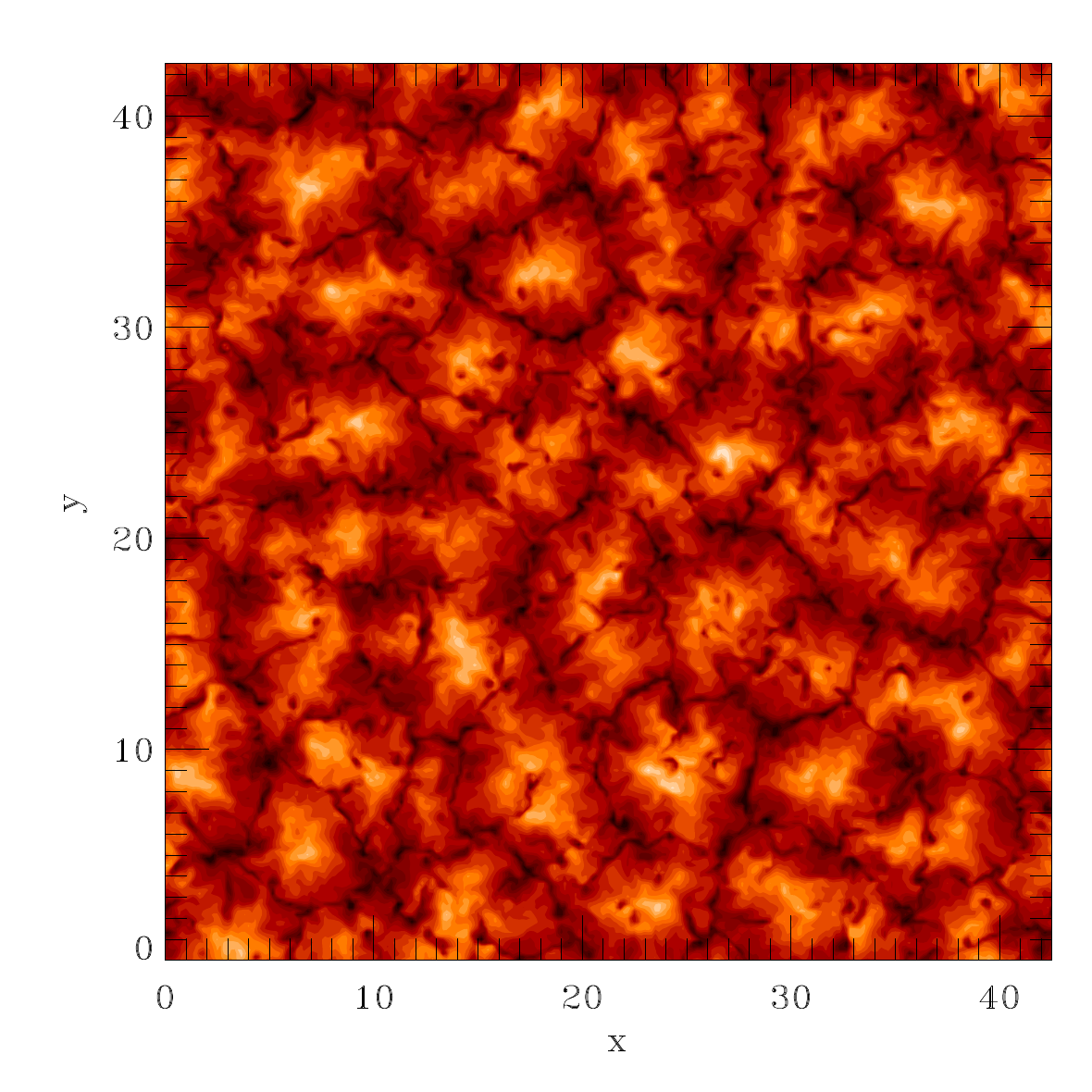}\hbox
  to -0.2cm{}
\includegraphics[width=0.43\linewidth]{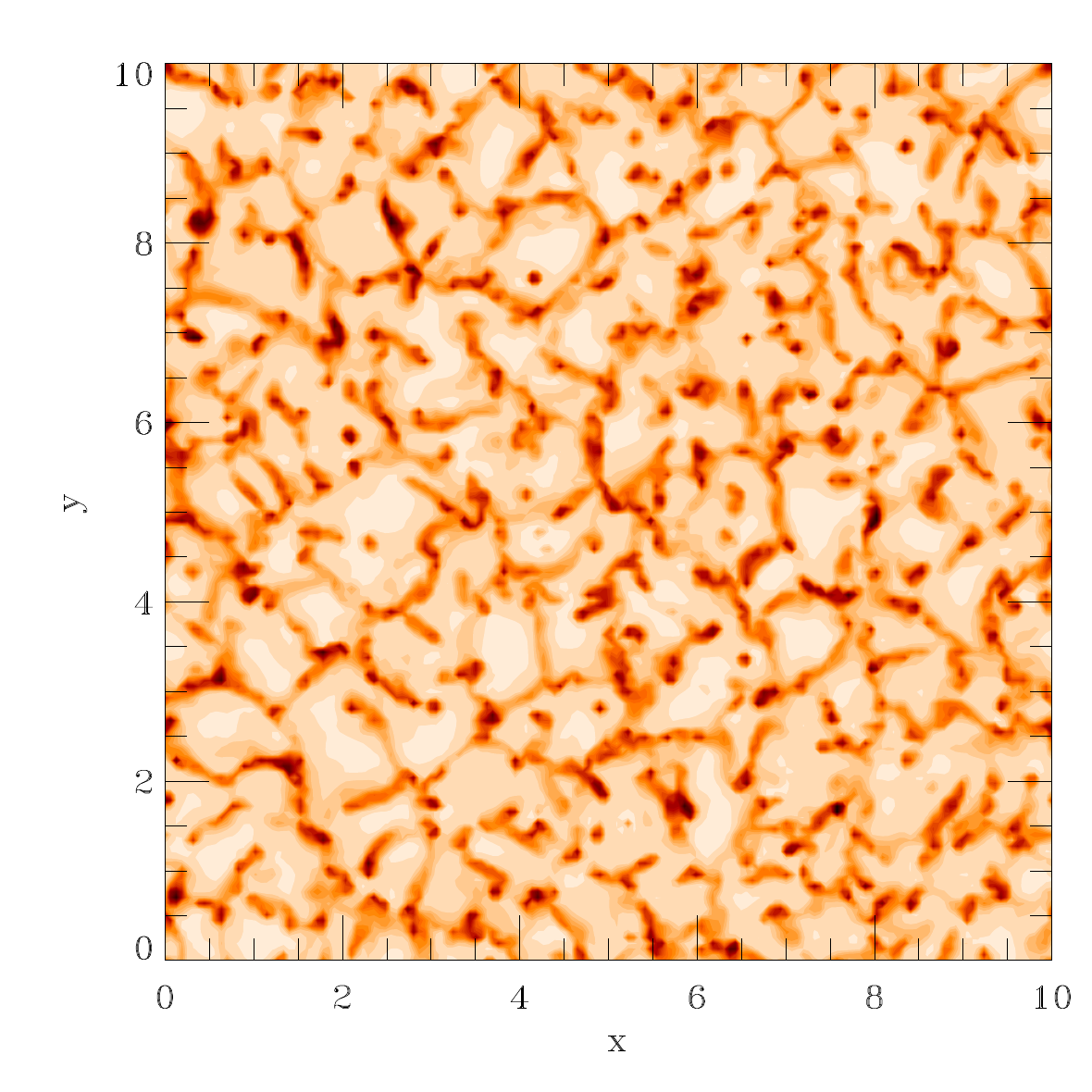}}
\vspace{-0.5cm}
\centerline{\hbox to -.7cm{}\includegraphics[width=0.43\linewidth]{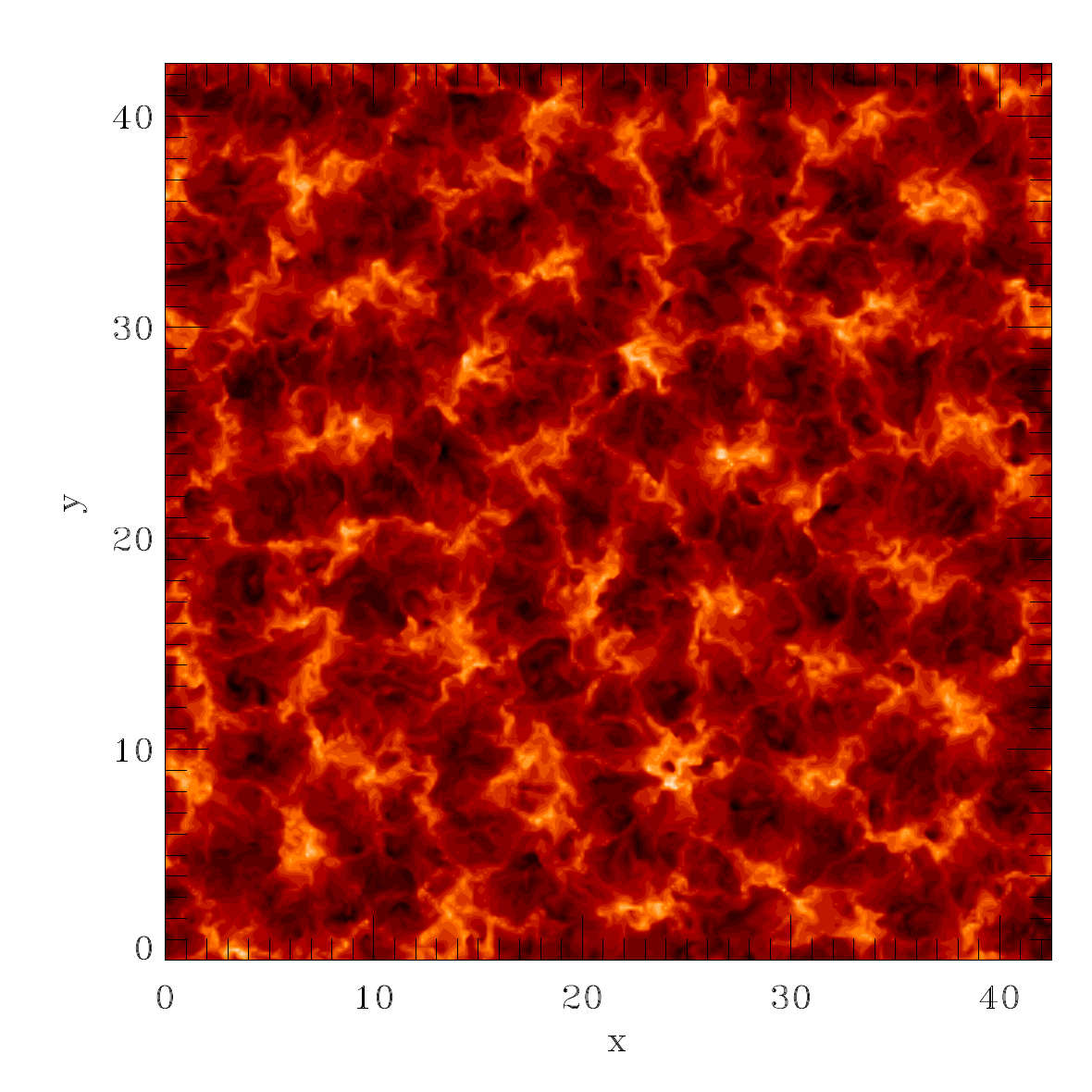}\hbox
  to -0.2cm{}
\includegraphics[width=0.43\linewidth]{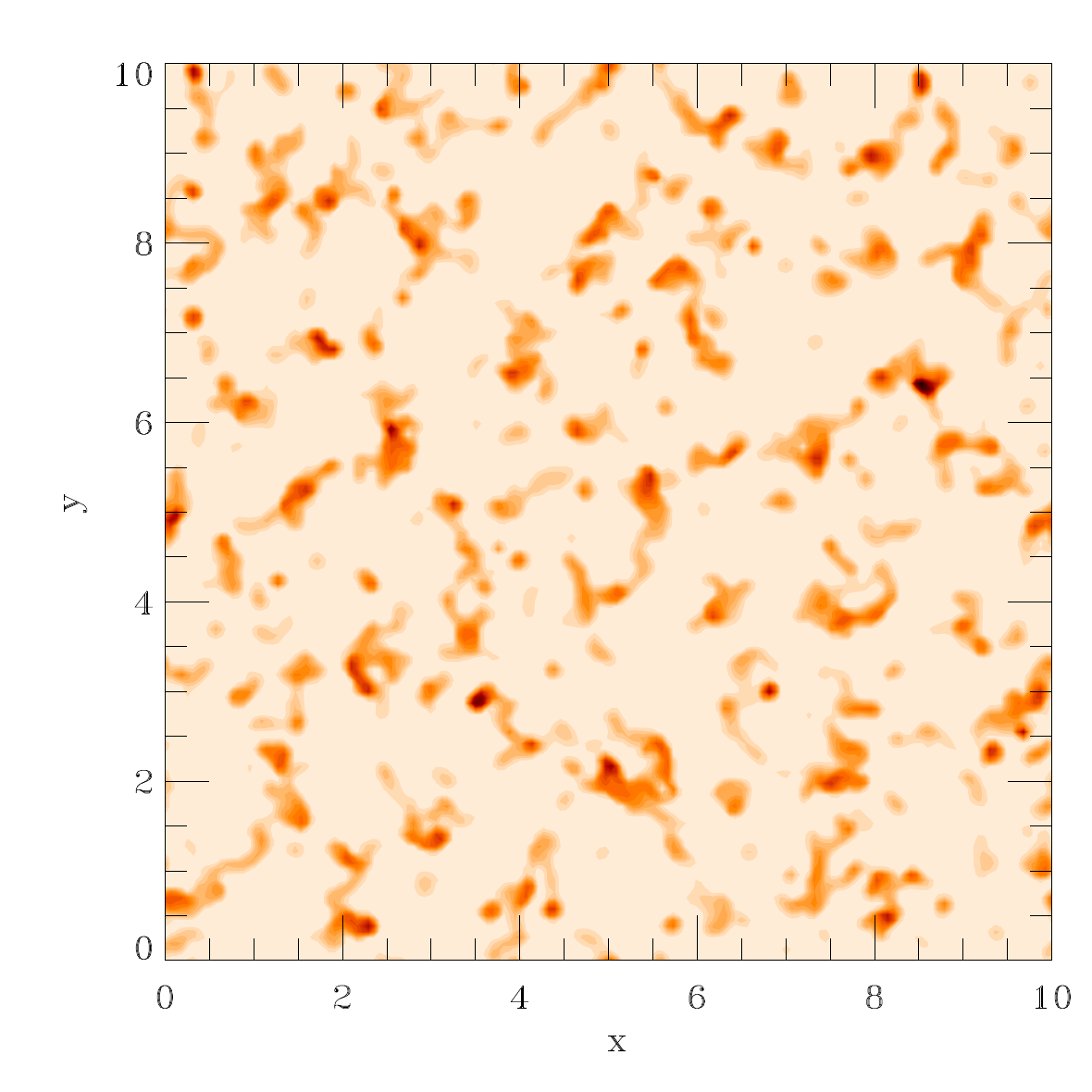}}
  \caption{Comparison between horizontal temperature maps in
    an idealized simulation of large-scale compressible convection in a
    stratified polytropic atmosphere (left, aspect ratio 42, see
    \cite{rincon05} for details) and horizontal temperature
    maps in a realistic simulation of large-scale solar-like
    convection (right, aspect ratio 10, see \cite{rieutord02} for details).
    Top: $z=0.99\,d$ and at optical depth $\tau=1$ respectively (surface).
    Middle: half-depth of the numerical domain. Bottom: bottom of the
    numerical domain. The emergence of the granulation pattern in the
    surface layers is clearly visible in both types of simulations, 
    on top of a larger-scale mesoscale dynamics extending down to deeper layers.}
  \label{figure:ideal}
\end{figure}

A strong feature of all three-dimensional, large-scale idealized
simulations using bottom and top wall boundaries is a pattern of
vigorous large-scale dynamical circulations, topped by a
smaller granulation-like pattern in the surface thermal
boundary layer. This is shown in Fig.~\ref{figure:ideal} (left).
These circulations are very likely the counterpart in large-scale
simulations of the thermal winds of aspect ratio unity seen in
simulations and experiments (Sect.~\ref{RB}). They were quickly
dubbed ``mesoscale structures'' in the solar physics context when they were
first reported, because their typical scale was larger than the scale of the
granulation pattern, yet their peak size was not quite as large as
that of supergranulation (relative to granulation).  This was also a
period when the question of the existence of mesogranulation
(Sect.~\ref{obsmeso}) as a separate physical phenomenon was widely
debated. As we will shortly see, it now seems increasingly clear that
these structures are to a large extent the counterpart of
supergranulation in the particular regime and geometry of these
simulations. These dynamical structures were first reported  by
\cite{cattaneo01} in large-scale Boussinesq simulations with an aspect ratio 
up to 20 and a Rayleigh number
$5\times 10^5$ (roughly 1000 times supercritical),  and were also
later observed in several other Boussinesq studies
\citep{hartlep03,parodi04,vharden08,pandey18}. 
Their typical correlation time in large-aspect ratio simulations is
much longer than the typical turnover time in the granulation boundary
layer, and their kinetic energy is also much larger than that contained in the
superficial granulation-scale motions. \cite{cattaneo01}, in the
spirit of the turbulence concepts described in
Sect.~\ref{theoryLSinst}, speculated that they may be the result of
a nonlinear inverse cascade or large-scale secondary instability of smaller-scale convective flows.

\begin{figure}[htbp]
\centerline{\includegraphics[width=\linewidth]{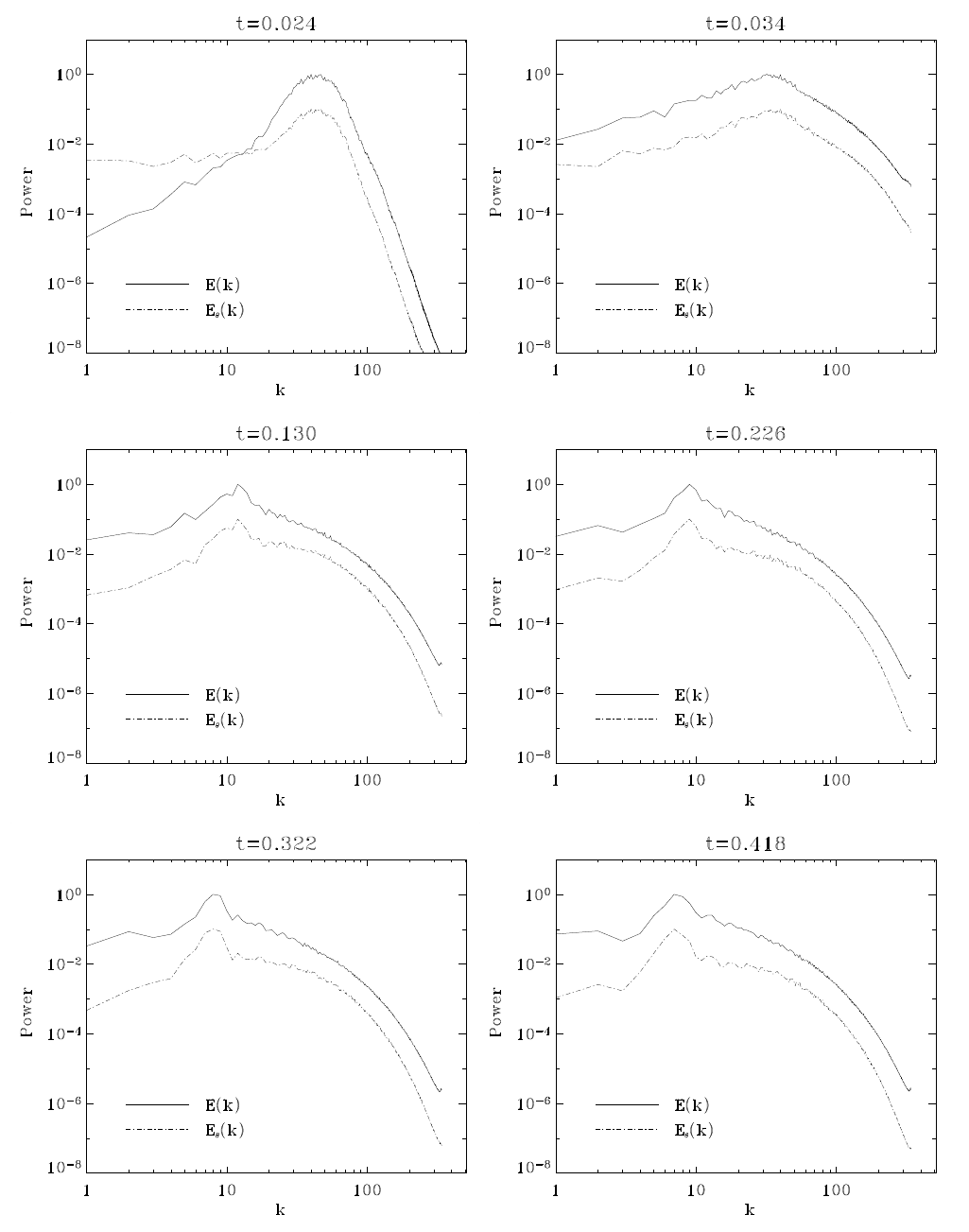}}
  \caption{From left to right and top to bottom: dynamical evolution
    of the spectra of temperature fluctuations $E_\theta(k)$ and
    kinetic energy fluctuations $E(k$) as a function of time and
    integer horizontal wave number $k$  and time $t$ (measured in
    vertical thermal diffusion units)  in the large-aspect ratio
    idealized simulation of turbulent convection with mild density
    stratification shown in Fig.~\ref{figure:ideal} (left),
    $Ra=3\times 10^5$ and $Pr=0.3$. $k=1$ corresponds to the
    horizontal size of the domain and $k=42$ to the vertical size of
    the domain. The peak scale of the spectra ($k\simeq 7$ towards the
    end of the simulation, much larger than the vertical size of
    the domain) correspond to the large-scale thermal structures
    visible in Fig.~\ref{figure:ideal} (extracted from
    \cite{rincon04}, Chap.~4.3, p.~85).}
  \label{figure:rincon04}
\end{figure}

\begin{figure}[htb]
\centerline{\includegraphics[width=0.8\linewidth]{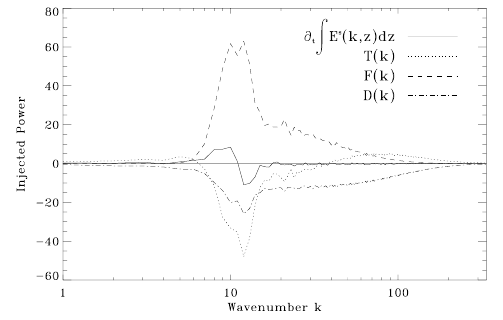}}
  \caption{Spectral-space energy budgets as a function of integer
    horizontal wavenumber $k$ in the large-aspect ratio simulations
    of turbulent convection described in Fig.~\ref{figure:ideal}
    (left) and
    Fig.~\ref{figure:rincon04}. Here, $F(k)$ is the buoyancy
    forcing term, $T(k)$ is the nonlinear transfer (cascade) term, 
    and $D(k)$ is the dissipation term \citep{rincon05}. The dominant
   balance in the peak injection range is between the positive
   buoyancy forcing term and the negative nonlinear transfer term,
   indicating that the large-scale dissipative structures 
   seen in Fig.~\ref{figure:ideal} (left) are powered
   by buoyancy. This injected power is dissipated by turbulent
   viscous dissipation mediated through spectral space by
   smaller-scale turbulent convective motions. \label{figure:rincon05}}
\end{figure}

The nonlinear dynamics and origin of these structures was investigated
in more detail by \cite{rincon05}  using fully compressible idealized
simulations with rigid walls, $Ra=3\times 10^5$ and a modest density
stratification, but with a very wide aspect ratio $\sim 42$. The main
conclusion of the analysis was that these large-scale flows dominate
the kinetic energy spectrum and are effectively driven by thermal
buoyancy. The  organization of such powerful flows at such scales was
found to be the result of a dynamical interplay between the linear
convection injection process and turbulent cascade and transport
processes. What happens is the following \citep{rincon04,rincon05}:
there is initially a broadband spectrum of linearly
unstable convection modes ($\sim 100$) in the simulation, but in the
early stages kinetic energy growth is much faster at the most
unstable scales of the system, whose horizontal scale is comparable to
the vertical scale of the system. This is of course expected from linear
analysis. Once in the nonlinear regime, however, the injection
of kinetic energy through thermal buoyancy is observed to continuously
drift to larger horizontal scales, with nonlinear interactions
cascading down the injected power down to smaller scales. This 
dynamics is illustrated in Fig.~\ref{figure:rincon04} and 
Fig.~\ref{figure:rincon05}. Physically, the smaller-scale
turbulence associated with the early dynamical saturation of the most
unstable modes acts as a
turbulent diffusion for the still-developing larger-scale
structures. Ultimately, the dominant scale appears to be set by a
balance between the rate of energy injection by buoyancy (set by the
Rayleigh number) and the rate of turbulent dissipation associated with
all the  saturated smaller-scale modes. Interestingly, both
laminar ``quasi-linear'' convection and nonlinear
interactions arguments  described in Sect.~\ref{theory} are 
phenomenologically relevant to this detailed numerical analysis,
albeit not in a straightforward way. In particular, while nonlinear
interactions play a big role in the dynamics, the large-scale
structures are definitely not driven by them and are, therefore, not 
due to inverse cascading or large-scale nonlinear instability.

As striking as it is in idealized simulations, the dynamics described
above, including the observation of a kinetic energy peak at scales
much larger than the most unstable linear scales, was much
less obvious in the first corresponding largish-aspect ratio realistic
simulations. An increase of the typical horizontal size of convective
structures with depth was for instance observed in aspect ratio 10
``realistic'' hydrodynamic simulations  by \cite{rieutord02}
(Fig.~\ref{figure:ideal} (right)), but no strong evidence was
reported  for particularly vigorous flows at scales larger than that
of granulation close to the surface (but only a spectrum of light
intensity at the surface, peaking at granulation scale, was
documented). Several possible explanations have been put
forward for this seeming discrepancy: one is that the small-scale dynamics
is not vigorous enough in these simulations, or is simply sufficiently
different given the presence of radiative transfer that the dynamics
is somewhat different from that in idealized simulations. Another
possibility is that boundary conditions play a key role in the
dynamical scale selection process, as most realistic simulations have an
open bottom boundary condition
and a strongly stratified atmosphere. It was notably 
pointed out by \cite{nordlund94} 
that using ``wall-type'' boundary conditions, as is standard in
idealized simulations, can significantly alter the shape of the
convective pattern, because walls allow for a return flow after
convective plumes smash down at the bottom. It may also be
that particular ``idealized'' (!) implementations of open bottom
boundary conditions used in most realistic (!) simulations
artificially suppress or quench large-scale convective motions that
would be present in the general case.

\subsubsection{State-of-the-art local hydrodynamic cartesian simulations\label{simlargelocal2}}
The first realistic numerical simulations including supergranulation scales
are due to \cite{benson06}, \cite{georgobiani07} and
\cite{stein09}. The latter used a 96~Mm wide and 20~Mm deep
three-dimensional numerical box but, just like
\cite{rieutord02}, they found a monotonic smooth increase
of the size of convective structures with depth, and no or very little
power enhancement at supergranulation scales in the surface power
spectrum. Similarly to \cite{spruit90}, they subsequently argued
that there was no reason why a particular scale should pop-up in the
continuum of scales present in the simulation (see \cite{nordlund09}
and \cite{georgobiani07} for representations of the power spectra of the
simulations). A gradual monotonic increase of the convection scale
with depth was also reported in an independent numerical study
by \cite{ustyugov08} in a 60~Mm wide and 20~Mm deep three-dimensional box.
The physics of ionisation of helium and hydrogen were included in the
model of \cite{stein09}, which allowed them to test for
the first time the first theoretical proposal for the origin of the
supergranulation by \cite{SL64} described in Sect.~\ref{theoryconv}.
Considering the gradual large-scale decrease of energy in the power
spectrum of their simulations, they concluded that the existence of
recombination layers of ionised elements cannot by itself explain
supergranulation.  This conclusion was further confirmed by
\cite{lord_etal14} using simulations in a box of size
$196\times 196 \times 49$~Mm$^3$ (the latter paper contains an
interesting discussion of the slightly surprising actual effects of
ionisation in the problem).

\begin{figure}[ht]
\centerline{\includegraphics[width=0.7\linewidth]{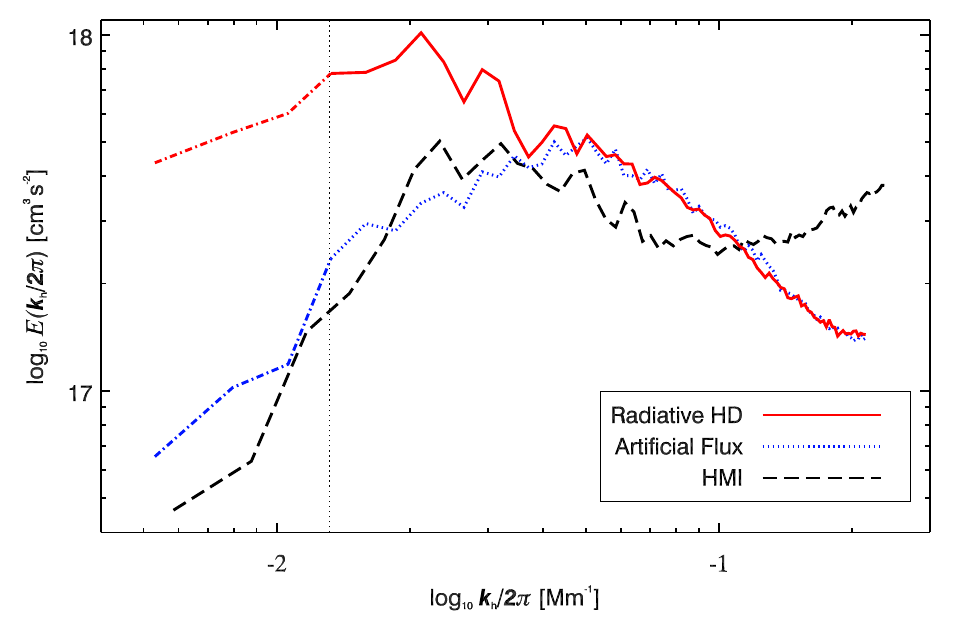}}
\caption{
Kinetic energy spectra in the radiative hydrodynamics simulations of
\cite{lord_etal14} (solid red line). The spectrum obtained by
reducing artificially the velocities in the deep layers is shown in
dotted blue line. The vertical line marks the scale of the largest
modes driven in the simulation. The black dashed line corresponds to
solar spectra obtained from SDO/HMI data and coherent structure
tracking (black dashed line).}
\label{figlord}
\end{figure}

The latest realistic local simulations of \cite{lord_etal14} appear to
show a clear excess of kinetic energy at large scales (see
Fig.~\ref{figlord}, red line), just like idealized simulations. The
reasons for this seemingly new convergence of the large-scale dynamical
properties of convection in realistic and idealized simulations
(assuming there was ever a big difference between the two) have not
yet been clearly discussed to the best of our
knowledge. Interestingly, a lot of the discussion in
\cite{lord_etal14}, as well as in subsequent studies
\citep[e.g.,][]{featherstone_hindman16,karak18}, is now focused on
there being too much energy at large scales in simulations in
comparison to the solar case.
They notably point out that much of the heat flux is
carried by the largest-scale flows in their simulations, which
demonstrate that these flows are strongly buoyantly driven, just like
the large-scale dynamics in idealized simulations described earlier.

Overall, it seems like all realistic and global simulations currently
overestimate the peak scale of convection and convective velocities at
large-scales in comparison to the solar case, while idealized
simulations such as that of \cite{cattaneo01} and \cite{rincon05}
underestimate them. Interestingly, \cite{lord_etal14} manage to
reproduce the Sun's kinetic energy spectrum by
using an artificial mechanism to carry the whole solar flux below a
critical depth (here 10~Mm). The consequence of this prescription is
to suppress the flux-transporting motions at the largest scales of the
box, resulting in the emergence of a peak scale in the  kinetic
energy spectrum comparable to that of supergranulation. This result
strongly suggests that understanding the asymptotic subtleties of
deeper-scale convection in the Sun is very important for a quantitative theory
of supergranulation. \cite{cossette_rast16} \citep[see also][]{kessar18}
have recently followed up on this idea with a series of idealized
convection simulations in a strongly stratified atmosphere composed of
different, prescribed superadiabatic surface layers matched to an
adiabatic interior. They find that  the vertical scale of the entropy
jump at the surface, which sets the entropy deficit of sinking buoyant
plumes (or ``entropy rain'', see \cite{brandenburg16}), has a major
effect on the peak ``supergranulation'' scale of the convection spectrum. 
We will frame these results in the context of a larger theoretical discussion 
in Sect.~\ref{discussion}.

While some of the results reviewed above suggest a strong,
purely non-rotating, hydrodynamic dependence of the supergranulation
scale on the modelling of the thermodynamic structure and heat flux of
the whole system, rotation and magnetic fields have also long been
thought to play a role in the supergranulation problem (see
Sect.~\ref{theory}). What do simulations tell us about these
interactions?

\subsubsection{Simulations with rotation}
As explained in Sect.~\ref{obsrot}, there is an increasing sense 
that a rotational connexion between the subsurface shear layer and
large-scale convection at scales comparable to or larger than
supergranulation exists in the Sun. Only a few local simulations have
specifically addressed the 
issue of the interactions between supergranulation and rotation. 
In an early attempt at simulating this problem in the Boussinesq
approximation in a numerical box elongated in the horizontal
direction, \cite{hathaway82} found that the presence of a tilted
rotation axis and generate a subsurface shear layer. The local
dynamics of angular momentum transport in turbulent convection has
since  been studied at much higher numerical resolution
\citep[e.g.,][]{brummell96,brummell98,kapyla04,brandenburg07} and,
while the focus of these papers is not specifically on
supergranulation, \cite{brandenburg07} argued that the travelling-wave
properties of supergranulation (Sect.~\ref{obsrot}) could be due to
the radial subsurface shear. In a related study, \cite{egorov04}
reported a good agreement between  the divergence-vorticity
correlations obtained from simulations of rotating convection and
those inferred from observations of the supergranulation flow field.

Based on new high-resolution global spherical simulations dedicated to
the supergranulation problem, \cite{featherstone_hindman16} have
recently argued that Coriolis
 effects may effectively quench large-scale, low Rossby number
 convection in the SCZ, resulting in a reduction of both large-scale
 heat transport and of the peak scale of convection at the surface, as
 observed in the simulations of \cite{lord_etal14} with artificial
 heat-flux reduction. This effect might also be indirectly connected
 to the non-rotating scale-selection effect put forward by
 \cite{cossette_rast16}, in the sense that the effects of rotation
 on convective heat fluxes can also indirectly affect the internal 
 and subsurface thermal structure on which the peak scale of the
 convection spectrum depends.

\subsubsection{MHD simulations\label{simmhd}}
Just like rotation, magnetic fields have long been recognized to
play a non-negligible dynamical role in solar convection. Many (mostly
local) large-scale local simulations have now been devoted to the
study of MHD convection at scales comparable to supergranulation and
to the process of network formation. These can be subdivided into
magnetoconvection simulations in an imposed mean magnetic field or
with a magnetic flux introduced  ``by hand'' at the beginning of the
run, and turbulent dynamo simulations, in which the magnetic field is
spontaneously generated by the turbulent convection flow starting
from an infinitesimal seed field. 

\begin{figure}[ht]
\vspace{0.5cm}

\centerline{\hspace{5pt}
(a)
\hspace{-5pt}
\parbox[t]{0.45\linewidth}{\vspace{-20pt}
\includegraphics[width=\linewidth]{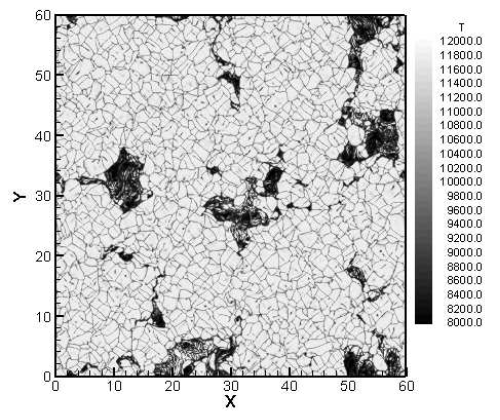}}
\hspace{5pt}
(b)
\hspace{-5pt}
\parbox[t]{0.45\linewidth}{\vspace{-20pt}
\includegraphics[width=\linewidth]{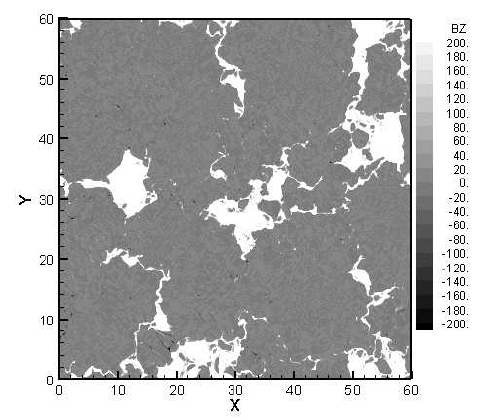}}}
\caption{
Horizontal maps of (a) temperature and (b) vertical magnetic field
fluctuations in the surface layers of local realistic simulations of
large-scale MHD convection \citep[image credits:][]{ustyugov09}.}
\label{figure:ustyugov}
\end{figure}

In the absence of a mean field threading the convection layer,
small-scale disordered magnetic fields consistently generated by
small-scale turbulent dynamo action get organized into a network
of ribbons and point-like magnetic flux concentrations, much like in
the quiet Sun \citep{cattaneo99,emonet2001,vogler07,favier14,rempel14, danilovic16}.
The typical scale at which this ``network'' forms in simulations
corresponds to that of the large-scale energetic motions
described earlier. A similar
phenomenology is observed in magnetoconvection
simulations in a weak mean (vertical or horizontal) field
\citep{ustyugov06,ustyugov08,ustyugov09,stein09b}. This
process is illustrated in Fig.~\ref{figure:ustyugov}, which shows a
large-scale realistic simulation of magnetoconvection with an imposed
50~G vertical field \citep{ustyugov09}. 
Only in the presence of a strong mean vertical field is
the peak horizontal scale of these turbulent convective motions 
strongly constrained and reduced by magnetic tension
\citep{tao98,cattaneo03}. The field strengths required  are generally
stronger than observed in most of the quiet Sun, but this kind of
effect may be particularly  noticeable in polar regions
\citep[e.g][]{tsuneta08}.

While an essentially passive magnetic field phenomenology provides 
the simplest explanation for the observed correlation between
supergranulation and the solar magnetic network, we argued in 
Sect.~\ref{magnetoconv} that a distribution of
spatially-intermittent magnetic fields organized into strongly
inhomogeneous structures (such as observed both in simulations and
observations) may dynamically affect convection. Numerical evidence 
for this remains limited, but \cite{ustyugov09} notably found
that local concentrations of strong magnetic flux seem to play an
important role in the scale-selection process in his simulations of
network formation with a weak but uniform mean field and may,
therefore, exert a significant dynamical feedback on supergranulation.
The simulation of \cite{hotta15} also provides an
 example on how small-scale turbulent field may affect convection and
  entropy mixing throughout the convection zone.
Other possible MHD effects briefly described in
Sect.~\ref{magnetoconv} and  in the discussion section of the first
published edition of this review \citep{rieutord10} remain more
speculative and have not been conclusively detected in simulations so far.

\section{Discussion and outlook}\label{discussion}
\begin{figure}[ht]
\centerline{\includegraphics[width=1.\linewidth]{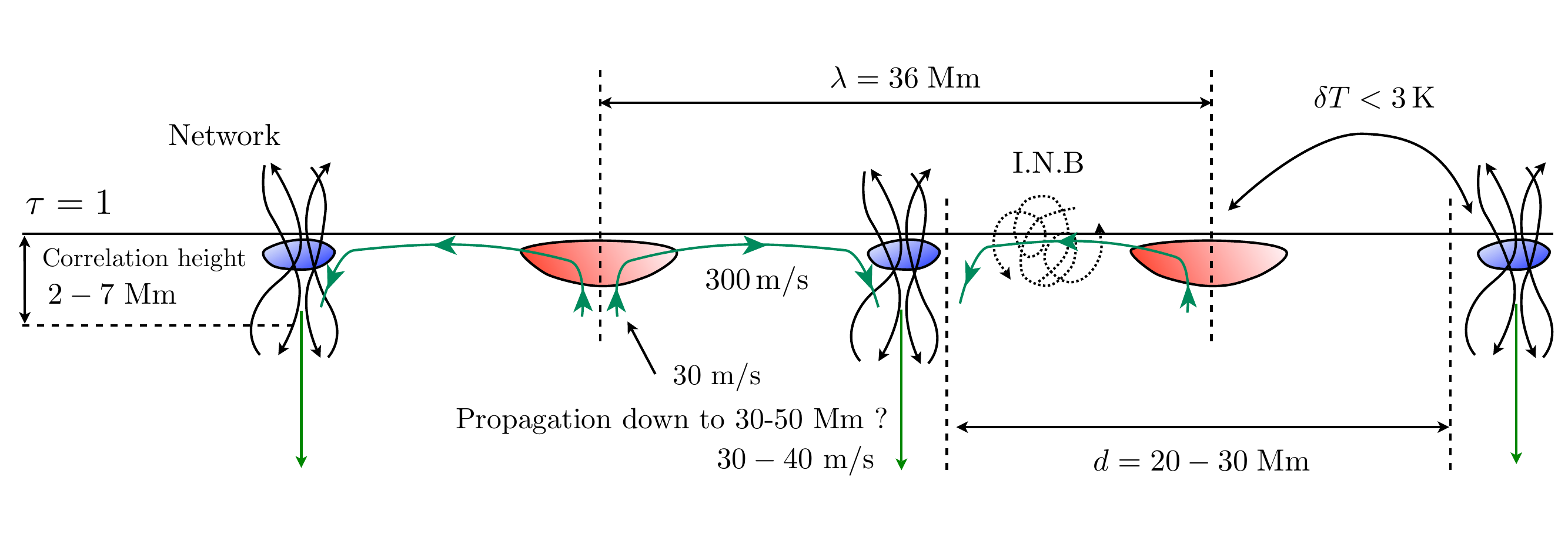}}
\caption{A schematic view of the supergranulation phenomenon as currently
  constrained by observations. $\lambda$ is the scale where the
  horizontal kinetic energy spectral density is maximum. $d$ is the
  diameter of ``coherent structures'' (supergranules). The red and
  blue patches depict the warm and cold regions of the flow. I.N.B
  denotes the internetwork magnetic field (the dichotomy between
  network and internetwork fields is probably not quite as clear as
  indicated in this drawing). The vertical structure and extent of the
  dynamics remains one of the main unknowns in this cartoon.}
\label{figure:super_pict_new}
\end{figure}

The physical and dynamical complexity of the supergranulation
problem is quite extraordinary: vigorous turbulent small-scale and
large-scale flows in a strongly stratified atmosphere, ionisation
physics, rotation, shear and tortuous magnetic-field geometries at
all observable scales. It is,
therefore, perhaps not suprising that both observers and theoreticians
have struggled for many decades after the initial historical
observational discoveries to identify and describe the essential processes
underlying this phenomenon. As shown in the previous sections,
research on the problem as strongly intensified in the last fifteen
years. This progress has largely been driven by massive
improvements in observational capacities and analysis techniques, as
well as in computing power. Before we discuss the emergent dynamical
picture and outline a few desirable and expected directions of future
research, let us first offer a compact recap of the current
observational knowledge on the problem against which theoretical and
numerical models have to be confronted. A pictorial representation of
the results is also provided in Fig.~\ref{figure:super_pict_new}.

\subsection{Summary of observations}\label{obsconc}
Supergranulation is a pattern of vigorous, mostly cellular-like flows
in the quiet solar photosphere and subphotospheric layers, detected 
through Doppler measurements, structure tracking, and helioseismic
analyses. The typical range of horizontal length scales of
supergranulation flows at the photosphere is 20--70~Mm, with a peak
at 36~Mm corresponding to the peak scale of the horizontal kinetic energy
spectrum of solar surface flows. This result is clearly established in
Fig.~\ref{figure:spectra} of Sect.~\ref{obs}.  Surface flows in
the meso to supergranulation range of scales are
strongly anisotro\-pic. A clear distinction must be made between the
horizontal and vertical components of the flow at the photospheric
level the horizontal, cell-like component of the flow (300--400~m/s) is
more than ten times more vigorous than the vertical upflow and
downflow components (20--30~m/s) detected respectively it the center
and periphery of supergranules, and the kinetic energy spectra of each
component look very different. The spectrum of the vertical component,
unlike the horizontal spectrum, may have a kink, but does not peak at
supergranulation scales. Instead, it increases down to granulation
scales. 

These different observations strongly suggest that
supergranulation-scale dynamics is the dynamical manifestation of a
vigorous turbulent energy-injection mechanism, which has many (but not
unambiguously all) of the observational hallmarks of thermal
convection. Different observations suggest that supergranules are
slightly warmer at their centre, but with a temperature drop smaller
than 3~K at the surface, possibly larger below the surface. 
Our understanding of the structure of subsurface flows in the
supergranulation range remains a bit fuzzy, but several recent
measurements seem to point to the existence of  dynamics at scales
comparable to or larger than that of supergranulation down to the
bottom of the near-surface shear layer, albeit with a shallow vertical
scale-height of variation close to the surface, of the order of 2--7~Mm.

Different studies suggest that supergranulation is influenced
by the global solar rotation, and may itself play an important role in the
establishement of the near-surface shear layer of the SCZ. Finally,
flows in the meso to supergranulation-scale range are strongly
correlated with the solar magnetic network and have a strong influence
of the distribution and advection of small-scale
magnetic fields up to network scales. Whether supergranulation itself
is constrained by magnetic forces cannot be easily asserted with
observations, although a few observations suggests that the emergence
of active regions, and stronger fields in general, may dynamically
affect it.

\subsection{Physics and dynamical phenomenology of supergranulation}\label{numconc}
There has been ample progress in the last ten to fifteen
years in our understanding of the phenomenology of the dynamics
of the solar surface in the range of scales relevant to the
supergranulation problem thanks to numerical simulations. At the very
least, there now appears to be much more numerical evidence than ten
years ago that supergranulation is buoyantly-driven, and is in
fact the energetically dominant convection scale on the large-scale
side of the injection range of the photospheric convection
spectrum. Simulations also increasingly show that the detailed
dynamical picture is significantly more complex than the classical
laminar picture of convection described in Sect.~\ref{theory},
although interestingly enough, the nonlinear organization of the
dynamics in the turbulent regime (illustrated for instance by the
broadband convection spectra in Fig.~\ref{figure:spectra} and the
large-scale nonlinear dynamics in the simulations described in
Sect.~\ref{simlargelocal}) appears to result in large-scale,
vigorous coherent motions reminiscent of laminar convection. 

Brute force observational and numerical progress on the problem has
recently been accompanied with several potentially-testable theoretical
developments and arguments regarding the phenomenology of turbulence
and convection at scales larger than granulation
\citep{lord_etal14,cossette_rast16,featherstone_hindman16,rincon_etal17}.
It was recently argued by \cite{rincon_etal17} on the basis of
the strong anisotropy of photospheric flows in the supergranulation to
granulation range that an appropriate description of
convection dynamics at the surface in this range of scales requires a
generalization of the classical isotropic Bolgiano-Oboukhov theory of
turbulent convection \citep{bolgiano59,oboukhov59,bolgiano62}
to the regime $kH\ll 1$, where $k$ is the
horizontal wavenumber of fluctuations and $H$ is a typical scale
height (the distance between plates in the Rayleigh-Benard experiment,
or a thermodynamic scale height in the stratified problem). A tentative
generalized theory of this kind can be derived using three key dynamical
assumptions: a dominant dynamical balance between buoyancy forces and
inertial terms in the momentum equation (as diagnosed in idealized
simulations, see Sect.~\ref{simlargelocal}), a constant flux of thermal
variance in spectral space in a well-mixed, nearly adiabatic turbulent
convection layer and a typical ``frustrated'' vertical scale of
variations of fluctuations independent of their horizontal scale and
of the order of $H$. The first two assumptions are part of the
standard Bolgiano-Oboukhov phenomenology, but the latter is specific
to the anisotropic regime $kH\ll 1$. Crucially, the theory predicts that the
horizontal kinetic energy spectrum continues to increase at scales
larger than granulation and the Bolgiano scale, while the vertical
kinetic energy decreases with increasing scale, in broad agreement
with observations. This result provides a possible partial way out of the
problem of the mismatch between the supergranulation scale and the
Bolgiano injection convection scale raised
Sect.~\ref{supergranulationpuzzle}: the theory suggests that the
latter is just a lower bound on the scale of the injection range, and
that an anisotropic, buoyancy-driven nonlinear Bolgiano-like injection
regime is possible at horizontal scales larger than the typical
vertical scale height of the domain (this regime has never been
investigated in laboratory experiments to the best of our
knowledge). Finally, the theory also predicts an
increase of temperature fluctuations with increasing horizontal
scale. This is in relative tension with the relatively weak
photospheric light intensity contrasts measurements reported in this
range of scales at the solar surface, but is on the other hand quite
consistent with a variety of numerical results, notably those of
\cite{rincon05}, \cite{lord_etal14} and \cite{cossette_rast16}
discussed in Sect.~\ref{numerics}. 

These preliminary theoretical predictions, considered jointly with the 
observation of a maximum in the kinetic energy spectrum at scales much larger
than granulation, raise several key questions: what sets the scale of
this maximum dynamically? why is supergranulation so prominent as a
flow pattern but not as a temperature pattern at the photospheric
level, and what is the spectrum of thermal fluctuations below the
photosphere? Numerical simulations are slowly getting to a
place where a much better understanding of the interactions and balance
between different relevant linear and nonlinear dynamical processes in this
range of scale becomes possible and these questions can be adressed.
A closely related issue is to reconcile the peak scales and
  amplitudes of the observed solar convection spectrum (the actual
supergranulation scale) with the dynamics of either global or local
simulations. In particular, why does the solar dynamics appear
  to have so little power on large scales ?

In the light of the numerical results reviewed in
Sect.~\ref{numerics}, there are at least three remaining credible
possibilities as to what sets the peak scale of the convection
spectrum and supergranulation: the internal thermodynamic structure
and the magnitude and thickness of the entropy jump in the surface
thermal boundary layer, which are directly constrained by the
production of heat in the Sun \citep{cossette_rast16,rincon_etal17,kessar18},
the interaction between slow, large-scale convection and rotation
\citep{featherstone_hindman16}, and the dynamical interactions between
convection and magnetic fields \citep[][see also discussion in
\cite{rieutord10}]{ustyugov09,stein09b}. The question of the amplitude
of thermal fluctuations at supergranulation scales has 
recently been discussed by \cite{cossette_rast16} and \cite{rincon_etal17}. 

\subsection{Outlook}\label{outlook}
There are many reasons to be cautiously optimistic about future
breakthroughs on the supergranulation problem, and it seems
increasingly possible that a robust phenomenology and even perhaps a
consistent nonlinear theory of large-scale convection in the SCZ can
be constructed in the next decade.

From an observational point of view, there has been a strong revival
of supergranulation studies in the last ten years with the launch of
Hinode and more recently SDO, and important progress has been made
on the characterization of both surface and subsurface
dynamics. Particularly encouraging are the ongoing efforts to improve
the characterization of subsurface dynamics with localhelioseismology, and
the newly acquired capability to study the surface dynamics from a
global perspective with either tracking or Doppler techniques (or both)
thanks to the highly-sampled SDO/HMI data. More observational results
of this kind are expected in the forthcoming years. These should
hopefully be accompanied with a better convergence between different
techniques and research groups, thereby enriching and consolidating
the existing corpus of observational constraints. Among many other
things, a better heloiseismic characterization of subsurface thermal
fluctuations at supergranulation scales and smaller scales would be
extremely valuable for this problem, and so would be a better, less
controversial characterization of large-scale convection flows  in
subsurface layers and in the deeper SCZ. There is still a lot of
fuzziness and disagreement between different  groups on these questions
and more work is required to settle them. The good news is that only a
fraction of SDO capacities seems to have been exploited so far on this front.
The potential of SDO has also almost not been exploited to look
for new observational clues of a possible dynamical relationship
between large-scale flows and magnetic fields in the quiet Sun.
There seems to be a lot of room left for new discoveries on this front too.

In-depth numerical investigations of the different dynamical
scenarios described in the previous paragraph  at even
higher-resolutions are also almost certainly going to be
carried out in the forthcoming years, but making significant further
progress is going to be challenging. First of all, as
explained in Sect.~\ref{numerics},
the actual thermodynamic profiles and entropy jumps established by
turbulent mixing in simulations are less extreme than in the SCZ
because the simulated regimes are not quite as asymptotic as in the
Sun.
Finally, the dynamical influence of magnetic fields in the
problem is also not easy to understand considering the complex
geometry and potentially insidious effects of solar magnetic fields. It is not
even clear that all the aspects of the MHD problem have yet been
properly recognized \citep[see, e.g., recent discussion
by][]{karak18}. Some of these issues may be very difficult to
address numerically, given the difficulty to simulate MHD in the low
magnetic Prandtl regime typical of the SCZ
\citep{schekochihin07,vogler07,graham09,rempel14}.

\begin{figure}[htb]
\raggedright(a)\\
\centerline{\includegraphics[width=0.85\linewidth]{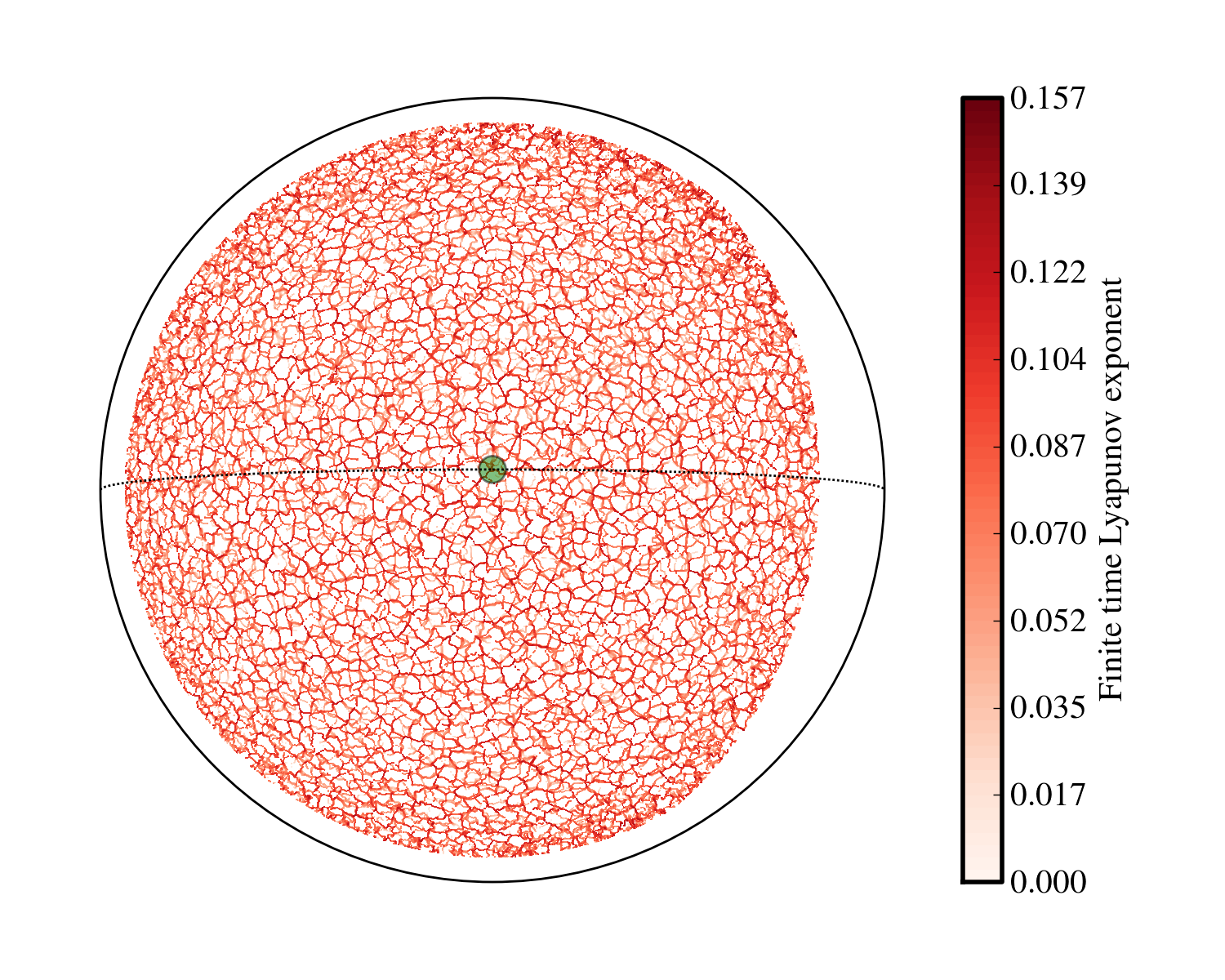}}
\raggedright(b)\\
\centerline{\hspace{-0.5cm}\includegraphics[width=0.6\linewidth]{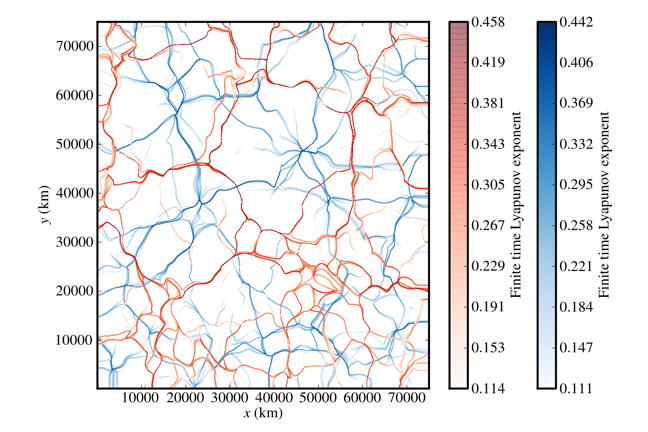}}
\caption{(a) Global distribution of 24-h-positive time FTLEs of solar
  surface flows (in inverse hour units) computed from horizontal velocity
  field maps derived from October 8, 2010 SDO data (one of the
  quietest periods of solar activity since the launch of SDO). The
  green circle diameter is 30~Mm. (b) Local distribution of
  FTLEs of solar surface flows computed from
  horizontal velocity field maps derived from August, 2007 Hinode
  data, (red: positive-time FTLEs, blue: negative-time FTLEs)
  (credits for both plots: F. Rincon  \& T. Roudier, so far unpublished).}
\label{figure:LCS}
\end{figure}

Progress on the supergranulation problem is not only interesting and
important in itself, but also from the perspective of understanding
the global solar dynamo, turbulent generation of solar magnetic
fields, and coronal heating. To emphasize this clearly and open the
subject for future discussions, we show in Fig.~\ref{figure:LCS}
solar surface distributions of finite time Lyapunov exponents (FTLE)
computed from recent Lagrangian tracer analyses of global and local maps of
horizontal photospheric velocity fieds derived from Hinode and SDO
observations using CST tracking. Such FTLE calculations, introduced in the
context of solar coronal heating by \cite{yeates12,chian14}, characterize
the Lagrangian transport properties of flows and make it possible to
image transport barriers, i.e., regions of
accumulation or rarefaction of passive tracers such as passive
magnetic fields, in the form of Lagrangian Coherent Structures
(LCS). These concepts have already found many applications
in other fields of physics such as oceanography and atmospheric
sciences \citep[e.g.,][]{lekien05,lehahn07,lekien10}. These two computations
reveal that the Sun is paved with supergranulation-scale lagrangian
coherent structures, and strikingly illustrate the Lagrangian process
of magnetic network formation. Based on this kind of analysis, it
is clear that supergranulation-scale
convection plays a major role in the global and local structuration
and dynamics of solar magnetic fields at the interface between the
solar interior and corona.

To conclude this review, let us note that solar surface convection is
one of the very few time and spatially-resolved examples of extremely
nonlinear dynamical fluid astrophysical phenomena. As we are increasingly
approaching a position where a detailed understanding and
characterization of this phenomenon seems possible, it is
certainly worth emphasizing that everything we can learn about 
it is likely to be strongly relevant and illuminating from a much
broader astrophysical and fundamental fluid dynamics perspective.

\begin{acknowledgements}
We would like to thank our many collaborators and colleagues, most
importantly Thierry Roudier, for sustaining our interest in the
supergranulation problem over the years, and for sharing their
insights on solar convection and magnetism with us. 
We are also grateful to the referees of the 2010 and 2018
  versions of this review for their detailed reading and comments on
  the long manuscript, and for pointing out several relevant references
  that we had overlooked.
\end{acknowledgements}



\bibliographystyle{spbasic}      
\bibliography{supergranulation_revised}   

\end{document}